\documentclass[a4paper,fleqn,usenatbib]{mnras}

\usepackage[T1]{fontenc}
\usepackage{ae,aecompl}
\usepackage{ulem}

\usepackage{graphicx}	
\usepackage{amsmath}	
\usepackage{amssymb}	

\usepackage{todonotes}

\usepackage{array}
\usepackage{booktabs}
\usepackage{color}
\usepackage{subfig}
\usepackage{verbatim}

\usepackage{upgreek}

\usepackage{times}

\newcommand{\hinv}{h^{-1}}
\newcommand{\beq}{\begin{equation}}
\newcommand{\eeq}{\end{equation}}

\newcommand{\nhat}{\hat{\textbf{n}}}

\newcommand{\mb}{\mathbf}

\newcommand{\HEALPIX}{{\textsc{Healpix}}}

\newcommand{\WMAPc}{{\slshape WMAP}}
\newcommand{\Planck}{{\slshape Planck~}}

\newcommand{\arcm}{\mathrm{arcmin}}

\newcommand{\Msol}{M_{\odot}}

\newcommand{\Mpc}{\mathrm{Mpc}}

\newcommand\gt[1]{\textcolor{gray}{#1}}

\newcommand{\simlt}{\lesssim}
\newcommand{\simgt}{\gtrsim}

\title[Constraining AGN feedback with the SZ effect]{Constraints on AGN feedback from its Sunyaev-Zel'dovich imprint on the cosmic background radiation}

\author[B. Soergel et al.]{
\parbox{\textwidth}{Bjoern Soergel$^{1,}$\thanks{E-mail: bsoergel@ast.cam.ac.uk},
Tommaso Giannantonio$^{1,2,3}$,
George Efstathiou$^{1}$,
Ewald Puchwein$^{1}$,
Debora Sijacki$^{1}$}\vspace{2mm}
\\
$^{1}$ Institute of Astronomy \& Kavli Institute for Cosmology,  University of Cambridge, Madingley Road, Cambridge CB3 0HA, UK\\
$^{2}$ Centre for Theoretical Cosmology, DAMTP, University of Cambridge, Wilberforce Road, Cambridge CB3 0WA, UK\\
$^{3}$ Universit\"{a}ts-Sternwarte, Ludwig-Maximilians-Universit\"{a}t M\"{u}nchen, Scheinerstr. 1, 81679 M\"{u}nchen, Germany
}

\pubyear{2016}

\begin{document}
\label{firstpage}
\pagerange{\pageref{firstpage}--\pageref{lastpage}}
\maketitle

\begin{abstract}
	We derive constraints on feedback by active galactic nuclei (AGN) by setting limits on their thermal Sunyaev-Zel'dovich (SZ) imprint on the cosmic microwave background (CMB). The amplitude of any SZ signature is small and degenerate with the poorly known sub-mm spectral energy distribution of the AGN host galaxy and other unresolved dusty sources along the line of sight.  Here we break this degeneracy by combining microwave and sub-mm data from \Planck with all-sky far-infrared maps from the AKARI satellite. We first test our measurement pipeline using the Sloan Digital Sky Survey (SDSS) redMaPPer catalogue of galaxy clusters, finding a highly significant detection ($>$$20\sigma$) of the SZ effect together with correlated dust emission. We then  constrain the SZ signal associated with spectroscopically confirmed quasi-stellar objects (QSOs) from SDSS data release 7 (DR7) and the Baryon Oscillation Spectroscopic Survey (BOSS) DR12. We obtain a low-significance ($1.6\sigma$)  hint of an SZ signal, pointing towards a mean thermal energy of $\simeq 5 \times 10^{60}$~erg, lower than reported in some previous studies. A comparison of our results with high-resolution hydrodynamical simulations including AGN feedback  suggests QSO host masses of $M_{200c} \sim 4 \times 10^{12}~h^{-1}M_\odot$, but with a large uncertainty. Our analysis provides no conclusive evidence for an SZ signal specifically associated with AGN feedback.
\end{abstract}

%
\begin{keywords}
Quasars: general -- galaxies: active -- galaxies: clusters: general  -- cosmic background radiation
\end{keywords}



\section{Introduction}
Active galactic nuclei (AGN), powered by accretion of material onto a supermassive black hole \citep{LyndenBell1969,Rees1984},
are amongst the most luminous objects in the Universe.

AGN can deposit enormous quantities of energy into their surroundings, driving large-scale outflows \citep[e.g.][]{Blandford1990,Silk1998, Fabian1999, King2003}. In fact, it is now widely accepted that AGN feedback plays a crucial role in our understanding of galaxy formation and evolution \citep{Kauffmann2000, DiMatteo2005, Croton2006, Bower2006, Sijacki2007, Schaye2010, Schaye2015, Sijacki2015, Dubois2016}. AGN feedback can also affect structure formation on scales larger than the immediate vicinity of the host galaxy. In clusters and groups of galaxies, AGN are believed to provide a heat source to balance radiative cooling of the hot gas in the central regions \citep[e.g.][]{Binney1995, Churazov2001, Churazov2002, Quilis2001, Sijacki2007, Puchwein2008, Dubois2010, McCarthy2011}.

From a cosmological perspective, AGN feedback can alter the shape of the matter power spectrum on small scales, 
affecting cosmological parameters determined from cosmic shear measurements (e.g.~\citealt{vanDaalen2011,Semboloni2011,Eifler2015}). 
Accurate modelling of the AGN feedback energetics is therefore essential 
when simulating the formation of structure in our Universe, in both high-resolution simulations of individual haloes (e.g.~\citealt{Bhattacharya2007})
and in simulations of cosmological volumes (e.g.~\citealt{Sijacki2007,Puchwein2008,Battaglia2010,Sijacki2015}).

In this paper, we use quasi-stellar objects (QSOs) as tracers of AGN activity. Because of their high luminosities, it is now possible
to construct large samples of QSOs from  wide-field spectroscopic surveys (e.g.~\citealt{Paris2016}), making this subclass of AGN 
well suited to statistical studies of AGN feedback energetics.

Traditionally most of the evidence for AGN feedback has been obtained from radio and/or X-ray observations (see e.g. \citealp{Burns1990, Boehringer1993, Carilli1994, Fabian2006, Forman2007}; and \citealp{Miley2008, Fabian2012} for reviews). It has now become possible to study gaseous outflows from individual AGN host galaxies at UV, optical and submillimetre wavelengths \citep[e.g.][]{Crenshaw2003, Rupke2011, Sturm2011, Maiolino2012, Tombesi2015}.

Another technique for constraining AGN feedback,  still in its nascent phase, is to exploit  the thermal Sunyaev-Zel'dovich (SZ) effect \citep{SZ1970,SZ1972}. When passing through the hot ionized gaseous halo of the AGN host galaxy, a small fraction of the cosmic microwave background (CMB) photons is scattered off electrons with high thermal velocities.
This leaves a characteristic, frequency-dependent signature in the CMB. If AGN feedback provides a  contribution 
to the thermal energy of the halo gas, this will affect the pressure profile which can, in principle, be detected 
via the SZ signature \citep{Chatterjee2007,Bhattacharya2007,Chatterjee2008,Scannapieco2008}.

In rich clusters of galaxies, the thermal energy from purely
gravitational heating is large enough that the SZ signature of
individual objects can be detected at high signal-to-noise by modern
CMB experiments
(e.g.~\citealt{Hasselfield2013,deHaan2016,Planckclusters2016}). However,
as we will show explicitly in Section \ref{sec:results}, the SZ
signal from AGN hosts is far too low to be detected in individual
systems. It is therefore necessary to use statistical techniques to
constrain the average AGN feedback energetics by, for example,
cross-correlating CMB maps with optically selected QSOs.
In this way, a tentative indication of a QSO feedback signal was
obtained by \cite{Chatterjee2009} by cross-correlating CMB maps from
the Wilkinson Microwave Anisotropy Probe (\WMAPc) with a photometric
QSO catalogue from the Sloan Digital Sky Survey (SDSS).

More recently, \cite{Ruan2015} stacked data from a Compton-$y$ map
(constructed by \citealt{Hill2013} via internal linear combination of the
\Planck 2013 maps) at the positions of spectroscopically confirmed
SDSS QSOs.  They reported a significant detection corresponding to a
thermal energy in the halo gas of $E_\mathrm{th} \simeq 10^{62}$~erg,
significantly larger than suggested by the AGN feedback models adopted in
hydrodynamical simulations (e.g.~\citealt{Sijacki2007,Battaglia2010}).
However, the interpretation of this signal as a detection of AGN
feedback has been disputed by \cite{Crichton2015} and
\cite{Verdier2015}, who argued that it was primarily caused by dust from
extragalactic sources, including the AGN host galaxy. An alternative
explanation has been proposed by \cite{Cen2015}, who claimed that the \cite{Ruan2015}
detection was caused by the cumulative, purely thermal, SZ signature from haloes
correlated with the AGN.

The analysis by \cite{Verdier2015} used \Planck 2015 maps and a
multi-frequency multi-component matched filter to extract the SZ
signal at the QSO positions. In addition to fitting to SZ, they
included models for synchrotron and dust emission. They found that most
of the correlated signal could be accounted for by dust, with no
significant detection of an SZ signature for QSOs at $z \simlt 2.5$.
For QSOs with $z \simgt 2.5$, however, they found evidence for a non-zero
SZ signal, with an amplitude corresponding to a total thermal energy in the ionized gas of $\simeq 3.5 \times 10^{61}$~erg (see Section~\ref{subsec:comparison} below),
much lower than reported by \cite{Ruan2015}.  
\cite{Verdier2015} concluded that it was not
possible to unambiguously attribute the SZ signal seen in the high-redshift
sample to AGN feedback because of the large uncertainties
associated with the halo gas scaling relations.

\cite{Crichton2015} performed a stacking analysis, cross-correlating a
radio-quiet subsample from the SDSS spectroscopic QSO catalogue with
CMB maps from the Atacama Cosmology Telescope (ACT) and sub-mm maps
from \textit{Herschel}-SPIRE.  This analysis also found evidence for a non-zero
SZ signal associated with the QSOs, corresponding to a thermal energy of $E_{\mathrm{th}} = (6.2 \pm 1.7) \times 10^{60} $ erg, or equivalently to a feedback efficiency of $f \sim 15\%$ for a typical
QSO activity timescale of $10^8$~yr.

However, the above-mentioned feedback constraints rely on marginalisation over
the amplitude and shape of the  dust spectral energy distribution (SED). 
As significant degeneracies exist between
dust and SZ parameters, especially when the fit is performed over a limited wavelength range, the SZ results can be biased by an inaccurate
determination of the dust properties.
Notably, the dust parameters deduced by \cite{Crichton2015} and \cite{Verdier2015} do
not agree. Both studies fitted a modified blackbody dust spectrum,
characterised by an amplitude, dust temperature $T_d$, and spectral
index $\beta_d$: \cite{Verdier2015} found
$T_d = ( 19.1 \pm 0.8) \, \mathrm{K}$ and $\beta_d = 2.71 \pm 0.13$	
for their full QSO sample ($0.1 < z < 5$) and their `dust only' filter,
with some evidence for a shift towards higher temperatures and lower
values of $\beta_d$ for QSOs at redshifts $z \simlt 2$. 
On the other hand,
\cite{Crichton2015} found $T_d = (40 \pm 3)  \, \mathrm{K}$ and $\beta_d = 1.12 \pm 0.13$
for their QSO sample covering the redshift range $0.5 -3.5$.\footnote{For reference, the \Planck analysis of Galactic dust emission
\citep{Planck2013_dust} gives 
$T_d = (20.3 \pm 1.3) \, \mathrm{K}$, $\beta_d=1.59 \pm 0.12$ 
over the high Galactic latitude sky ($\vert b \vert >
15^\circ$), parameters that are quite typical of nearby normal
galaxies \citep{Clemens2013}, and close to those of the cosmic infrared
background \citep{Mak2016}.}
As there is partial overlap in the QSO
samples used in the \cite{Verdier2015} and \cite{Crichton2015}
analyses, it seems unlikely that the disparity in the dust parameters
is reflecting a difference in the QSO selection or their average
physical properties. However, since dust emission dominates the
cross-correlation signal in both analyses, any claim of a non-zero SZ
signal is contingent on an accurate dust emission model.

In this paper, we use the \Planck 2015 microwave and sub-mm
maps in combination with all-sky far-infrared (FIR)
data at $90 \, {\mathrm{\upmu m}}$ from the AKARI satellite.  
With this additional spectral coverage on the falling, high-frequency part
of the dust SED, it is possible to
constrain the parameters of the dust emission model more robustly.
This allows us to break the degeneracy between the dust parameters and
the inferred SZ amplitude, providing a more reliable constraint on the
average AGN feedback energetics.

\begin{figure*}
	\centering
	\includegraphics[width=2\columnwidth]{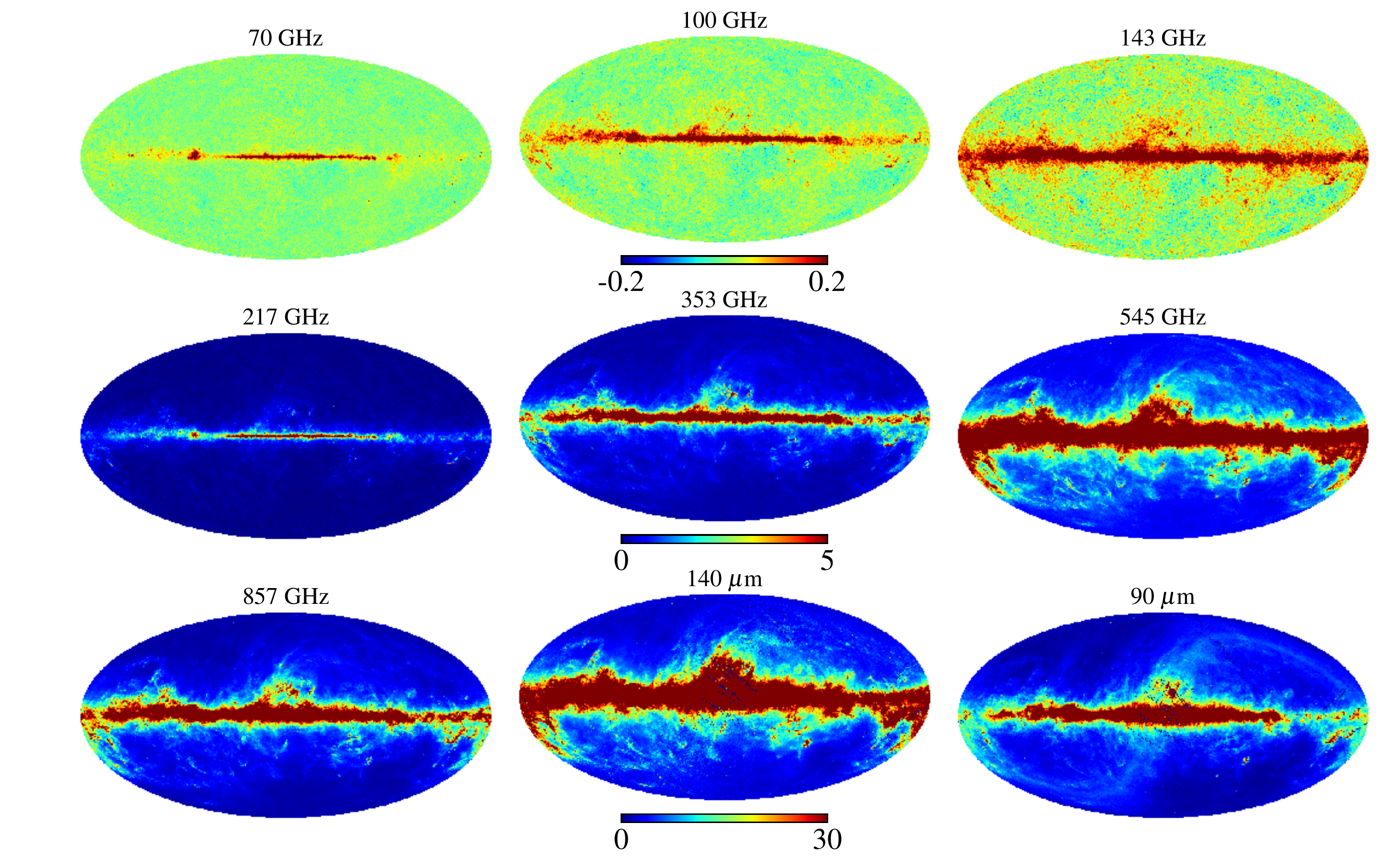}
	\caption{CMB and FIR maps (in MJy/sr): we show here the seven \Planck maps (LFI 70~GHz and HFI 100-857~GHz), as well as the AKARI 140 and 90 $\upmu$m maps. The 140 $\upmu$m channel was not used in the analysis due to larger calibration uncertainties, and is shown for illustration purposes only. All three maps in each row have the same colour scale, but the scale changes between the rows. The extent of the Galaxy in the various maps gives a good visual impression of the SED of Galactic dust; the SED of the cosmic infrared background (CIB) is relatively similar (e.g.~\citealt{Mak2016}), but redshifted.}
	\label{fig:allmaps}
\end{figure*}

\begin{table*}
	\begin{tabular}{ccccccccccccccc}
		\toprule
		&  \multicolumn{7}{c}{\textbf{\Planck}}  && \multicolumn{4}{c}{\textbf{AKARI}} && \textbf{\gt{IRIS}} \\
		\midrule
		Band centre (GHz)   &    70 &   100 &   143 &   217 & 353 & 545 & 857 && \gt{1,870} & \gt{2,140} & 3,330 & \gt{4,310} & & \gt{3,000} \\
		Band centre ($\upmu$m)& 4,280 & 3,000 & 2,100 & 1,380 & 849 & 550 & 350 &&  \gt{160} &    \gt{140} & 90   & \gt{65} && \gt{100} \\
		Beam FWHM  ($\arcm$) & 13.31 & 9.68 & 7.30 & 5.02 & 4.94 & 4.83 & 4.64  && \gt{$\simeq 1.47^\ast$} & \gt{1.47} & 1.30 & \gt{1.05}  && \gt{4.3}\\ 
		Calibration uncertainty$\dagger$ & 1\% & 1\% & 1\% & 1\% & 1\% & 6\% & 6\%  && \gt{$>56.1\%$} & \gt{$>45.4\%$} & $15.1\%$ & \gt{$>20.1\%$} && \gt{13.5\%} \\
		\bottomrule
	\end{tabular}
	\caption{Summary of the map properties relevant for our analysis. Due to their large and partially unknown calibration uncertainties at low flux densities, we discard the AKARI bands displayed in grey, and only use the AKARI 90 $\upmu$m channel for our main analysis. ($^\ast$The resolution of the AKARI 160 $\upmu$m channel was not measured during the calibration process, but is expected to be close to the 140 $\upmu$m band; see \citealt{Akari_calibration}. $\dagger$ The calibration uncertainty is assumed to be fully correlated between the \Planck 70 to 353 GHz, 545 and 857 GHz, and IRIS 100 and AKARI 90 $\upmu$m bands, respectively.)}
	\label{tab:maps}
\end{table*}

\begin{figure}
	\centering
	\includegraphics[width=\columnwidth]{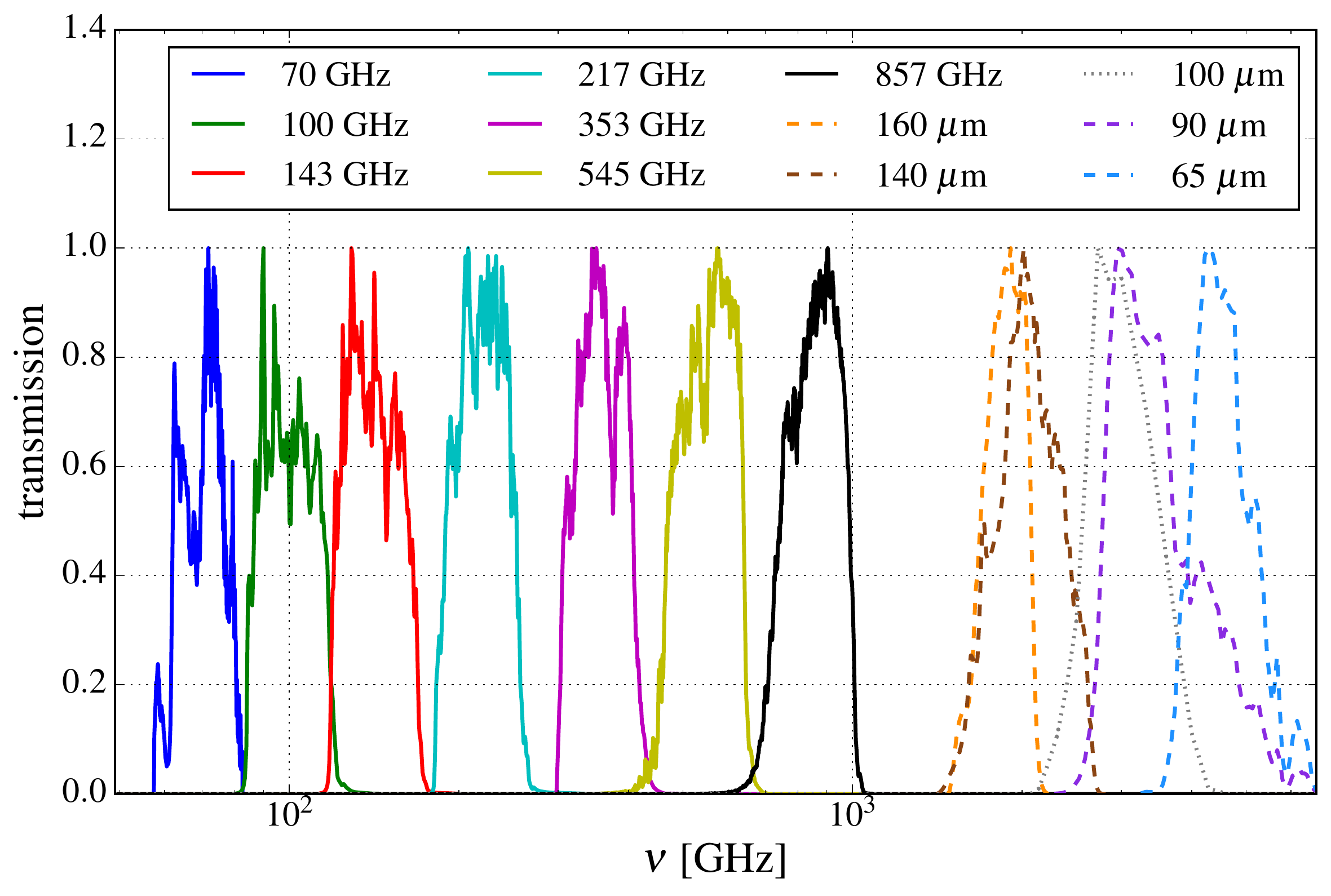}
	\caption{Instrumental bandpasses: we show the detector responses (normalized to a peak amplitude of unity) for the microwave and sub-mm data used in our analysis. Solid lines denote the \Planck bandpasses, whereas dashed (dotted) lines show the AKARI (IRAS) transmission curves.}
	\label{fig:bandpasses}
\end{figure}

Our paper is organized as follows. Section~\ref{sec:data} describes the
various data products used in this work. Section~\ref{sec:meth} provides
a detailed description of our analysis methods;  in particular, we
discuss the filtering of the input maps, the SED measurement, and our
multi-component modelling and fitting of the QSO
emission. Section~\ref{sec:results} presents our main results. As an
important consistency check of our methodology, we demonstrate using
optically-selected galaxy clusters that our pipeline leads to an
unambiguous detection of the SZ signal together with a correlated
dust signal. We then report our results for the QSO
cross-correlations. Section~\ref{sec:discussion} discusses our results
and compares then against high-resolution hydrodynamical simulations
of both QSO host haloes and clusters that incorporate AGN feedback. Our
conclusions are summarized in Section~\ref{sec:concl}.  
In the course of
this project, we further experimented extensively with IRAS maps over the
frequency range $12-100 \, \upmu {\rm m}$ and with AKARI maps at $160$, $140$ and
$65 \, \upmu {\rm m}$. Our reasons for excluding these maps from our main analysis
are described in Appendix~\ref{app:dataquality}.

\section{Data sets}
\label{sec:data}

\subsection{Planck CMB maps}
The \Planck satellite has surveyed the full sky in nine microwave and sub-millimetre bands, with the Low Frequency Instrument (LFI, \citealt{Planck2015_LFI}) contributing three bands between 30 and 70 GHz, and the High Frequency Instrument (HFI, \citealt{Planck2015_HFI1,Planck2015_HFI2}) featuring six channels between 100 and 857 GHz.
Here we use the LFI 70 GHz map and the six HFI single-frequency maps\footnote{These maps were publicly released by the \Planck Collaboration as part of their 2015 data release and are available in the
\HEALPIX~(\url{http://healpix.sourceforge.net/}) format with $N_\mathrm{side} = 2048$ at the Planck Legacy Archive: \url{http://pla.esac.esa.int}.};
the two other LFI maps (30~GHz, 44~GHz) do not add additional information because of their significantly poorer angular resolution (30~GHz) and higher noise level (44~GHz); see e.g.~\citet{Planck2015_LFI,Planck2015_HFI1}.
For a detailed description of the time-ordered data filtering, mapmaking and calibration process we refer to the HFI and LFI papers, and just summarize the properties of the maps relevant for our analysis in Table~\ref{tab:maps}. For completeness, Fig.~\ref{fig:allmaps} shows a plot of the maps used in this analysis, and Fig.~\ref{fig:bandpasses} shows the respective instrumental bandpasses.

The \Planck maps up to 353 GHz used the 
orbital dipole of the CMB for calibration, whereas the 545 and 857 GHz maps were calibrated on planets \citep{Planck2015_overview}.
Here we assume the following conservative and simplified model for the calibration uncertainties: 
for the bands up to 353 GHz, we assume 1\% absolute calibration uncertainty correlated between the individual bands.
Similarly, we assume a 6\% uncertainty for the 545 and 857 GHz bands, as in \cite{Mak2016}.
These calibration uncertainties are conservative estimates, but adopting tighter
errors would have negligible effects on the results presented in this paper.

The broad spectral range makes the \Planck maps well suited to investigating a possible
Sunyaev-Zel'dovich signature at the lower frequencies, with the higher frequencies providing constraints
on dust emission.

However, for all \Planck maps the beam is significantly larger than the angular extent of the typical QSO host galaxy
(for reference, a length of proper size 0.1~Mpc at $z=2$ subtends an angle of $\simeq 0.2\arcmin$);
this is especially true for the \Planck low-frequency channels that are important for the SZ signature.
Nonetheless, although we are not able to resolve
individual sources, the \Planck maps still allow us to constrain the average emission properties of the QSO sample.

\subsection{AKARI far-infrared data}
\label{sec:akari}

At the median QSO redshifts of $z \simeq 2$ and for typical dust parameters, we expect the dust SED to peak at around $\sim 1$~THz.
Therefore, additional data at frequencies $\nu \gtrsim 1$~THz adds valuable information with which to fix
the dust emission parameters.
In turn, this allows a more reliable extrapolation of the dust SED down to low frequencies
at which we might hope to see an SZ signal.

To complement the \Planck maps at higher frequencies, we use all-sky FIR data 
from the AKARI satellite.
For additional tests we have also used data from the Infrared Astronomical Satellite (IRAS, \citealt{IRAS}), as reprocessed by \cite{IRIS} in the 
form of the IRIS maps. However, we do not use the IRIS maps in our
main analysis for reasons discussed in  Appendix~\ref{app:dataquality}.
We summarize the key properties of the FIR data in Table~\ref{tab:maps} and show the maps and instrumental bandpasses in Figs.~\ref{fig:allmaps} and \ref{fig:bandpasses}, respectively.

The AKARI satellite has surveyed almost the entire far-infrared sky in four bands with central wavelengths between 160 and 65 $\mathrm{\upmu m}$.\footnote{The full-sky AKARI maps at \HEALPIX~$N_\mathrm{side} = 4096$ resolution are available from the \textit{Centre d'Analyse de Donn\'ees Etendues}: \url{http://cade.irap.omp.eu/dokuwiki/doku.php?id=akari}.}
In comparison to IRAS/IRIS, 
the nominal noise levels are broadly similar, but AKARI has a higher angular resolution of $1-1.5\arcmin$ compared to the $4.3\arcmin$ resolution of the
100 $\mathrm{\upmu m}$ IRIS map.
The relatively high resolution of AKARI is particularly useful to disentangle `actual' point sources from objects that are merely unresolved in the high-frequency \Planck and the IRIS 100 $\mathrm{\upmu m}$ maps.
The data processing, mapmaking and calibration of the AKARI maps
is described in detail by \citet{akari} and \citet{Akari_calibration}.

The AKARI map at 90 $\mathrm{\upmu m}$ has the lowest noise level 
and is the only AKARI map for which a reliable calibration 
is available down to low intensities (see Table~\ref{tab:maps} and \citealt{Akari_calibration}). 
The other AKARI bands have large and uncertain calibration errors at low intensities, making their robust inclusion into the analysis problematic.
If appropriate calibration uncertainties were folded into the analysis, these bands would carry little statistical weight. 
If, however, they were included with underestimated calibration errors, they could indeed bias our results (as we demonstrate in Appendix~\ref{app:impact}). 
We therefore include only the 90 $\mathrm{\upmu m}$ map in our main analysis.

\subsection{SDSS QSO catalogue}
\begin{figure}
	\centering
	\includegraphics[width=\columnwidth]{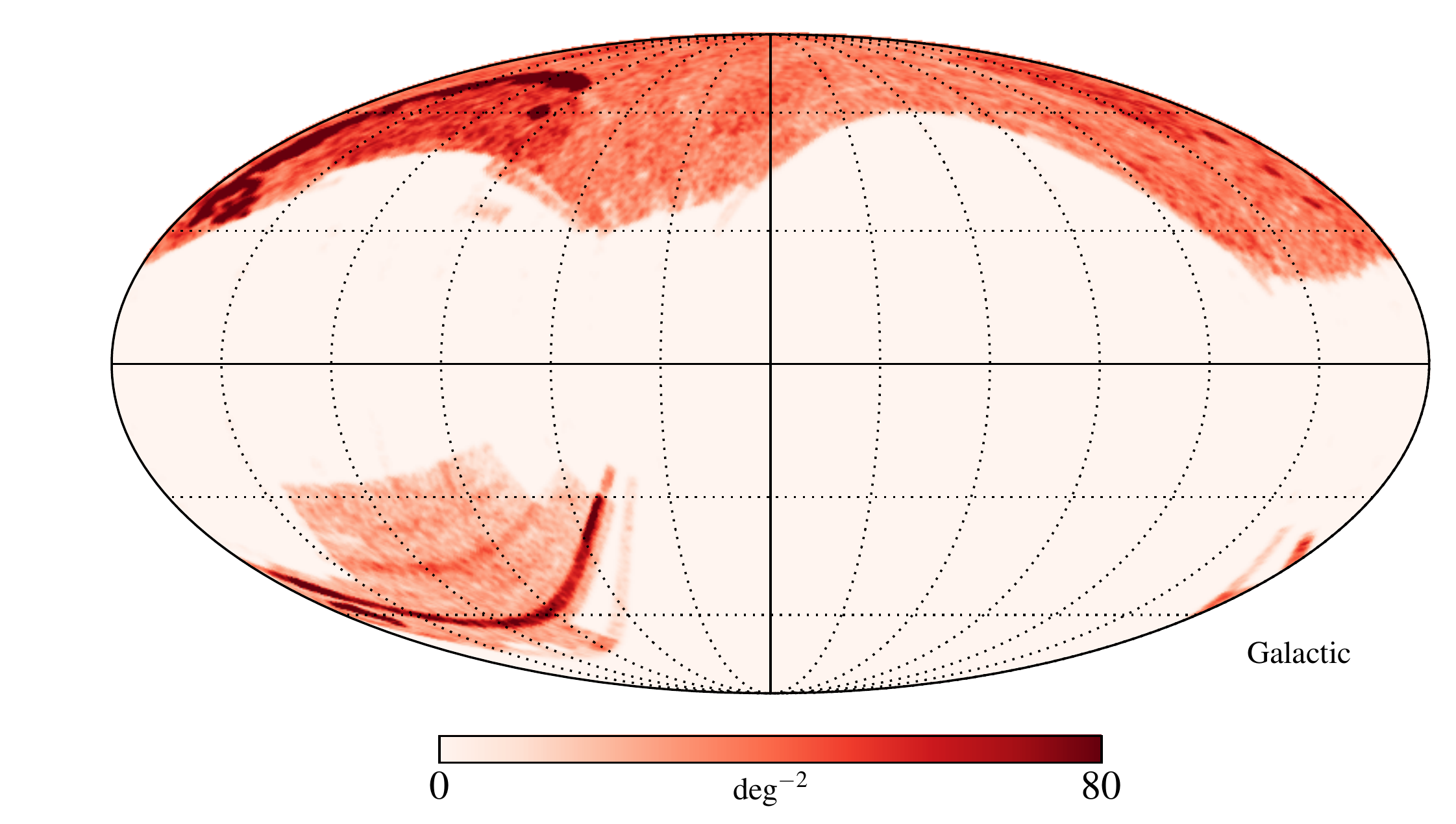}
	\caption{Sky coverage of the merged SDSS QSO sample in Galactic coordinates: for plotting purposes we have smoothed the QSO density on a scale of $1 \, \deg$.}
	\label{fig:qsodensity}
\end{figure}

In this paper we make use of catalogues of spectroscopically confirmed QSOs created from SDSS-II \citep{York2000} and SDSS-III \citep{Eisenstein2011} Baryon Oscillation Spectroscopic Survey (BOSS, \citealt{BOSS}) data. 
In particular, we use the QSO catalogues from the seventh (DR7, \citealt{Schneider2010} and twelfth (DR12, \citealt{Paris2016}) SDSS data releases.\footnote{The catalogues are publicly available at \url{http://classic.sdss.org/dr7/products/value_added/qsocat_dr7.html} and \url{http://www.sdss.org/dr12/algorithms/boss-dr12-quasar-catalog/}, respectively.}
The QSO target selection process is described in detail by \cite{Ross2012}; here we simply summarize the points relevant to this work.
Whereas the DR7 sample mostly contains `low' redshift ($z \lesssim 2.5$) objects, the DR12 sample specifically targeted QSOs with $z > 2.15$ \citep{Paris2016}. 
However, as a result of a colour degeneracy in the target selection from photometric data, lower-redshift QSOs were also observed, leading to a secondary maximum in the redshift distribution around $z \simeq 0.8$.
From these two catalogues, we create a merged sample. 
It is worth noting that there is a non-zero overlap between the two catalogues, mostly because previously confirmed $z > 2.15$ QSOs were re-observed for DR12 \citep{Ross2012}.
When combining the two samples, we remove any duplicates. 
The sky coverage of the merged sample is shown in Fig.~\ref{fig:qsodensity}.

\begin{figure}
	\centering
	\includegraphics[width=\columnwidth]{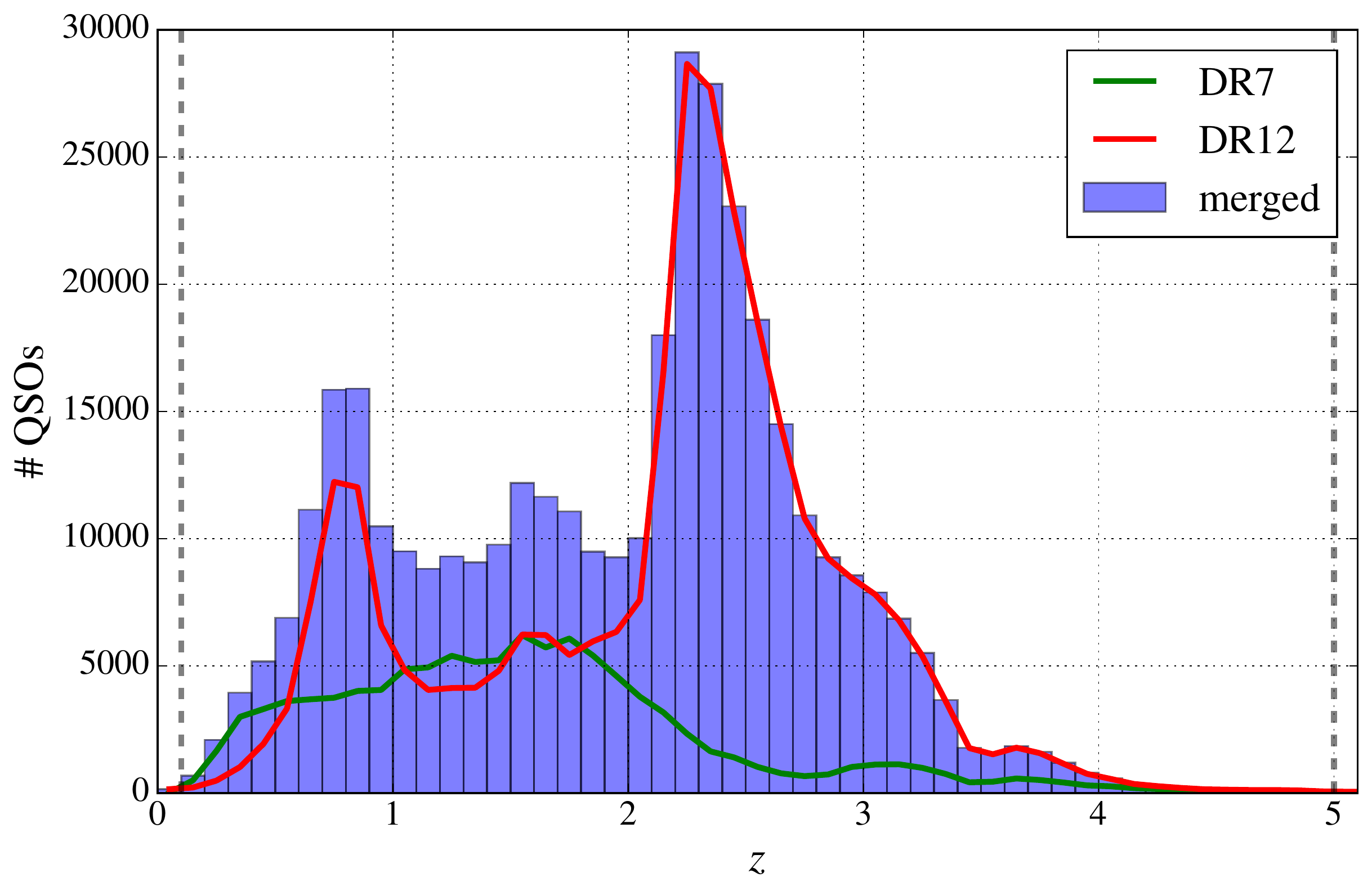}
	\caption{Redshift distribution of the QSO samples: the red and green solid lines refer to the individual DR7 and DR12 samples, whereas the blue histogram shows the merged sample used in the main analysis (duplicates have been removed). The dashed vertical lines at the edges denote the redshift cuts used in the main analysis.}
	\label{fig:qso_zhist}
\end{figure}

Both catalogues provide several redshift estimates. 
These include the SDSS pipeline $z$-estimate, as well as results obtained via principal component analysis, the position of specific emission lines, and visual inspection.
Any of these redshift estimates is accurate enough for our purposes.
For the main analysis we use the SDSS pipeline redshifts, but have verified that  our results do not change if we use any of the other methods.
Fig.~\ref{fig:qso_zhist} shows the redshift distribution of the DR7 and DR12 QSO samples and the combined sample. 
Since we are interested in the average emission properties of the QSOs, we remove the sparsely populated low- and high-$z$ tails of the redshift distribution ($z<0.1$ and $z >5$). 
These cuts remove less than a percent of the sample.
For our main analysis we use the full redshift range $0.1 < z < 5$ of the merged sample, containing 377,136 spectroscopically confirmed QSOs.
To test for redshift evolution of the measured SZ and dust parameters, we also split the sample into 
five redshift bins with approximately equal number of QSOs (see Section~\ref{sec:zsplit}).

\subsection{RedMaPPer galaxy clusters}
For validation and testing purposes, we also use a catalogue of galaxy clusters identified in the SDSS DR8 data with the \textit{red-sequence Matched-filter Probabilistic Percolation} (redMaPPer) cluster finder \citep{Rykoff2013}.
Most of the optically identified clusters do not have an SZ signal that is strong enough for a blind detection in the \Planck data.
By stacking them, however, we should obtain a significant SZ detection, which will serve as a test for our measurement pipeline.
For similar stacking analyses of clusters and CMB data see e.g. \cite{Afshordi07,Diego10,Bartlett11,Hajian13}.

Briefly, the redMaPPer cluster finding algorithm uses a spectroscopic training set to calibrate red-sequence models at various redshifts, which are then used to iteratively assign membership probabilities to galaxies in the vicinity of a given cluster candidate.
The resulting number of member galaxies --- the optical richness $\lambda$ --- has been found to provide a low-scatter mass proxy (e.g.~\citealt{Rykoff2012,Saro2015}); in addition the redMaPPer algorithm provides accurate photometric redshift estimates.
The original SDSS DR8 catalogue has been compared to CMB data from \Planck by \cite{Rozo2014a}.
Subsequently, the algorithm has been further developed and applied to Dark Energy Survey data \citep{Rykoff2016}; \cite{Saro2015}, \citet{Soergel2016}, and \citet{Saro2016} have used these catalogues for SZ studies in conjunction with CMB data from the South Pole Telescope (SPT) SZ survey.

In this work, we use an updated version (v6.3) of the SDSS catalogue that includes the improvements to the algorithm described by \citet{Rykoff2016}.\footnote{The catalogue is publicly available at \url{http://risa.stanford.edu/redmapper/}.}
Prior to the cuts we apply during the analysis, the catalogue contains 26,111 clusters with $\lambda \gtrsim 20$ in the redshift range is $0.1 \lesssim z < 0.6$, corresponding to a lower mass threshold of $M_{500} \gtrsim 10^{14}~M_\odot$ \citep{Saro2015}.
The median richness and redshift of the catalogue are $\lambda_m \simeq 33$ and $z_m \simeq 0.37$, respectively.

\section{Analysis methods}
\label{sec:meth}

\subsection{Selection of a clean sample}
\label{subsec:samplesel}
A crucial part of this analysis is the selection of a sample that is not affected by the various possible contaminants such as point sources, radio emission at low frequencies, and Galactic dust at higher frequencies. Here we describe and motivate these cuts and discuss their impact on the sample size.

\subparagraph{Radio-loud cut for QSOs:}
Some SDSS QSOs are radio-loud (see e.g.~\citealt{Kellermann1989}), 
that is, they show synchrotron emission in the radio bands. 
Typically, the synchrotron contribution to the microwave SED can be modelled as a falling power law in frequency, thus contributing mainly at the lowest frequencies used in the analysis. Radio emission has the potential to hide an SZ signature, as its positive contribution to the low-frequency SED can partially cancel a negative SZ signal.
We therefore remove from the sample all QSOs that have a radio counterpart detected in the Faint Images of the Radio Sky at Twenty-Centimeters (FIRST) survey \citep{FIRST} with a nominal source detection threshold of 1 mJy at 1.4 GHz;
for details about the matching process see \citet{Paris2016}.
The FIRST sky coverage mostly overlaps with the SDSS/BOSS footprint; however small regions at the edges were not covered. 
To avoid contamination from radio sources in these regions, we additionally remove all QSOs in parts of the sky that have not been observed by FIRST.
These two cuts reduce our sample size by approximately 14\%. 

Even after the application of this cut, it is still possible for radio sources below the FIRST detection threshold to contribute to our flux density measurement at the lowest frequencies.
We derive an estimate for the contamination by undetected radio sources in Section~\ref{subsec:radiotest}, and explicitly test for it by fitting a radio component to our measured SED.

\subparagraph{Radio contamination for clusters:}
As cluster galaxies can also host radio sources (see e.g.~\citealt{Gupta2016}), we have tested removing clusters from the redMaPPer sample if they are close to a FIRST-detected radio source
However, because of the spatial extent of clusters and the large \Planck beams, this leads to a large number of chance associations.
For this reason, we do not perform a `radio-loud' cut for the clusters. 
It is therefore possible that residual synchrotron contamination might affect the lowest frequency, as discussed in Section~\ref{subsec:results_clusters} below. Because of the much higher SZ signal, synchrotron contamination is a much smaller problem for the clusters than for the QSOs.

\subparagraph{Mask:}
We next remove objects in regions with high contamination by Galactic dust. 
In our main analysis we only include objects within the 40\% of the sky with the lowest Galactic dust emission, as quantified by the masks produced by the \Planck Collaboration.
AS SDSS/BOSS mostly covers the Northern and Southern Galactic caps, this conservative cut only reduces our sample size by approximately $10\%$.
To ascertain that our results do not depend significantly on the choice of mask, we repeat our analysis with an even more conservative mask (using 30\% of the sky), and a slightly less conservative mask (using 50\% of the sky).
In both of these tests, we obtain results that are consistent with the main analysis, demonstrating that contamination by Galactic dust is not a significant issue.

\subparagraph{Point source cut:}
Resolved point sources in the \Planck sub-mm bands typically have flux densities of $\sim 1$ Jy, with the brightest sources reaching  10 Jy or more 
(e.g.~\citealt{Planck2013_compactsources}); this is two to three orders of magnitude larger than the typical sub-mm flux densities of a QSO host ($\simeq 10~\mathrm{mJy}$, see Section~\ref{sec:resqso} below). Even a relatively small number of point sources associated with QSOs in our sample can therefore affect our results.

To identify point sources, we use the \Planck 2015 Catalogue of Compact Sources \citep{Planck2015_compactsources}, the AKARI Far-Infrared Surveyor (FIS) Bright Source Catalogue \citep{AKARI_pointsources}, and for additional tests also the IRAS Point Source Catalogue \citep{IRAS_pointsources}.

We also include unconfirmed point source candidates and sources with unreliable flux measurements from the point source catalogues.
We then remove all QSOs that are closer than $3 \times \theta_\mathrm{FWHM}$ to a point source in the respective catalogue in any of the bands used for our analysis; this conservative selection removes approximately 20\% of the remaining QSOs.
It is worth noting that the high resolution of the AKARI 90~${\mathrm{\upmu m}}$ band enables us to perform a more stringent point source removal than possible with the \Planck bands alone.

\subsection{Map filtering}
\label{sec:filtering}
Having defined a QSO sample, we turn to estimating the flux
densities in the CMB and IR maps at the QSO positions.  The first step
is to construct a matched spatial filter (e.g.~\citealt{Haehnelt1996,
	Schaefer2006}) that is designed to recover the QSO
signal in the presence of much larger contaminants (primary CMB, dust
emission, instrumental noise).
For an object centred at $\nhat_0$, we write the data as
\beq
d(\nhat) = A \tau(|\nhat-\nhat_0|) + N(\nhat) \, ,
\eeq
where $A$ is the unknown amplitude of a known (beam-convolved) profile with rotational symmetry, $\tau(\theta)$, and $N$ are all other signals (which we consider as effective `noise').
We perform the filtering in the spherical harmonic domain, which avoids
having to project submaps into an approximate flat-sky geometry.

Using the symmetry of the input profile and the convolution theorem on the sphere, the harmonic coefficients of the filtered map, $a_{\ell m}^\mathrm{filt}$, are related to the unfiltered ones according to
(e.g.~\citealt{Schaefer2006})
\beq
a_{\ell m}^\mathrm{filt} = \sqrt{\frac{4\pi}{2\ell+1}} \Psi_{\ell 0} \times a_{\ell m}^\mathrm{unfilt} \equiv {F}_\ell \times a_{\ell m}^\mathrm{unfilt}  \, , 
\label{eq:convtheorem}
\eeq
with
\beq
\Psi_{\ell 0} = \left( \sum_\ell \frac{\tau_{\ell 0}^2}{N_\ell}\right)^{-1} \times \frac{\tau_{\ell 0}}{N_\ell} \, .
\eeq 
Here $\Psi_{\ell 0}$ and $\tau_{\ell 0}$ are the $m=0$ harmonic coefficients of the filter function  and input profile,
respectively. All $\Psi_{\ell, m \neq 0}$ and $\tau_{\ell, m \neq 0}$ vanish due to the symmetry; $N_\ell$ is the `noise' power spectrum.

As the QSO contribution to the full power spectrum is completely negligible compared to the primary CMB (at low frequencies), foregrounds (mostly dust emission at higher frequencies), and instrumental noise, we set $N_\ell = C_\ell^\mathrm{tot}$ and measure $C_\ell^\mathrm{tot}$ directly as the power spectrum of the unfiltered input maps. 
We use the same mask (apodized with a Gaussian with 2 deg FWHM) as for the QSO sample selection above, and correct for the effect of the mask and the pixel window function on the measured $C_\ell^\mathrm{tot}$.

We further include an additional smooth high-pass filter,
precluding any large-scale features ($\ell \lesssim 300$) from biasing
our flux density measurements.  This also removes any monopole from
the maps, removing any additive overall calibration uncertainty.  The
filtering is furthermore limited to $\ell_\mathrm{max} = 3
N_\mathrm{side} -1$, where $N_\mathrm{side} = 2048$ for \Planck and
IRIS maps, and $N_\mathrm{side} = 4096$ for AKARI.  We also ensure
that the filter ${F}_\ell$ smoothly rolls off to zero at
$\ell_\mathrm{max}$ to avoid ringing artefacts in the filtered maps.

\subparagraph{Filtering for QSOs:}
Except for the lowest-$z$ objects (that we have already removed from our sample), the SZ and IR emission by QSO host haloes is completely unresolved in the \Planck maps.
\citet{Verdier2015} constructed a filter assuming a Navarro-Frenk-White (NFW, \citealt{navarro96}) profile matched to a halo with $M_{500} = 10^{13} \, \Msol$ at $z=2$, resulting in
$\theta_{500} = 0.27\arcmin$.
This is not necessarily an optimal choice, as indeed most studies point to somewhat lower QSO host masses (see Section~\ref{sec:sims} below); nonetheless it is significantly smaller than the \Planck beam at any frequency ($\simeq 5-13 \arcmin$).
Even at the higher spatial resolution of AKARI ($\sim 1.5\arcmin$), the QSO host haloes still remain unresolved. 
Therefore we assume in this work that the input profile for QSOs $\tau(\theta)$ is simply given by the beam (normalized to unit amplitude in pixel space). 

In this case, the matched filter reduces to a simple Wiener filter on the sphere (e.g.~\citealt{Tegmark1998,Chatterjee2009}) with
\beq
{F}_\ell = \left[\sum_\ell \frac{(2 \ell +1)\tilde{B}_\ell^2}{4\pi C_\ell^\mathrm{tot}} \right]^{-1}\frac{\tilde{B}_\ell}{C_\ell^\mathrm{tot}} \, ,
\eeq
where 
$\tilde{B}_\ell = 2\pi\sigma^2 B_\ell$,
$B_\ell$ is the instrumental beam in harmonic space 
and $\sigma = \theta_\mathrm{FWHM} / \sqrt{8 \ln 2}$
(the extra normalization factor occurs because we have normalized the profile to unit amplitude in pixel space).

\begin{figure}
	\centering
	\includegraphics[width=\columnwidth]{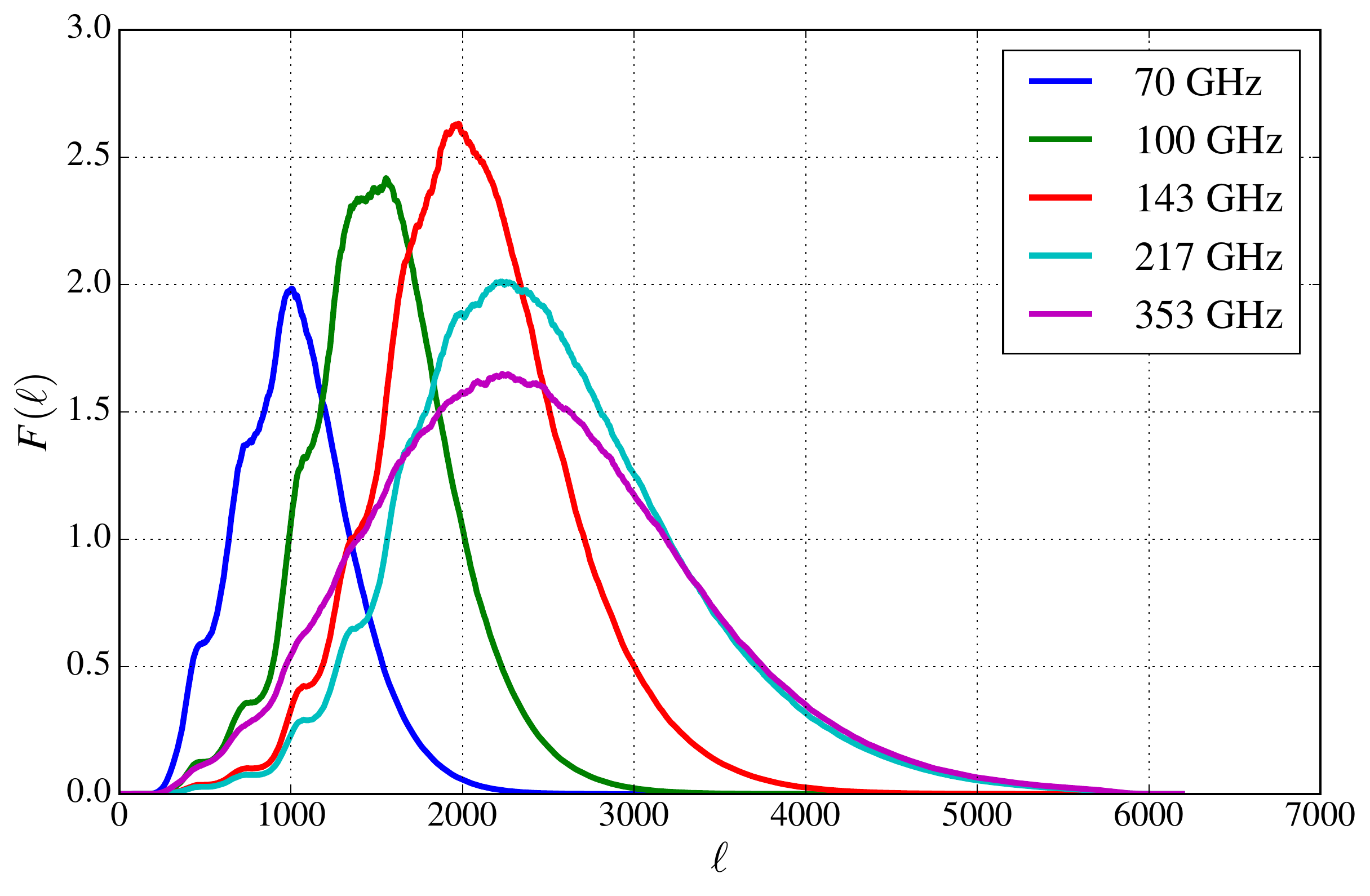}
	\caption{Filtering: we show here the filter kernels for the
	unresolved QSOs in the first five \Planck bands, which are those sensitive to a potential SZ signal. 
	From low to high frequencies, the resolution of the \Planck bands improves; therefore the	peak of the filtering kernels are shifted to higher $\ell$.
	For plotting purposes only, we have smoothed the kernels with a running average of $\Delta \ell = 50$. 
	}
	\label{fig:filters}
\end{figure}

This construction of the filter is model-independent since we do not need to assume a specific profile.
As an example for the resulting filters, we show in Fig.~\ref{fig:filters} the filtering kernels for the first five \Planck bands from 70--353 GHz; these are the bands with sensitivity to the SZ signal.

\subparagraph{Filtering for clusters:}
The redMaPPer clusters that we use as a test of our analysis pipeline are 
partially resolved in the higher-frequency \Planck maps and mostly resolved in the AKARI maps.
To account for this, we modify the way we construct the filter as follows:
we use a projected isothermal $\beta$-profile \citep{cavaliere76} with the index $\beta$ (not to be confused with the dust spectral index $\beta_d$) set to unity, i.e.
\beq
\tau(\theta) = \left(1+\theta^2/\theta_c^2\right)^{-1} \, ,
\eeq
as an input profile for the filter. 
This is a widely used choice, both for blind SZ cluster detection (e.g.~\citealt{Bleem2015}), and matched-filter estimates for the CMB temperature (or intensity) at the positions of optically selected clusters (e.g.~\citealt{Soergel2016}).

The profile is scaled with the projected core radius $\theta_c$; here we use $\theta_c = 1\arcmin$. For a cluster sample with similar richness limits, but with slightly higher median redshift $z \simeq 0.5$,
\cite{Soergel2016} found $\theta_c = 0.5\arcmin$ to be a good match to the actual cluster profile, so that $\theta_c = 1\arcmin$ should be reasonably well matched in our case.
As we mostly intend the clusters to be a test case for our SZ extraction algorithm, we do not attempt to further optimize this choice or to construct a more sophisticated (e.g.~adaptive) matched filter.
Nonetheless we have checked that our recovered emission parameters do not depend strongly on this choice by repeating the filtering with $\theta_c = 0.5\arcmin$ and $\theta_c = 2\arcmin$.

Using this particular profile has the advantage that there is an analytic expression for $\tau_{\ell 0}$, the spherical harmonic coefficients of the input profile. 
This can be seen as follows:
(1) In the flat sky case, the Hankel transform of zeroth order of the profile is given by $\theta_c^2 K_0(k \theta_c)$, where $K_0$ is the modified Bessel function of the second kind and $k$ is the wavenumber.  
(2) In the case of a function with circular symmetry (such as our input profile), the Hankel transform of order zero reduces to the 2D Fourier transform (modulo a normalization constant of $2\pi$).
(3) Generalizing this to the sphere (see equation~\ref{eq:convtheorem}),
we compute $\tau_{\ell 0}$ via
\beq
\tau_{\ell 0} = 2\pi \theta_c^2 \sqrt{\frac{2\ell +1}{4\pi}} \times K_0(\ell \theta_c) \, .
\eeq
(4) We finally account for the effect of the instrumental beam with $ \tau_{\ell 0} \rightarrow \tau_{\ell 0} \times B_l$.

This analytic construction of the filter has the advantage that we do not need to compute spherical harmonic transforms of the input profile numerically, making the filter construction both computationally efficient and robust against numerical errors.

\subsection{Flux density measurement}
\label{FD_measurement}	

The flux density contributed by a QSO or cluster $i$ is then 
\beq
S_\nu^i = I_\nu^\mathrm{filt}(\nhat_i) \times \int \mathrm{d}\Omega \,\tau_\nu(\theta) \,, 
\label{eq:fluxdens}
\eeq
where $I_\nu^\mathrm{filt}(\nhat_i)$ is the pixel value in the filtered map at its position $\nhat_i$ and $\tau_\nu$ is the filter profile for the band $\nu$.
In the case of the unresolved QSOs $\tau_\nu(\theta) = B_\nu(\theta)$, so that the integral in equation~\ref{eq:fluxdens} simply yields the beam area.
For the clusters, we have normalized the $\beta$-profile to unit amplitude in real space before convolving with the instrumental beam for the respective frequency. Therefore the matched filter returns the amplitude of the unconvolved profile, which is thus also used in the integral in equation~\ref{eq:fluxdens}.
In this case, we have to evaluate the integral numerically,  
truncating the integration at a suitably high value
$\theta_\mathrm{max}$. In this analysis, we choose
$\theta_\mathrm{max} = 5 \theta_c = 5\arcmin$, but we find that
our results are insensitive to the precise choice of $\theta_\mathrm{max}$.

From the individual $\{S_\nu^i\}$ we then compute the mean flux densities at every frequency and their frequency-frequency covariance matrix as
\beq
\bar{S}_\nu = \left\langle S_\nu^i \right \rangle_i \, , \quad C_{\nu,\nu\prime} = \frac{1}{N_\mathrm{o}(N_\mathrm{o}-1)} \sum_i^{N_\mathrm{o}} (S_\nu^i - \bar{S}_\nu)(S_{\nu\prime}^i - \bar{S}_{\nu\prime}) \, ,
\label{eq:meancov}
\eeq
where $N_\mathrm{o}$ is the number of QSOs or clusters in the sample.
We have also estimated $C_{\nu,\nu\prime}$ from jack-knife or bootstrap resampling from the respective catalogue. The results from both of these approaches are in good agreement with the covariance estimated via equation~\ref{eq:meancov}.

It is worth noting that we do not use an inverse-variance weighting for estimating the mean flux density in equation~\ref{eq:meancov}; this is because the noise properties of our all-sky maps are relatively uniform once the Galactic plane is masked.
The only notable exceptions are the ecliptic poles, where the
\Planck scanning strategy produces caustic-like structure in the
number of observations per pixel. These `caustic' regions are prone to
systematics such as destriping errors; so upweighting them via strict
inverse-variance weighting could potentially cause a bias.

\begin{figure}
	\centering
	\includegraphics[width=0.85\columnwidth]{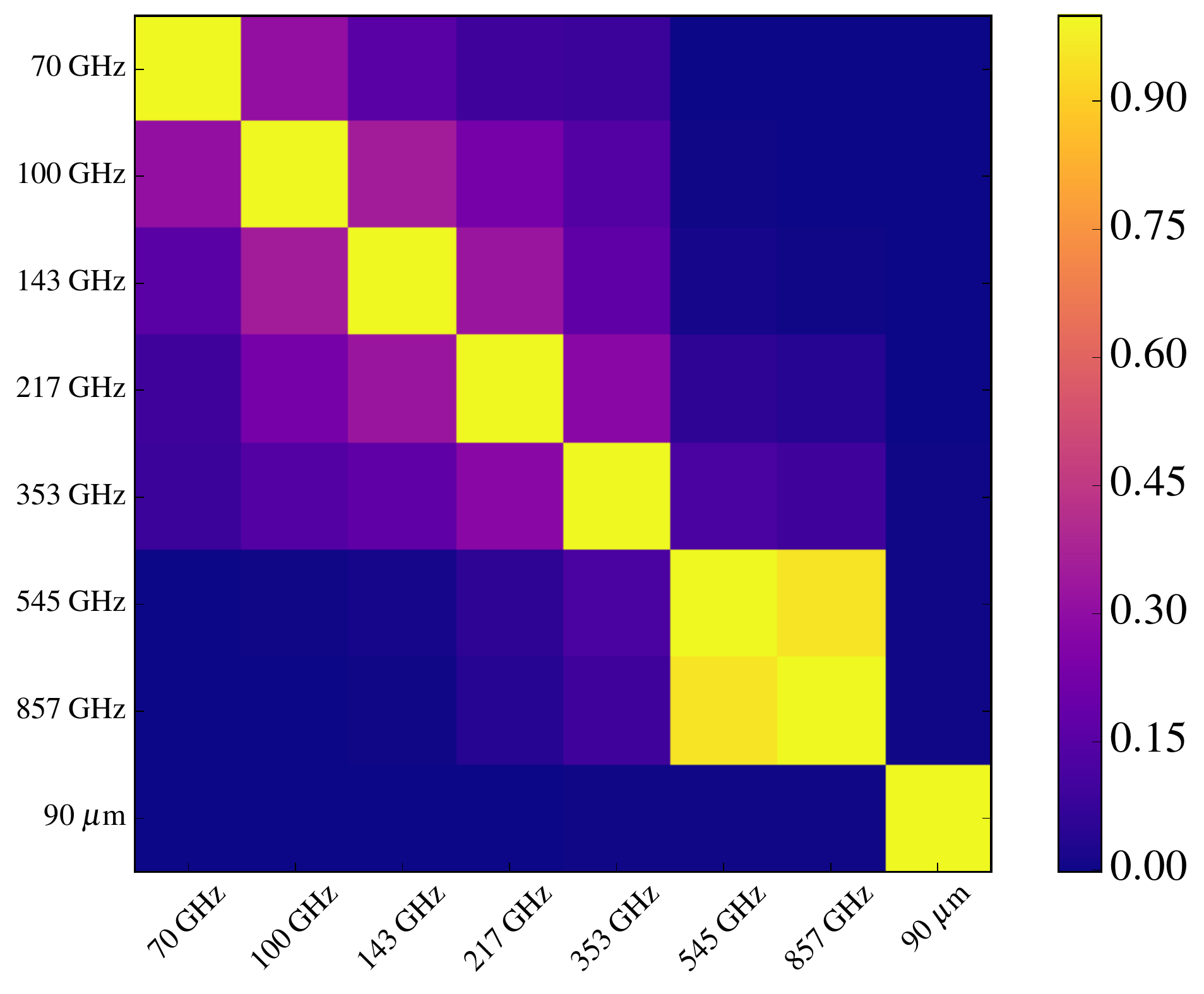}
	\caption{Correlation matrix for the QSO flux density measurement. 
	}
	\label{fig:corrmat_quasars}
\end{figure}

We finally add the contributions to the covariance matrix from the absolute and relative calibration of the individual bands,
which we have modelled as described in Section~\ref{sec:data}.
For visualisation purposes, we define the correlation matrix
$
R_{\nu \nu\prime} \equiv C_{\nu \nu\prime}/\sqrt{C_{\nu \nu}C_{\nu\prime \nu\prime}} 
$
and show it in Fig.~\ref{fig:corrmat_quasars} for the merged QSO sample.
The off-diagonal correlations partially originate from the inclusion of correlated calibration uncertainties.

\subsection{Emission modelling}
\label{sec:modelling}

We now describe how we model the flux density data with a multi-component emission model.

\subparagraph{SZ effect:}
In the non-relativistic approximation, the thermal SZ contribution to the total flux density can be written as (e.g. \citealt{Birkinshaw1998,Carlstrom2002})
\beq
S_\nu^\mathrm{SZ} = Y \times I_0 \, g(x) \, ,
\eeq
where $I_0 = 2 \left(k_B T_\mathrm{CMB}\right)^3/(hc)^2$; the SZ has a distinctive frequency dependence of the form 
\beq
g(x) = \frac{x^4 e^x}{\left(e^x -1\right)^2} \left[ x \coth\left(\frac{x}{2}\right) -4 \right] \, ,
\eeq
where $x \equiv h \nu/k_B T_\mathrm{CMB}$ is a dimensionless frequency.
The SZ template itself, $I_0 \, g(x)$, has units of specific intensity; its amplitude, $Y$, has 
units of solid angle, 
and it corresponds to the average integrated Compton-$y$ parameter
\beq
Y = \left\langle \int \mathrm{d}\Omega_i \, y(\nhat_i) \right\rangle_i \, \quad \text{with} \quad y=\int \mathrm{d}l \, n_e(\mb{r}) \, \sigma_T \, \frac{k_B T_e(\mb{r})}{m_e c^2}    \, .
\label{eq:SZ}
\eeq
In the above equation, $n_e$ and $T_e$ are the electron number density and temperature, and $\sigma_T$ is the Thomson cross section.

Our QSO sample spans a large range in redshift, therefore the projected angular sizes vary in our sample. 
Even in the absence of any intrinsic evolution of the QSO host properties, this could lead to a redshift evolution of the integrated Compton-$y$ parameter of individual objects, $Y_i$.
Variations in the intrinsic properties of the host haloes can be disentangled from size variations by binning the QSO data in narrow redshift bins (see Section~\ref{sec:zsplit} below).
This can, of course, increase the statistical noise. To co-add the data over a wide redshift range, we therefore define the rescaled (or intrinsic) parameter (e.g.~\citealt{Planckint_hotgas})
\beq
\hat{Y} = Y \times E(z)^{-2/3} \left[\frac{d_A(z)}{500 \, \Mpc} \right]^2 \,,
\label{eq:rescale_y}
\eeq
where $d_A(z)$ is the angular diameter distance to redshift $z$ and $E(z) = \left[ \Omega_m (1+z)^3  + \Omega_\Lambda \right]^{1/2}$ is the dimensionless Hubble expansion rate.
Unless otherwise noted, we use a flat $\Lambda$CDM cosmology with $\Omega_m = 0.3$ and $h=0.7$ to compute these quantities.
In the case of a purely self-similar scaling \citep{Kaiser1986} and fixed halo mass, $\hat{Y}$ is independent of redshift.
For the actual measurement we absorb this rescaling in the template, such that the amplitude of a recovered SZ signal directly measures $\hat{Y}$.

It is worth noting that $\hat{Y}$ measures the SZ signal integrated over a cylinder along the line of sight.
Under the assumption of a specific gas pressure profile, it is possible to convert this into a spherically integrated quantity, which is however not directly observable.
As gas pressure profiles for the low-mass haloes that typically host QSOs are not  well known, we have kept our 
analysis as model-independent as possible, and therefore we do not perform such a conversion; this should be kept in mind when comparing our results to those of previous studies.
We  note further that for the comparison to simulations presented in Section~\ref{sec:sims} such a conversion is not 
necessary either, as it is possible to directly predict the cylindrically integrated signal from the simulations.

Once we have measured $\hat{Y}$, we can relate it to the thermal energy content of the hot halo gas as follows:
for an individual halo at redshift $z$, the corresponding thermal energy is given by
\beq
E_\mathrm{th}(z) = \frac{3 m_e c^2}{2\sigma_T} \left(1+\frac{1}{\mu_e}\right) \, Y(z) \, d_A^2(z) \, ,
\label{eq:Eth_z}
\eeq
where $\mu_e$ is the average particle weight per electron and
$Y(z)$ is estimated from the mean redshift-independent $\hat{Y}$ via equation~\ref{eq:rescale_y}.
The mean thermal energy of our sample, accounting for the redshift distribution, is then given by
\beq
\bar{E}_\mathrm{th} = \int dz \, \frac{dN}{dz}(z) \, E_\mathrm{th}(z) \, ,
\label{eq:Eth}
\eeq
where $dN/dz(z)$ is the normalized redshift distribution of the sample.

\subparagraph{Dust emission (CIB):}
The second contribution to our model is the sub-mm emission from the QSO host galaxy or other galaxies along
the same line of sight. 
This is mostly optical and UV light that has been absorbed by dust grains and re-emitted in the infrared or sub-mm.
Essentially all of this emission is unresolved, hence it is also known as the cosmic infrared background (CIB).

We model the CIB with a single-component modified blackbody (e.g.~\citealt{Blain2002}), 
\beq \label{eq:dustbb}
S_\nu^\mathrm{dust} = 
A_{\nu_0}^\mathrm{CIB} \times 
\left[\frac{\nu(1+z)}{\nu_0}\right]^{\beta_d} \, B_\nu \left[\nu(1+z),T_d\right] \, ,
\eeq
where $B_\nu(\nu,T)$ is the Planck spectrum, $\beta_d$ is the dust spectral index, and $T_d$ is the dust temperature.
We normalize the dust emission at a frequency $\nu_0 = 353\,$GHz. Its amplitude, $A_{\nu_0}^\mathrm{CIB}$, has units of solid angle, analogously to $Y$.
As we are mainly interested in a potential SZ signature, we treat $A_{\nu_0}^\mathrm{CIB}$ as a purely phenomenological parameter quantifying the dust emission, and do not relate it to quantities such as the dust mass by adopting a specific model.
We note that the parameters governing the dust model can be significantly degenerate in SED fits:
the parameters $\beta_d$ and $T_d$ follow a well-known degeneracy, such that typically only the combination $T_d \beta_d^{0.6}$ 
is well constrained by the data \citep[e.g.][]{Verdier2015}.

The standard modified blackbody template of equation~\ref{eq:dustbb} is a simple parametrization of the complex dust physics. With the limited number of frequencies used in this paper, it is difficult to test more complex models, such as a two-temperature dust component (see e.g.~Appendix~C of \citealt{Crichton2015}).

\subparagraph{Dust emission (Galactic):}
Galactic dust emission is also well approximated by a modified blackbody \citep{Planck2013_dust}. 
This makes it hard to disentangle Galactic dust from the CIB based on their spectral 
properties only (e.g.~\citealt{Planck2015_dust,Planckint_dust,Mak2016}). However, emission from Galactic dust is uncorrelated with the QSO sample,  adding  noise but not bias to our measurement of the CIB amplitude. Furthermore, we focus 
on the northern and southern Galactic caps, where we do not expect a strong contamination by Galactic dust. We therefore
make no attempt to model Galactic dust, but instead test the sensitivity of our results by 
repeating our analysis with different masks (see Section~\ref{subsec:samplesel}).

\subparagraph{Synchrotron emission:}
We have removed all QSOs with a known radio counterpart, and those outside the footprint of the FIRST radio observations, from our sample. 
In Section~\ref{subsec:radiotest} we estimate that the contribution from radio sources below the FIRST detection threshold is $\lesssim 0.03$ mJy at 100 GHz; this is significantly smaller than the $1\sigma$ uncertainties on the lowest-frequency points.
We therefore do not expect a significant level of contamination by synchrotron emission. 
Nevertheless, we test for residual synchrotron contamination by fitting an additional component with
\beq
S_\nu^\mathrm{radio} = A_{100}^\mathrm{radio} \times \left(\frac{\nu}{100 \, \mathrm{GHz}}\right)^{-\alpha} \, ,
\eeq
where the spectral index $\alpha$ is typically found to be between 0.5 and 1 (e.g.~\citealt{Planck2016_radio}).
Here we either fix the spectral index to the most conservative choice $\alpha = 0.5$,
or fit it jointly with the radio amplitude $A_{100}^\mathrm{radio}$.

\subparagraph{Other contributions:}
There are several other sources of anisotropies in the microwave and sub-mm sky, in particular the primary CMB at the lower frequencies, the Poisson component of the CIB at the higher frequencies, and instrumental noise at all frequencies.
All of these are uncorrelated with the positions of the QSOs. 
Primary CMB and instrumental noise can be positive or negative, therefore they cancel out when averaging over all QSOs, adding noise but no bias to our flux density measurements. 
On the other hand, the CIB Poisson component is always positive and could in principle bias our measured SED.
In practice, we find that after filtering the maps as described in Section~\ref{sec:filtering} above, this contribution is negligible.
We explicitly demonstrate this by replacing the QSO positions with randomly drawn points with the same distribution on the sky as the original objects.
These are then processed with the same pipeline; we show the results in Appendix~\ref{app:randompoints}.
We note that this test also demonstrates that our results for the Compton-$y$ parameter are not affected by a stacking bias.

\subparagraph{Full model:}
The model for the emission of QSOs at redshift $z$ and frequency $\nu$ is thus 
\beq
S(\nu,z,\mathbf{p}) = \sum_c S^c(\nu,z,\mathbf{p}^c) \, ,
\eeq
where the sum over $c$ labels components. 
For the main analysis $c = \{\mathrm{SZ},\mathrm{dust}\}$, while for additional tests we also consider radio emission. 
Furthermore, $\mathbf{p}^c$ contains the parameters for component $c$ and $\mathbf{p}$ is the full parameter vector that we are aiming to constrain; i.e.~$\mathbf{p} = \{\hat{Y}, A_{\nu_0}^\mathrm{CIB}, \beta_d, T_d \}$ for the main analysis. For the radio contamination test we also include $A_{100}^\mathrm{radio}$ and $\alpha$. 

In this analysis, we do not measure the emission at a fixed central frequency $\nu$, but rather over a range of frequencies defined by the instrumental bandpasses.
In our case, the latter are relatively wide, especially for the AKARI bands, some of which span almost a factor of two in frequency (see Fig.~\ref{fig:bandpasses}).	
Additionally, the modified blackbody dust spectrum rises steeply for $ 300 \lesssim \nu \lesssim 800$ GHz, and then falls again sharply for $\nu \gtrsim 1$ THz.
For these  reasons, we must include the instrumental bandpasses $R(\nu)$ in our model prediction.

As the models for the individual components depend on $z_i$, one in principle has to evaluate them for every single QSO. 
To speed up the evaluation of the model, we group the QSOs into $N_{z-\mathrm{bins}}$ redshift bins, evaluate the model at the redshifts of the bin centres, and weight the bins with the redshift distribution $dN/dz$.
We have tested that with $N_{z-\mathrm{bins}} = 100$ the binned model only differs from the full model at the sub-percent level, which is well below the statistical uncertainties.

Our full model is thus
\beq
S_\nu(\mathbf{p}) = \int dz \, \frac{dN}{dz}(z) \int d\nu^\prime R(\nu^\prime) \sum_c S_\nu^c(\nu^\prime,z,\mathbf{p}^c) \, .
\eeq

\subsection{Model fitting}
Now we have all the ingredients in place to estimate the parameters of the QSO emission model.
We write the posterior probability distribution of the parameters in Bayesian fashion as
\beq
\mathrm{P}(\mb{p}|\bar{\mb{S}}_\nu) \propto \mathrm{P}(\bar{\mb{S}}_\nu|\mb{p}) \mathrm{P}(\mb{p}) \, ,
\eeq
where $\bar{\mb{S}}_\nu$ is the vector of mean observed flux densities, $\mathrm{P}(\bar{\mb{S}}_\nu|\mb{p})$ is the likelihood and $\mathrm{P}(\mb{p})$ is the prior assumed on the parameters.
In our case, we assume Gaussian likelihoods, so that
\beq
- 2 \ln \mathrm{P}(\bar{\mb{S}}_\nu|\mb{p}) =  \left[\mb{S}_\nu(\mb{p}) - \bar{\mb{S}}_\nu\right]^T C^{-1} \left[\mb{S}_{\nu}(\mb{p}) - \bar{\mb{S}}_{\nu}\right] \, .
\eeq
We restrict all parameters to positive values to avoid unphysical results; apart from this restriction, we choose the priors to be flat and uninformative.
We then sample from the posterior using the \texttt{emcee} implementation \citep{emcee} of an affine-invariant Markov Chain Monte Carlo (MCMC) ensemble sampler \citep{Goodman2010}.
To assess the convergence of our chains, we use the Gelman-Rubin criterion \citep{Gelman1992,Brooks1998}.

We then report the results for the QSO emission parameters and their $\pm 1\sigma$ uncertainties as the 50th, 84th and 16th percentile of the one-dimensional (i.e.~marginalized over all other parameters) posterior distribution of the respective parameter.
If a parameter is consistent with zero at the $2\sigma$ level, we also report the $2\sigma$ upper limit, defined as the 95th percentile of the respective 1D posterior.

\section{Results}
\label{sec:results}

\subsection{Galaxy clusters}
\label{subsec:results_clusters}

\begin{figure*}
	\centering
	\includegraphics[width=1.5\columnwidth]{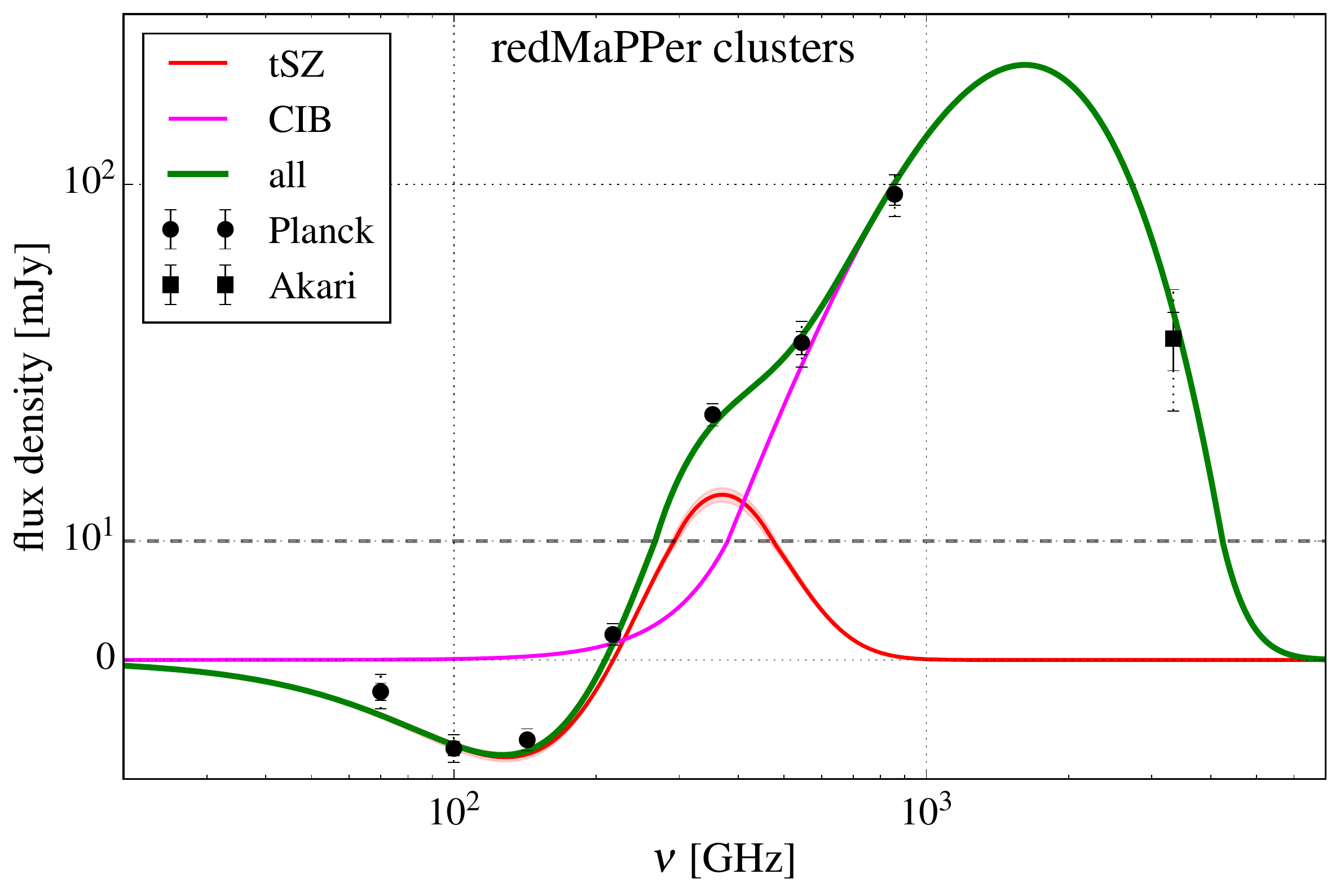}
	\includegraphics[width=1.5\columnwidth]{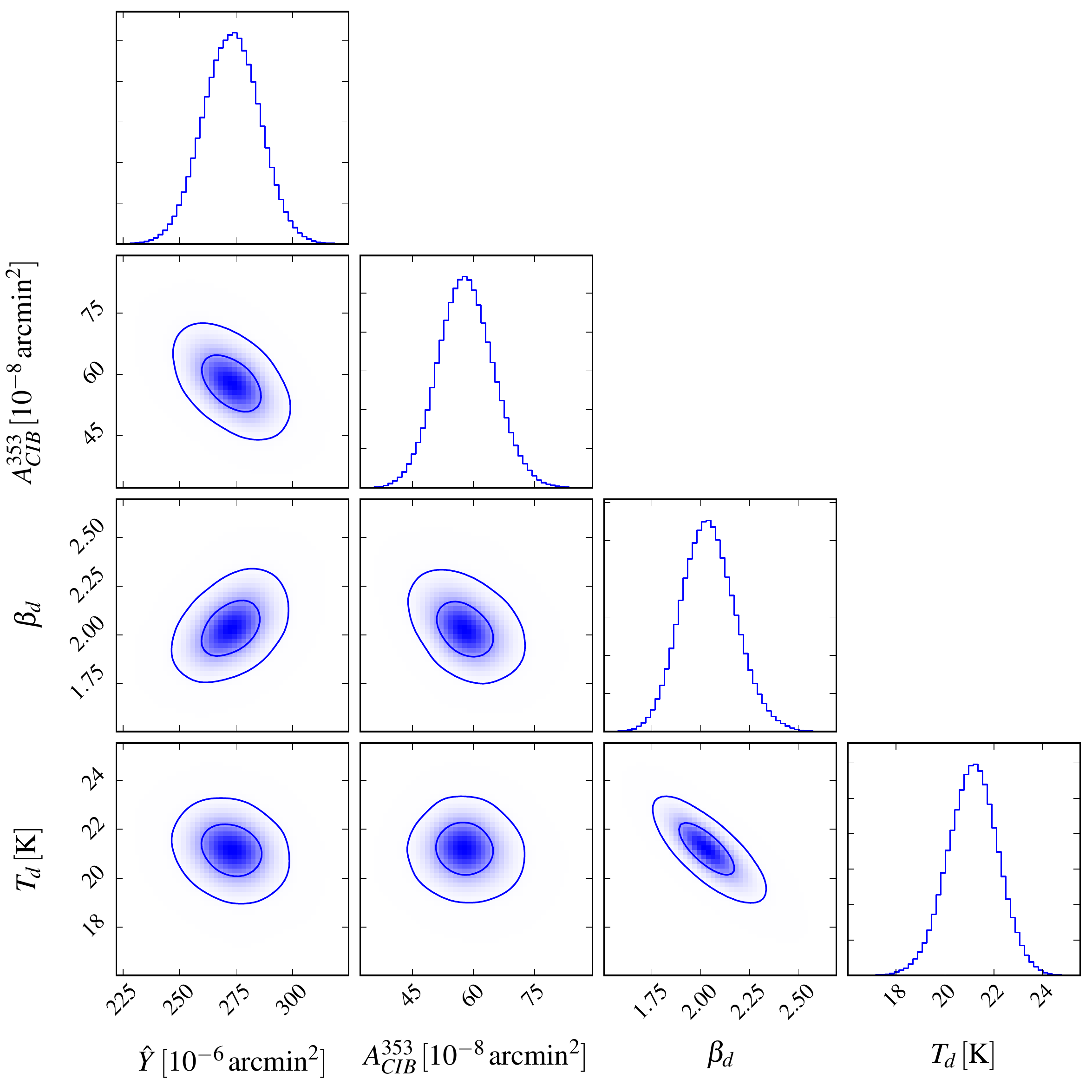}
	\caption{Results for the \textit{RedMaPPer galaxy clusters}. \textbf{Top:} we show here the measured flux densities and the best-fitting model together with its individual components. For clarity, we have included both the $1\sigma$ (solid) and the $2\sigma$ (dashed) error bars on the measured flux densities.
		Furthermore, the red shaded region shows the $1\sigma$ uncertainty on the amplitude of the SZ signal.
		Note that the vertical scale is linear for flux densities below $10~\mathrm{mJy}$, and then switches to a logarithmic scale; the dashed horizontal line marks the transition.	
		\textbf{Bottom}: results of the MCMC parameter estimation. The panels on the diagonal show the marginalized 1D posteriors on the four parameters, whereas the off-diagonal panels show marginalized 2D constraints. Here, the colour scale denotes the density of MCMC samples, while the contours show the $68\%$ and $95\%$ confidence regions. Note that $\hat{Y}$ and $A_{\nu_0}^\mathrm{CIB}$ are displayed in units of $10^{-6}$ and $10^{-8}~\arcm^2$, respectively, to allow for a direct comparison with the QSO results in Fig.~\ref{fig:bestfit_model_quasars} below. 
		}
	\label{fig:bestfit_model_clusters}
\end{figure*}

We begin with discussing the results from the RedMaPPer clusters, 
which we provide as a test of our analysis pipeline, demonstrating that we are able to recover a significant SZ signal associated with optically selected clusters, together with physically sensible dust emission parameters.
We show in Fig.~\ref{fig:bestfit_model_clusters} the flux densities measured from the cluster sample together with our best-fitting model from the MCMC analysis, and the constraints on the emission parameters.

For the integrated, rescaled Compton-$y$ parameter of the clusters we obtain
\beq
\hat{Y} = (2.73 \pm 0.12) \times 10^{-4} \, \arcm^2 \, ,
\eeq
which is a highly significant detection ($>$$20\sigma$), as expected.
From equation~\ref{eq:Eth}, this corresponds to a mean thermal energy in the hot halo gas of
\beq
\bar{E}_\mathrm{th} = (2.14 \pm 0.10) \times 10^{62} \, \mathrm{erg} \, .
\eeq
Furthermore, we measure an amplitude of the CIB dust emission
$ A_{\nu_0}^\mathrm{CIB} = 5.80^{+0.68}_{-0.64} \times 10^{-7} \, \arcm^2$
and modified blackbody parameters of
\beq
\beta_d = 2.04^{+0.14}_{-0.13} \, , \quad
T_d = (21.1 \pm 1.0)~\mathrm{K}\, .
\eeq
These results are also summarized in Table~\ref{tab:results}.
The strong correlated emission in the sub-mm bands indicates that the dust emission from the galaxies within redMaPPer clusters is clearly detected in our data. 
The levels of sub-mm emission that we find are compatible with those found by \cite{Bleem_thesis} when stacking maps from the SPT-SZ survey and \textit{Herschel}-SPIRE at the positions of optically selected clusters from the Blanco Cosmology Survey \citep{Bleem2015Blanco}.
Evidence for dust in clusters has also been reported by \mbox{\cite{Chelouche2007}} and \cite{Gutierrez2014}.

Our dust emissivity index $\beta_d$ is in good agreement both with the theoretical expectation of $\beta_d \sim 1.5-2$ in the FIR (e.g.~\citealt{Blain2002,Franceschini2000}) and the value used to fit sub-mm SEDs of individual galaxies (e.g.~\citealt{Calzetti2000}).
Furthermore, the dust temperature of $ \simeq 20~\mathrm{K}$ is 
consistent with the temperature seen in normal quiescent galaxies \citep[e.g.][]{Smith2012},
as we would expect for member galaxies of the relatively low-redshift SDSS RedMaPPer clusters.

Our simple two-component model nominally does not provide a good fit to all data points: we find $\chi^2_\mathrm{red} = 16.0/4$ when evaluated at the parameters given by the 50-th percentile of the MCMC chain.\footnote{Note that this is not the nominal minimum $\chi^2$, but should be close to it in the case of Gaussian posterior distributions.}
However, this unusually high value is driven largely by the 70~GHz data point, which deviates from the best-fitting model by $\simeq 3\sigma$.
This is mainly caused by radio emission from cluster members, which is known to be a potential bias for SZ studies (e.g.~\citealt{Bleem_thesis,Gupta2016}).
The value of $\chi^2_\mathrm{red}$ decreases significantly if we also fit for the amplitude of a radio component ($\chi^2_\mathrm{red} = 8.6/3$) or perform a radio cut that removes all clusters within $3\arcmin$ of a FIRST-detected radio source ($\chi^2_\mathrm{red} = 7.1/4$).
However, the former ignores the uncertainty on the radio spectral index (see Section~\ref{sec:modelling}), whereas the latter removes roughly half of our clusters (mostly due to chance associations) and introduces an additional selection effect (see Section~\ref{subsec:samplesel}).
In these cases, the constraint on $\hat{Y}$ (or $\bar{E}_\mathrm{th}$) is affected at the $\sim 10\%$ level. We therefore include an additional systematic uncertainty when comparing our results to simulations in Section~\ref{sec:sims}.
Another option would be to completely discard the 70~GHz data point for the cluster analysis. This would also lead to a significantly improved $\chi^2_\mathrm{red}$, while not changing our results significantly.

\subsection{QSOs}
\label{sec:resqso}
\begin{figure*}
	\centering
	\includegraphics[width=1.5\columnwidth]{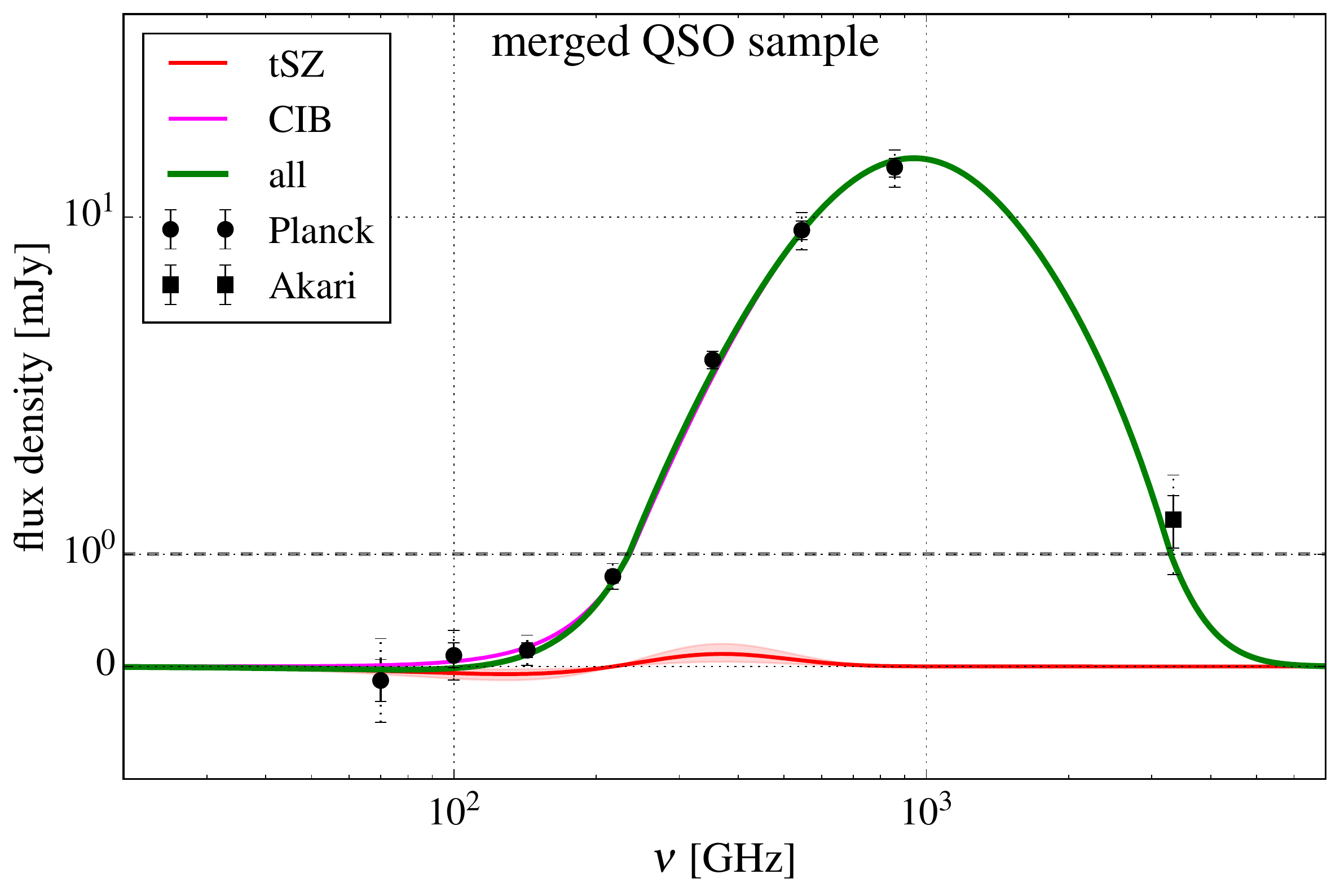}
	\includegraphics[width=1.5\columnwidth]{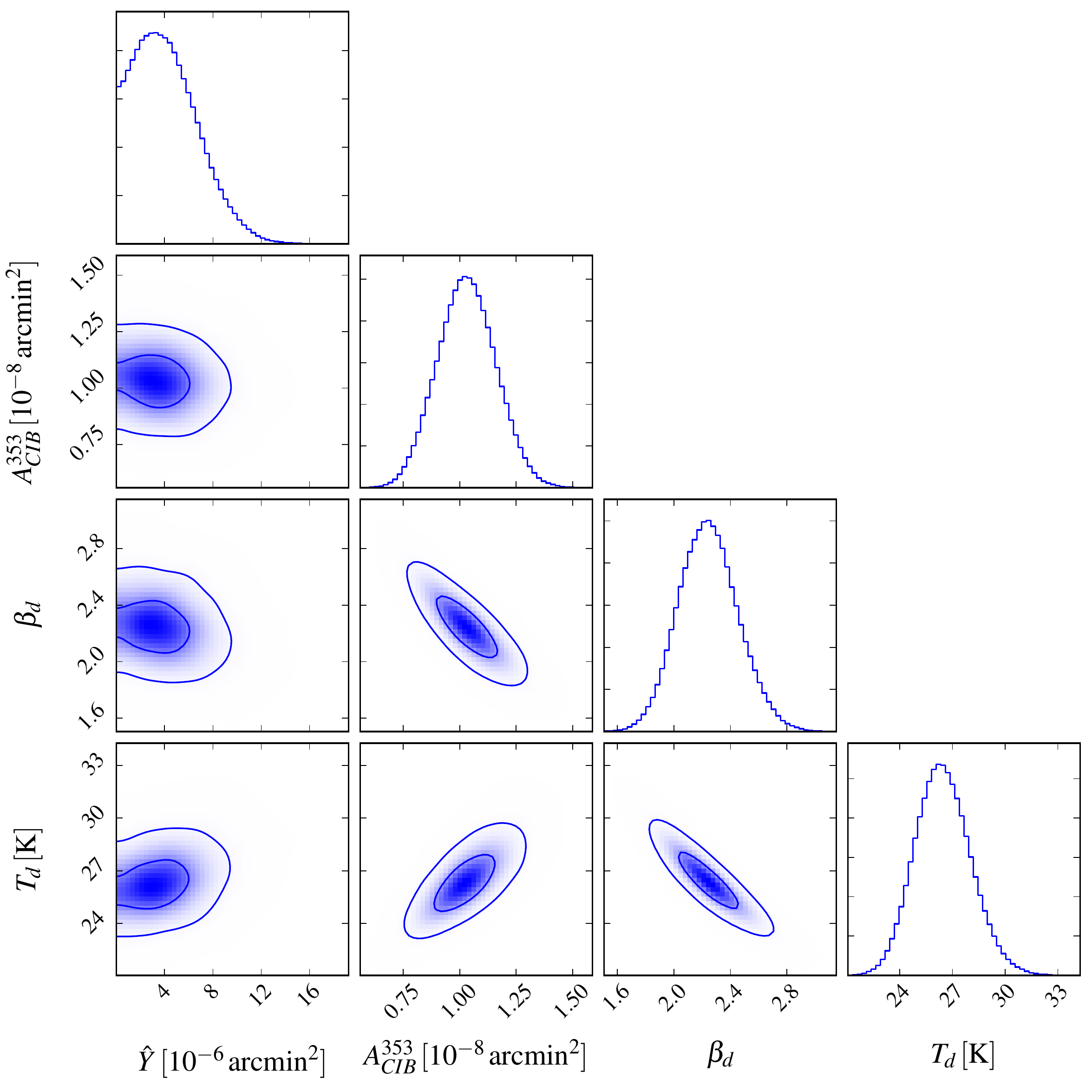}
	\caption{
	Results for the \textit{merged QSO sample}.
	\textbf{Top:} we show here the flux density measurement and best-fitting emission model for the merged QSO sample. 
	The vertical scale is linear for flux densities below $1~\mathrm{mJy}$, and logarithmic above this value.
	As before, the red shaded region shows the $1\sigma$ confidence region for the SZ contribution to the measured SED. Unlike the case for clusters, there is only weak evidence for an SZ contribution to the measured SED.
	\textbf{Bottom:} this panel shows the MCMC results for the QSO sample. 
	It is worth noting that there is no strong degeneracy between $\hat{Y}$ and the dust parameters, showing that our result on $\hat{Y}$ is robust against uncertainties in the dust solution.  
	}
	\label{fig:bestfit_model_quasars}
\end{figure*}

We proceed with the constraints on the QSO emission model, which are the main result of this paper.
Similarly to Fig.~\ref{fig:bestfit_model_clusters} for the clusters, Fig.~\ref{fig:bestfit_model_quasars} shows the flux density measurement, the best-fitting model, and the MCMC results for the merged QSO sample.
Here our dust+SZ model provides an acceptable fit to the data ($\chi^2_\mathrm{red} = 7.0/4$ when evaluated at the 50-th percentile as before).

We measure an SZ amplitude of
\beq
\hat{Y} = 3.7^{+3.0}_{-2.3} \, \times \, 10^{-6} \, \arcm^2 \, ; \quad \hat{Y} < 8.6 \times 10^{-6} \, \arcm^2 \, (95\%) .
\eeq
The data indicate a mild preference for a non-zero SZ signal, but its amplitude is consistent with zero at the $1.6\sigma$ level.
This translates into a thermal energy of
\beq
\bar{E}_\mathrm{th} = 5.2^{+4.2}_{-3.3} \, \times \, 10^{60}~\mathrm{erg} \, ; \quad \bar{E}_\mathrm{th} < 12 \times 10^{60}~\mathrm{erg} \, (95\%) \, .
\eeq

These numbers are significantly smaller than those reported by \cite{Ruan2015}, and more in line with the feedback energetics suggested by hydrodynamical simulations.
Section~\ref{sec:discussion} below presents a more detailed discussion of the implications  of these results for QSO feedback energetics and a comparison to previous studies.

Regarding the dust parameters, we measure an amplitude of 
$ A_{\nu_0}^\mathrm{CIB} = ( 1.03 \pm 0.13 ) \times 10^{-8} \, \arcm^2$;
the modified blackbody parameters are
\beq
\beta_d = 2.25 \pm 0.21 \, , \quad
T_d = (26.4 \pm 1.6) \, \mathrm{K} \,.
\label{eq:qsodustresult}
\eeq
These QSO results are also summarized in Table~\ref{tab:results}.
Here we find a dust emissivity index that is marginally higher than that for the clusters, but it is consistent with $\beta_d = 2$ at the $\simeq 1\sigma$ level.
The best-fitting dust temperature determined from the QSOs is significantly higher than that for the clusters.
This is perhaps not surprising, as the dust in the QSO host galaxy may be heated by the UV and optical emission from the central QSO.
In addition, as the typical redshift of the QSOs is $z \simeq 2$ --- close to the peak epoch of cosmic star formation (e.g.~\citealt{Madau2014}) ---
the typical QSO host galaxy should have a larger fraction of O/B-type stars than a quiescent low-$z$ cluster galaxy, providing additional UV/optical flux.
Both of these effects will lead to an increase of the dust temperature compared to normal galaxies. We note, however, that the dust temperature of equation~\ref{eq:qsodustresult} is lower than the temperature of  $T_d \sim 40 - 50 \, \mathrm{K}$ seen in optically luminous, higher-redshift QSOs. \citep[e.g.][]{Priddey2001,Omont2003}.

For this main result, we have only used the bands that have a well-determined calibration and satisfy a successful stacking null test (see Sections~\ref{sec:akari} and Appendix~\ref{app:randompoints}, respectively).
In Appendix~\ref{app:impact} we repeat the analysis adding in additional AKARI and IRAS bands. In this case, the AKARI bands with uncertain calibration pull the dust solution away from the best-fit values determined above. 
This leaves more room for an SZ signal, but at the same time provides a poor overall fit to the data. 
We provide this example in order to demonstrate the sensitivity of the SZ result to an accurate dust solution.

\subsection{QSOs: constraints on residual radio emission}
\label{subsec:radiotest}

\begin{table*}
	\def\arraystretch{1.3} 
	\begin{tabular}{lccccccc}
		\toprule
		& $z_m$ & $\hat{Y}~[10^{-6}~\mathrm{arcmin}^2]$ & $\bar{E}_\mathrm{th}~[10^{60}~\mathrm{erg}]$ & $\bar{E}_\mathrm{th}^\mathrm{95\%}~[10^{60}~\mathrm{erg}]$& $A_{\nu_0}^\mathrm{CIB}~[10^{-8}~\mathrm{arcmin}^2]$ & $T_d~[\mathrm{K}]$ & $\beta_d$ \\		
		\midrule		
		redMaPPer clusters & 0.37 & $273 \pm 12$ & $214 \pm 10$ &  $<229$ & $58.0^{+6.8}_{-6.4}$ & $21.1 \pm 1.0$ & $2.04^{+0.14}_{-0.13}$ \\
		\midrule		
		merged QSO sample & 2.15 & $3.7^{+3.0}_{-2.3}$ & $5.2^{+4.2}_{-3.3}$ & $< 12$ & $1.03 \pm 0.13 $ & $26.4 \pm 1.6 $ & $2.25 \pm 0.21$ \\		
		\hspace{10pt}+ radio amplitude  & 2.15 & $5.0^{+3.3}_{-2.9}$ & $7.0^{+4.7}_{-4.1}$ & $< 15 $ & $0.96 \pm 0.13$ & $25.9^{+1.7}_{-1.5}$ & $2.33 \pm 0.23 $ \\		
		\bottomrule
	\end{tabular}
	\caption{We summarize here our best-fitting model parameters from Section~\ref{sec:results}. Except for the 95\% upper limit on $\bar{E}_\mathrm{th}$, all error bars are $\pm 1 \sigma$, obtained from the 16th and 84th percentile of the marginalized posterior distributions.
		Note that the conversion between $\hat{Y}$ and $\bar{E}_\mathrm{th}$ depends on the redshift distribution of the sample.}
	\label{tab:results}
\end{table*}

We have already removed all QSOs associated with radio sources above the FIRST 1.4 GHz detection threshold of $S_0 = 1$~mJy, which removes $f_\mathrm{det} \simeq 5\%$ of the QSOs.
In our main analysis, we have therefore assumed that synchrotron emission does not contribute significantly to our flux density measurements.
In this section we check this assumption by estimating the contribution to our flux density measurements from faint radio sources with FIRST flux density $S_{F} < S_0$.

Naively one could approach this by simply fitting an additional radio component to the flux density measurements.
However, only the two to three lowest-frequency points are sensitive to radio emission; therefore the SZ and synchrotron amplitudes are highly degenerate, introducing the risk of overfitting.  
This was also noted by \cite{Verdier2015}, who found that their three-component multi-frequency filter could not distinguish between SZ and a synchrotron component with `negative' amplitude, and vice versa.
We therefore first estimate an upper limit for residual radio contamination from the emission properties of QSOs with a radio counterpart detected in FIRST. 
We find that the expected level of radio contamination is well below the statistical uncertainties.
For completeness, we then repeat our MCMC analysis including a radio component, with a prior for the radio amplitude based on our estimate of residual radio contamination.

We begin by estimating the radio luminosity function of the SDSS/BOSS QSOs that have a radio counterpart detected in FIRST, finding that it is well approximated by 
\beq
\frac{\mathrm{d}N}{\mathrm{d}S_\mathrm{F}} \propto S_\mathrm{F}^{-1.4} \equiv n(S_{F})\quad \text{for} \quad S_0 < S_\mathrm{F} \lesssim 200~\mathrm{mJy} \, ,
\label{eq:radiolum}
\eeq
where the flux density range up to $S_\mathrm{max} = 200$~mJy contains more than 95\% of all detected radio counterparts in our sample; for larger flux densities 
the luminosity function cuts off exponentially.
We assume that the scaling of equation~\ref{eq:radiolum} approximately holds true for $S_\mathrm{F} < 1~\mathrm{mJy}$ as well.
Then the ratio of QSOs associated with radio sources below the FIRST detection threshold to those with detected counterparts, $r_\mathrm{undet}$, and the mean flux density of the undetected sources, $\bar{S}_F^\mathrm{undet}$, are given by
\beq
r_\mathrm{undet} \simeq \frac{ \int_{S_\mathrm{min}}^{S_0} \mathrm{d}S_{F} \, n(S_{F}) }  { \int_{S_0}^{S_\mathrm{max}} \mathrm{d}S_{F} \, n(S_{F}) } 
\,, \quad 
\bar{S}_F^\mathrm{undet} \simeq \frac{ \int_{S_\mathrm{min}}^{S_0} \mathrm{d}S_{F} \, n(S_{F}) S_{F}}  { \int_{S_\mathrm{min}}^{S_0} \mathrm{d}S_{F} \, n(S_{F}) } \, , 
\eeq
where $S_\mathrm{min}$ is the lower flux density limit that we consider for our estimate of radio contamination. 
For concreteness we use $S_\mathrm{min} = 0.01$~mJy, leading to $r_\mathrm{undet} \simeq 6$ and $\bar{S}_F^\mathrm{undet} = 0.12$~mJy.
The lower limit was chosen to be around 10\% of the $1\sigma$ uncertainty of our 100~GHz flux density measurement; even fainter radio sources would contribute negligibly to our measurement.

The radio spectrum of synchrotron sources detected in microwave bands can typically be approximated with a broken power law $\nu^{-\alpha}$, which is relatively flat ($\alpha \simeq 0$) for $\nu < \nu_\mathrm{br}$  and steepens ($\alpha \simeq 0.5$) for $\nu > \nu_\mathrm{br}$, with $\nu_\mathrm{br} \simeq 50$~GHz (e.g.~\citealt{Planck2016_radio}). 
Using these values, we estimate the residual radio contamination at 100~GHz as
\beq
S^{100}_\mathrm{radio} \simeq f_\mathrm{det} \, r_\mathrm{undet}\, \bar{S}_F^\mathrm{undet} \times \left(\frac{100~\mathrm{GHz}}{\nu_\mathrm{br}}\right)^{-0.5} \simeq  0.03~\mathrm{mJy} \, .
\label{eq:radiocont}
\eeq
We consider this a conservative estimate, for the two following reasons: 
(1) We have not accounted for a flattening of $n(S_{F})$ at the faint end.
(2) By construction our radio counterparts are associated with high-redshift sources, and therefore likely at higher $z$ than the sample studied in \citet{Planck2016_radio}. 
If the rest-frame spectra of these objects are nevertheless broadly comparable, we have likely underestimated the `downscaling' in frequency in equation~\ref{eq:radiocont}. 
Both of these effects would cause us to overestimate $S^{100}_\mathrm{radio}$.
We thus estimate the radio contamination at 100~GHz to be $< 0.03$~mJy, which is at most a third of the purely statistical uncertainties on our flux density measurement.
We therefore conclude that residual synchrotron emission does not contribute significantly to our measurement and can safely be neglected.

As an additional test, we repeated our MCMC analysis with an additional radio component. Based on the 
estimates presented above,
we chose a prior $0 < A_{100}^\mathrm{radio} < 0.1~\mathrm{mJy}$. 
The results of the three-component fits are given in Table~\ref{tab:results}.
The constraint on $A_{100}^\mathrm{radio}$ is weak and dominated by the prior choice ($A_{100}^\mathrm{radio} < 0.08~\mathrm{mJy}$ at 84\% C.L.).
We find no evidence for a radio component, in agreement with our estimate in equation~\ref{eq:radiocont}.
The addition of a radio component has almost no effect on the parameters of the dust component, but does, however, slightly degrade our constraints on the SZ component.
We now measure
$E_\mathrm{th} = 7.0^{+4.7}_{-4.1} \times 10^{60}~\mathrm{erg}$, 
or an upper limit of $E_\mathrm{th} < 15 \times 10^{60}~\mathrm{erg}$ (95\% C.L.).

We have also tested allowing a completely free radio amplitude (i.e.~without the physically motivated prior from above), and/or to fit the radio spectral index $\alpha$ jointly with the amplitude.
In none of these cases we have found evidence for radio contamination or a significant detection of an SZ signal, consistently with our physical argument given above.

\subsection{QSOs: redshift evolution}
\label{sec:zsplit}

\begin{figure*}
	\centering
	\includegraphics[width=0.9\columnwidth]{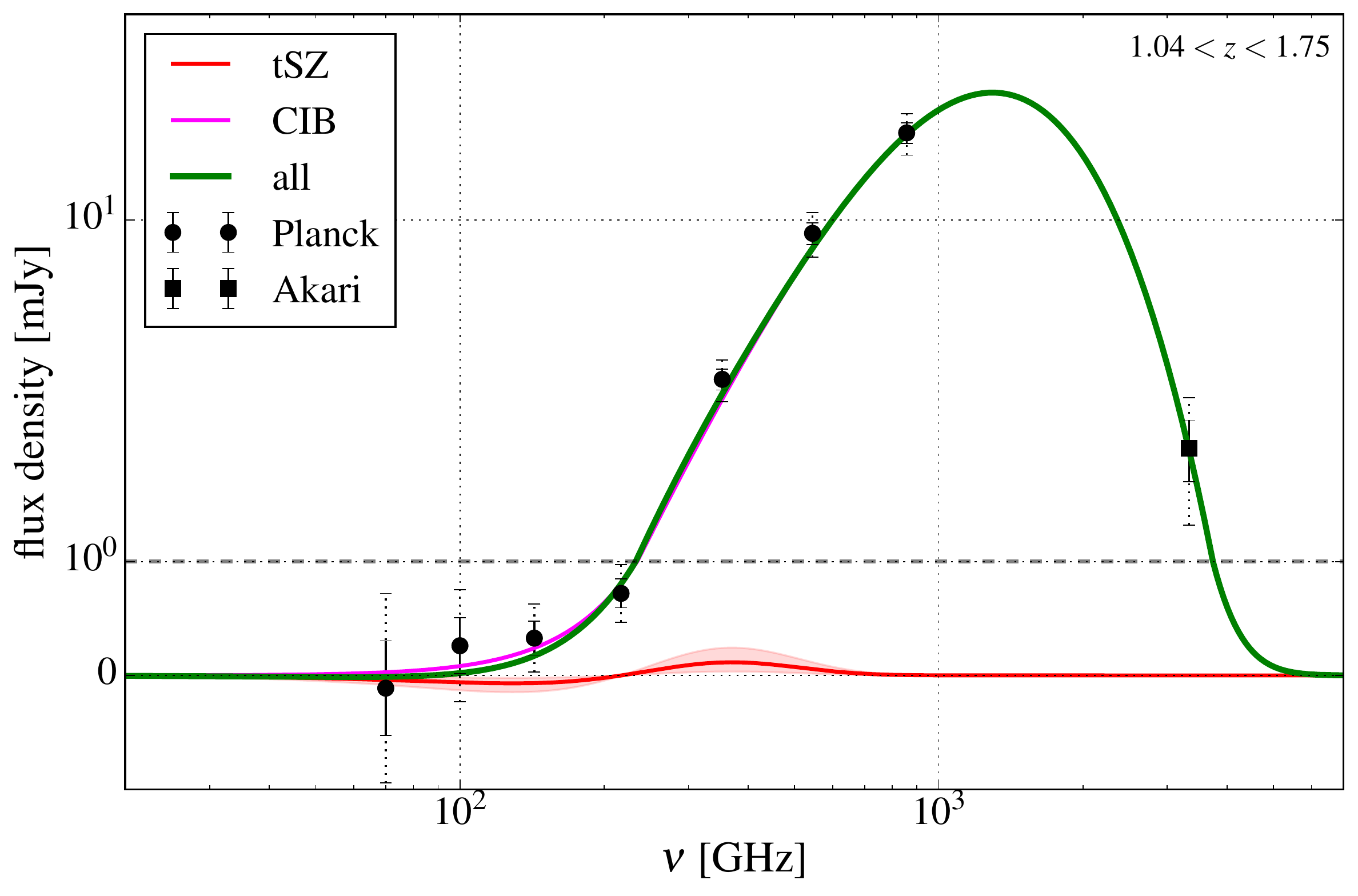}
	\includegraphics[width=0.9\columnwidth]{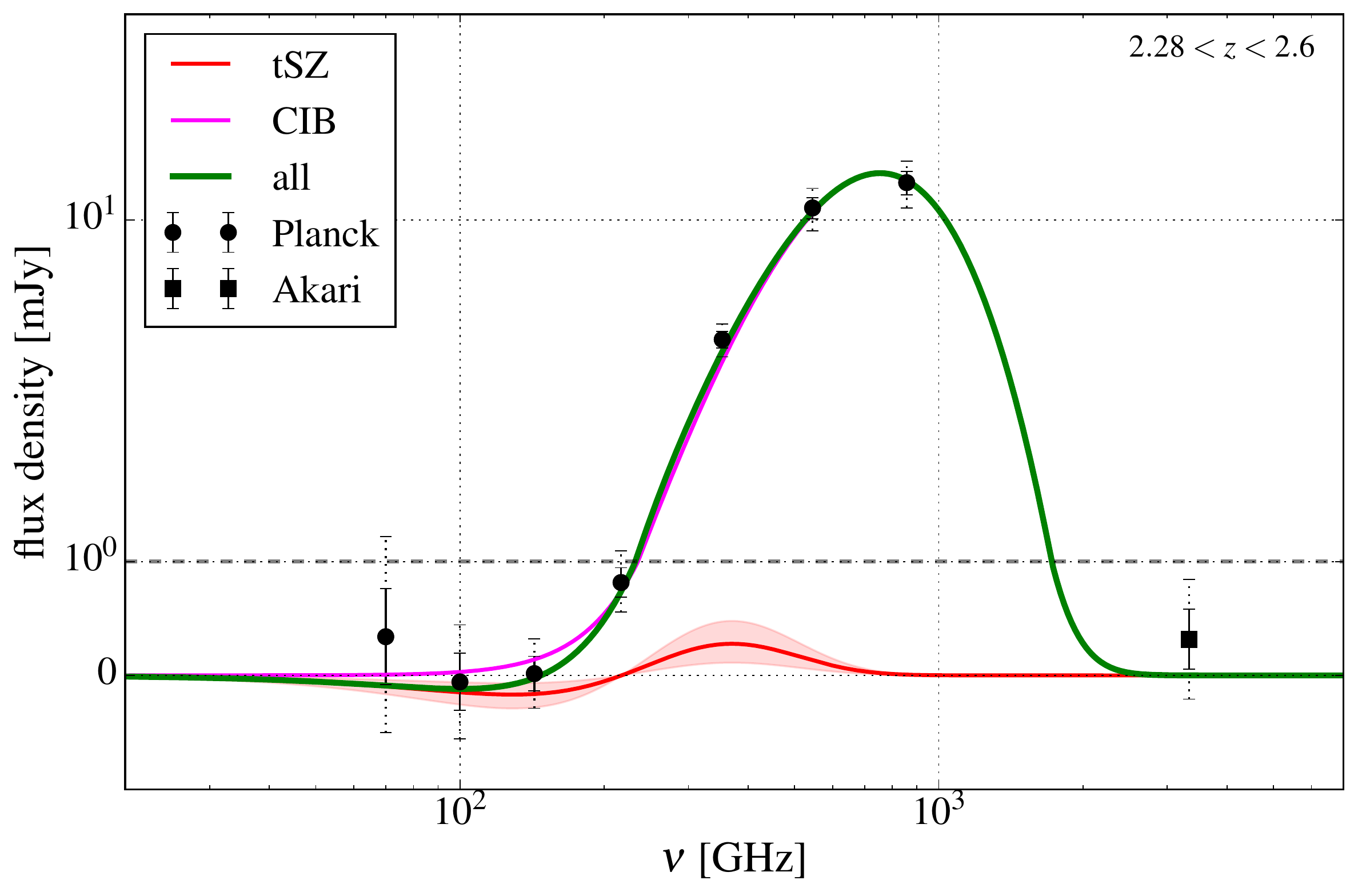}
	\caption{Redshift split: we show here two representative examples of the measured SED and best-fitting model.
	\textbf{Left}: Second redshift bin ($1.04 < z < 1.75$).
	\textbf{Right}: Fourth redshift bin ($2.28 < z < 2.6$). 
	The redshifting of the SED is clearly visible when comparing the two panels. 
	}
	\label{fig:zsplit_24}
\end{figure*}

\begin{figure}
	\centering
	\includegraphics[width=\columnwidth]{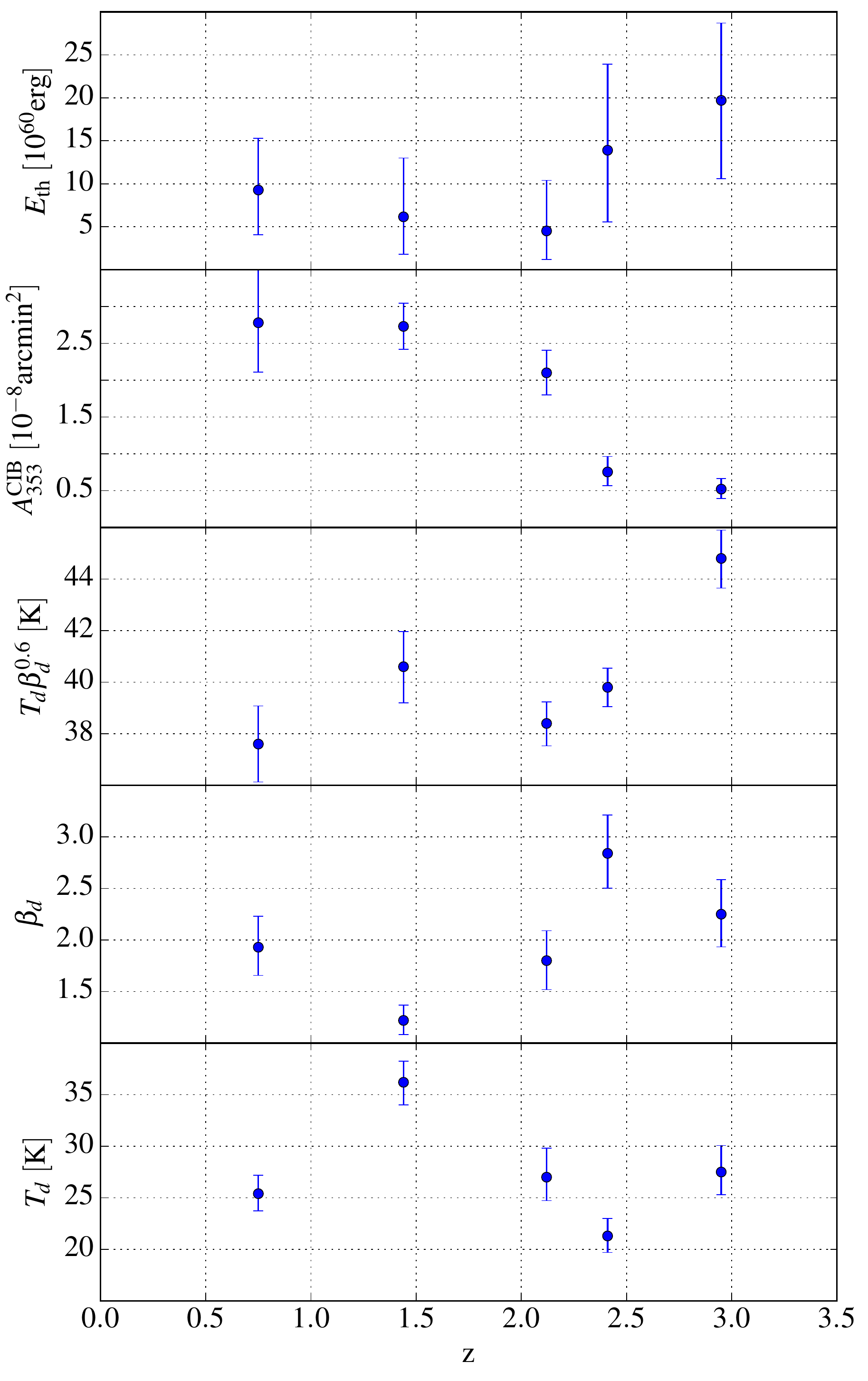}
	\caption{Redshift split: we show here the redshift evolution of the parameters of the emission model (plotted at their respective median $z$). From top to bottom, the panels refer to the total thermal energy, CIB amplitude, the well-constrained dust parameter combination $T_d\beta_d^{0.6}$, and further $\beta_d$ and $T_d$ separately. 
	}
	\label{fig:zsplit_new}
\end{figure}

In the analysis presented so far, we have rescaled the Compton-$y$ parameter so that it is independent of redshift in the case of purely self-similar scaling (see equation~\ref{eq:rescale_y}). 
Nonetheless, especially given the large redshift range of the QSO sample, there could still be a redshift evolution caused by deviations from self-similarity.
Furthermore, any change in the AGN activity over cosmic time would also affect the measured thermal energies. 
Therefore we proceed to split the QSO sample into five redshift bins with approximately equal number of objects, leading to splits at $z=\{1.04,1.75,2.28,2.6\}$.
We then repeat the measurement of the stacked SED and of the MCMC sampling; as examples, we show the results of the second and fourth $z$-bin in Fig.~\ref{fig:zsplit_24},
and we present in Fig.~\ref{fig:zsplit_new} the full redshift evolution of the measured emission parameters.

Similarly to the main sample, we also do not obtain a strong detection of the SZ signal in any redshift bin; 
only in the highest $z$-bin the significance is $\gtrsim 2 \sigma$.
We further observe some redshift evolution in the dust parameters, with a particularly low $\beta_d$ and high $T_d$ in the second redshift bin.
However, the combination $T_d \beta_d^{0.6}$, which is best constrained by the data, stays relatively stable (except for the highest $z$-bin).
Qualitatively, a similar trend in the dust parameters was also found by \cite{Verdier2015}. Quantitatively, however, they reported a much larger variation between the low-$z$ and high-$z$ dust solutions, which is in some tension with the physical expectation of $\beta_d \simeq 2$ 
in the low-frequency limit of optically thin dust emission (e.g.~\citealt{Blain2002}). 

The evolution of the dust parameters could indicate changes in the dust properties, which could potentially be linked to the redshift dependence of the star formation rate and/or AGN accretion rate.
However, our data does not allow for any more conclusive statements on this hypothesis.

\section{Interpretation and Discussion}
\label{sec:discussion}

In this section, we first compare our results for both clusters and QSOs to hydrodynamical simulations 
that include AGN feedback. We then discuss the implications of our measurements for QSO host halo masses and feedback energetics. Finally, we compare our results and conclusions with those of previous studies.

\subsection{Comparison to simulations}
\label{sec:sims}

\begin{figure*}
	\centering
	\includegraphics[width=1.8\columnwidth]{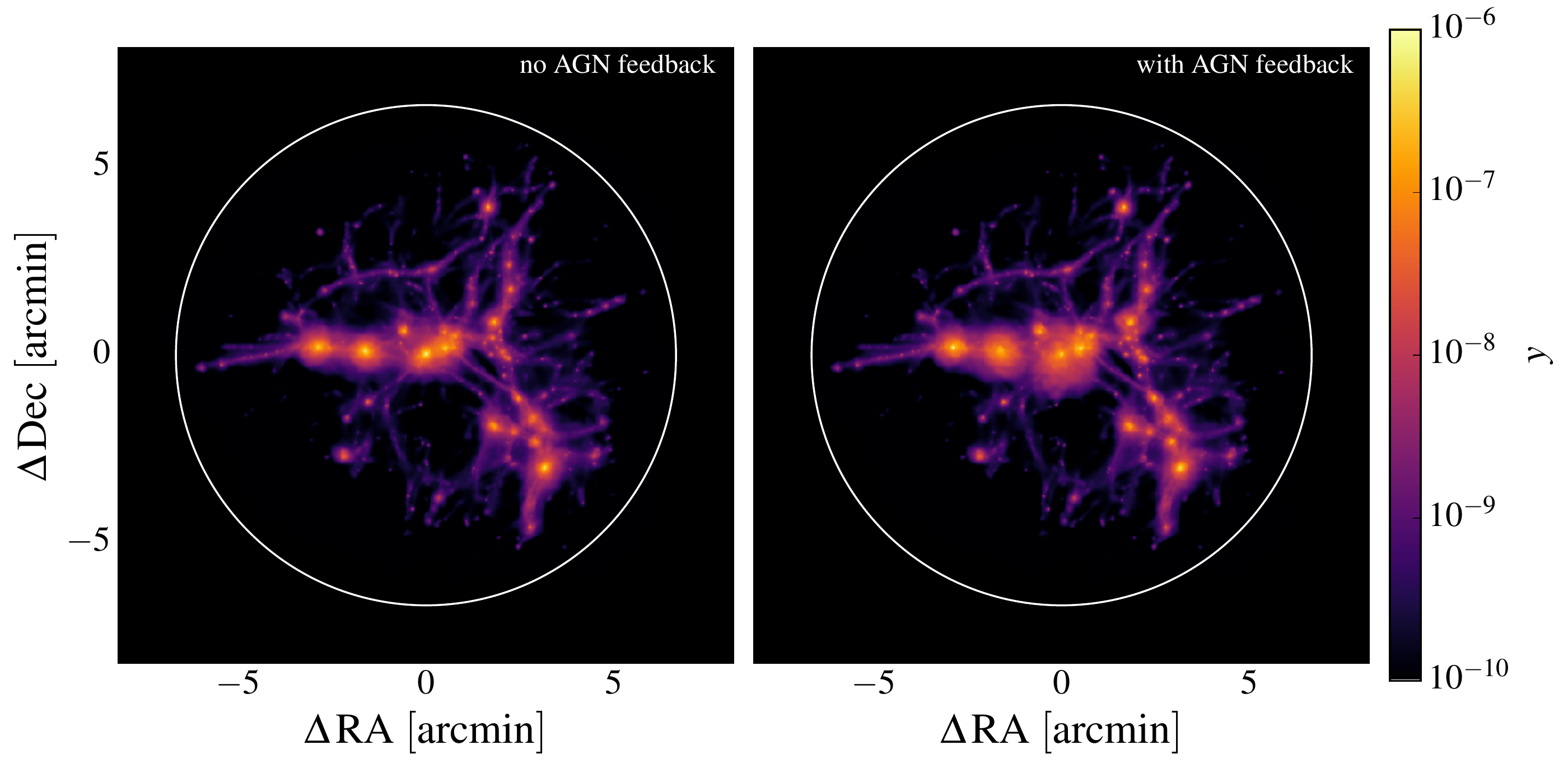}
	\caption{Projected Compton-$y$ maps: we show here the maps obtained from the $z=2.07$ snapshot of a halo with $M_{200c} = 1.4 \times 10^{12}~\hinv \Msol$. The left panel shows the case without AGN feedback, whereas the right-hand side includes AGN feedback. While we see evidence of gas being pushed towards the outskirts of the main halo, the overall thermal energy only increases by $20\%$ in this representative example. The white circle denotes the $\mathrm{FWHM = 13.3\arcmin}$ beam of the \Planck 70~GHz band, which is broadly representative of the `aperture' probed by our analysis.}
	\label{fig:sim_bh_nobh}
\end{figure*}

We use high-resolution zoom-in simulation of individual haloes spanning a mass range from $5 \times 10^{12}~\hinv M_\odot$ to $4 \times 10^{14}~\hinv M_\odot$ at $z=0$, 
covering the expected masses for QSO hosts and the lower end of the galaxy cluster mass scale.
The simulations have been performed with the smoothed particle hydrodynamics (SPH) code P-GADGET3 (which is based on GADGET-2, \citealt{Springel2005}) using a zoomed initial conditions technique \citep{Tormen1997}.
\footnote{The simulations we use have an input cosmology that is slightly different from our fiducial cosmology ($\Omega_m = 0.25$, $\Omega_\Lambda = 0.75$, $h=0.73$), but given the uncertainties in our estimates of thermal energy and mass this does not affect our comparison significantly.}

These simulations have previously been used to demonstrate the influence of AGN feedback on X-ray group and cluster scaling relations \citep{Puchwein2008} and on the different stellar components of clusters \citep{Puchwein2010}.
AGN feedback in the simulations is modelled as described by \cite{Sijacki2007}.
All of the simulations were run twice, resulting in a control set of simulations that include cooling and star formation but no AGN feedback together with a second set in which AGN feedback was switched on. 
This makes it relatively easy to disentangle the effects of AGN feedback from purely gravitational heating.

Since these simulations were run, there have been incremental changes to
the AGN feedback model (e.g.~\citealt{Sijacki2015}) and significant
developments in cosmological numerical hydrodynamics. To check that
our results are not affected significantly by these changes, 
we compute the integrated Compton-$y$ parameter and total thermal energy for one massive cluster (the AREPO-IL run in \citealt{Sembolini2016}) simulated with the 
moving-mesh code AREPO \citep{Springel2010}, using the same AGN feedback model as that adopted in the
Illustris simulation \citep{Vogelsberger2014, Genel2014}.  The resulting thermal
energy agrees with the scaling relation determined from the P-GADGET3 simulations
shown in Fig. \ref{fig:M_Eth} below.  
Changes in the hydrodynamic scheme and AGN feedback model are therefore
unlikely to affect the conclusions presented in this
section, especially given the large uncertainties of the observational
results presented in this paper.

Depending on the halo mass, the diameter of the high-resolution zoomed-in region ranges from $\sim$7 to 30 comoving Mpc$/h$:
this is much larger than the QSO host halo, as shown in Fig.~\ref{fig:sim_bh_nobh}, and it therefore contains a large fraction of the surrounding correlated large-scale structure.
Due to the use of zoom-ins, our SZ estimate from the simulations will miss any contribution from the line of sight outside this region, which would otherwise be present in a full cosmological lightcone.
However, we show explicitly in Appendix~\ref{app:randompoints} that our measurement on the data is not affected by stacking bias from uncorrelated structure along the same line of sight.
Given the present level of accuracy, a precise quantification of 
residual contribution from correlated structure outside the zoom region
is beyond the scope of this work.

To  compare with our observational results, we choose the simulation snapshot that is closest to the median redshift of the corresponding sample, leading to $z_\mathrm{snap} = 2.07$ for the QSOs, and $z_\mathrm{snap} = 0.36$ for the clusters.
We then create $512\times512$ pixel maps of the Compton-$y$ parameter from the zoom simulations by projecting the gas particles along the line of sight, which we choose to be parallel to one axis of the simulation box. 
Perpendicular to the line of sight, we project all particles within a square centred on the respective halo of interest.
For the latter, we use a fixed physical size of $6~\hinv\Mpc$.

Following \cite{Springel2001}, we make use of the fact that for every gas (SPH) particle in the simulation,
its contribution to the integrated Compton-$y$ parameter is proportional to its pressure times volume, given by 
$pV = (\gamma - 1)X_H \mu \, x_e \, m \, u$. 
Here $\gamma$, $X_H$, $\mu$, and $x_e$ are the adiabatic index, primordial hydrogen fraction, mean molecular weight and electron-to-hydrogen number density ratio (the latter is traced dynamically in the simulations);  $m$ and $u$ are the mass and internal energy per unit mass of the particle.
We  evaluate the integral in equation~\ref{eq:SZ} by projecting all particles as described above, weighting the contribution of particle $i$ to the map $y(\mathbf{x})$ with $(pV)_i \times W(|\mathbf{x}_i-\mathbf{x}|)$, where $W$ is the projected SPH smoothing kernel (e.g.~\citealt{Springel2001}).

Fig.~\ref{fig:sim_bh_nobh} shows the resulting $y$-maps, both with and without AGN feedback, for a halo at $z=2.07$ with a mass of
$M_{200c} = 1.4 \times 10^{12}~\hinv\Msol$
(defined as the mass that is enclosed within a sphere that has 200 times the critical density of the universe at the given redshift).
In this representative example, AGN feedback affects the distribution and temperature of the gas around the central halo, leading to a more extended and more diffuse Compton-$y$ signature. However, the total thermal energy is only increased by $20\%$, an effect too small to be seen in our data
since we find only marginal evidence for an SZ signal in the QSO sample.

As we use different filters for the clusters and the QSOs, we need to process the $y$-maps in different ways, 
as described below. 
\begin{figure*}
	\centering
	\includegraphics[width=2\columnwidth]{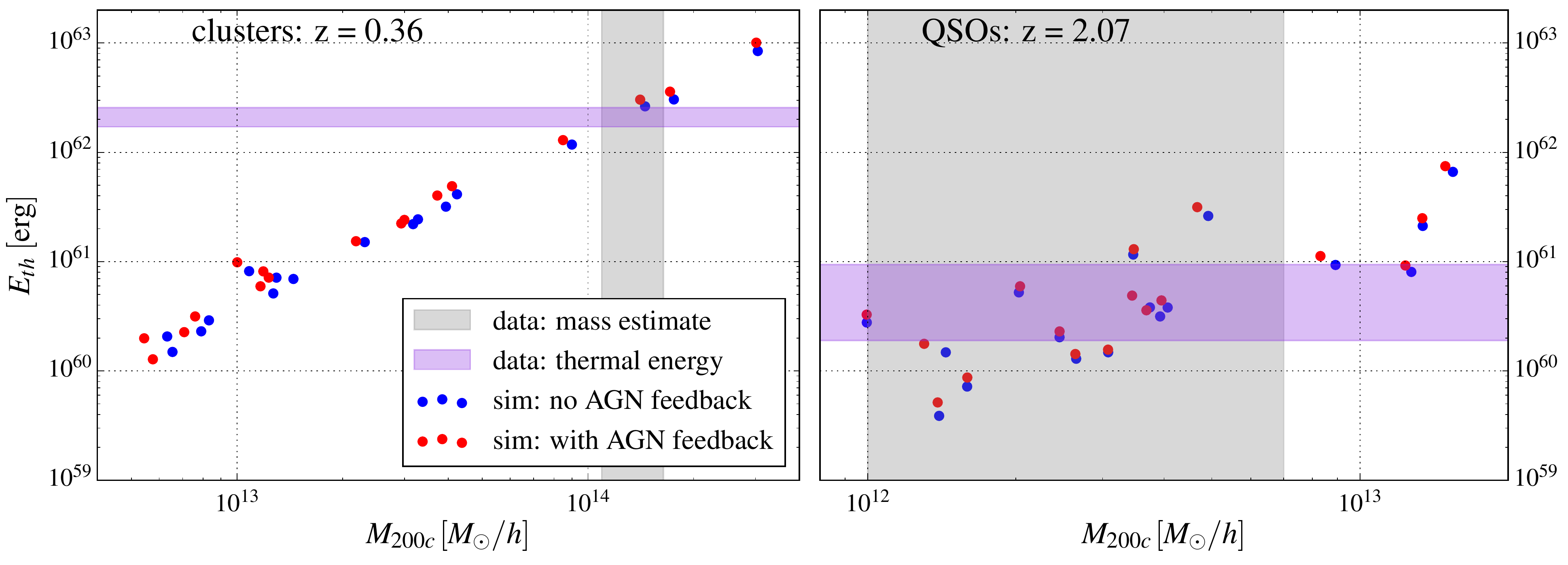}
	\caption{Thermal energy as estimated from the simulations: the blue points denote the results from the control-run simulations without AGN feedback; red are simulations of the same haloes with AGN feedback. 
	\textbf{Left:} we show here the result at the median $z$ of the redMaPPer cluster catalogue (low redshift). 
	The purple horizontal band shows our estimate of $E_\mathrm{th}$ from the redMaPPer clusters, where we have conservatively assumed $20\%$ uncertainty to account for systematic effects such as the choice of the filter profile
	(see Section~\ref{sec:filtering}) and radio contamination (see Section~\ref{subsec:results_clusters}).
	The grey vertical region shows an estimate of the typical 
	$M_{200c}$ of the clusters.
	\textbf{Right}: this panel displays the progenitors of the same objects, at the median redshift of the merged QSO sample.
	Again, the purple and grey shaded bands show our estimate of $E_\mathrm{th}$, and the range of QSO host masses from the literature, respectively.}
	\label{fig:M_Eth}
\end{figure*}

\subparagraph{Low redshift (clusters):}
For the clusters, we filtered the maps using a
$\beta$-profile matched filter with $\theta_c = 1\arcmin$; to
estimate the total $Y$, we have then integrated the profile out to
$5 \times \theta_c$.  For the comparison to simulations, we
therefore integrate the $y$-maps obtained from the low-redshift
snapshot at the median $z$ of the redMaPPer clusters within a
circular aperture of the same radius. We then use
equation~\ref{eq:Eth_z} to convert the integrated Compton-$y$
parameter to the total thermal energy. 
The left-hand panel of Fig.~\ref{fig:M_Eth} shows the dependence of $E_\mathrm{th}$ on the
halo mass.

We find a tight scaling relation between halo mass and
thermal energy over a large range in halo masses, with a slope that is approximately consistent with the self-similar relation $E_\mathrm{th} \propto M^{5/3}$ \citep{Kaiser1986}. 
At larger halo masses ($M_{200c} \gtrsim 3 \times 10^{13}~\hinv\Msol$),
AGN feedback leads to a small increase in the overall thermal energy.  At lower
masses, however, AGN feedback leaves the total thermal energy almost
unchanged, while reducing the halo mass.  This arises because strong AGN
feedback can remove some of the gas from low-mass haloes. At higher
mass, the gas stays bound and receives extra heating from the
feedback.

To compare our results from the data to the simulations, we convert richness to mass using the relation from \cite{Simet2016}, who have calibrated the masses of $z \lesssim 0.3$ SDSS redMaPPer clusters from stacked weak lensing measurements. 
\cite{Melchior2016} have extended this to $z \sim 0.6$ using Dark Energy Survey (DES) Science Verification cluster and lensing data, finding good agreement with the \cite{Simet2016} relation.\footnote{\cite{Saro2015,Saro2016} have used clusters detected both in the SPT-SZ survey and DES to calibrate the masses from the SZ detection significance, also yielding a broadly consistent mass calibration.}
The redMaPPer clusters used for our measurement have a
median richness of $\lambda \simeq 33$. 
Including a factor $M_{200c}/M_{200m} \simeq 0.8$, this leads to a
typical halo mass of $M_{200c} \simeq 1.4 \times 10^{14}~\hinv M_\odot$.
To account for statistical and systematic uncertainties in this estimate, we conservatively assume a $20\%$ uncertainty on this value.

We have measured a
thermal energy of 
$E_\mathrm{th} \simeq 2.1 \times 10^{62}~\mathrm{erg}$ for the SDSS redMaPPer clusters,  
which is fully consistent with the scaling relation inferred from the simulations.
This agreement
demonstrates that our pipeline is able to robustly estimate the
thermal energy of a sample of objects from CMB and sub-mm maps,
and that the results are compatible with theoretical expectations.

\subparagraph{High redshift (QSOs):}
We now turn to the high-redshift snapshots at the median $z$ of the
QSO sample.  Here the QSO emission is completely unresolved by
the \Planck maps and so we simply integrate the entire simulated
$y$-map, before using equation~\ref{eq:Eth_z} to compute the thermal
energy. The results are shown in the right-hand panel of
Fig.~\ref{fig:M_Eth}.
As for the low-redshift clusters, we see a scaling relation
between $E_\mathrm{th}$ and halo mass, but with higher scatter. This increased
scatter is caused by the different environments of the simulated haloes, which
contribute to $E_\mathrm{th}$ estimated within the \Planck beam (see Fig.~\ref{fig:sim_bh_nobh}).
We have explicitly verified this by integrating the $y$-map only within the projected $R_{200}$ of the respective halo, finding a significantly lower scatter in the scaling relation.

AGN feedback slightly increases the thermal energy over the
entire mass range by about $(17 \pm 6)\%$, with no significant dependence on mass. Unlike for the case of low-mass haloes at low
redshift where AGN feedback leads to an appreciable halo mass
reduction due to the powerful ``radio mode" feedback, here this is not the case
as energetically less efficient ``quasar mode" feedback is operating
(for further details see \citealt{Sijacki2007}).

As we are not sampling the higher-luminosity part
(i.e. $\log \frac{L_{\rm bol}}{L_\odot} \gtrsim 12$) 
of the QSO luminosity function very well because of the limited size of our sample, 
we have performed several tests to establish that our comparison between observed and simulated QSOs is adequate. For every black hole in the simulation accreting in the radiatively efficient way, we have estimated its bolometric luminosity as $L_\mathrm{bol} = \epsilon \dot M c^2$,
where $\dot M$ is the current accretion rate and we have set the radiative efficiency to $\epsilon = 0.1$.
We have then computed the luminosity function and have checked that it is sufficiently well
populated at QSO luminosities of $\log \frac{L_{\rm bol}}{L_\odot} \gtrsim 11.5$. Repeating the
same procedure at $z \sim 3$, we find that
the luminosity function is well populated out to $\log \frac{L_{\rm bol}}{L_\odot} \simeq 12.5$,
consistent with the expectation of a higher accretion rate at this redshift. However, we find
that the impact of AGN feedback on the total thermal energy is lower at $z \sim 3$ than at $z \sim 2$
and that the relative increase in $E_{\rm th}$ does not strongly depend on the current accretion rate
of the central black hole. This indicates that $E_{\rm th}$ measured from the SZ effect
probes the integrated past accretion history, thus making our comparison to observations meaningful.

Mass estimates for AGN host haloes are usually derived from
clustering measurements or halo occupation distribution (HOD)
modelling and vary considerably, with values ranging from
$10^{12}~\hinv M_\odot$ \citep{White2012}, to $3-4 \times
10^{12}~\hinv M_\odot$ \citep{Croom2005,Richardson2012}, and up to
$\sim 7 \times 10^{12}~\hinv M_\odot$ \citep{Porciani2004}.  Most
studies find no evidence for significant evolution of the host mass
with redshift (note however the tentative indication for an upturn
in the typical mass at $z \sim 3.2$ reported by
\citealt{Richardson2012}).
Galaxy formation simulations suggest that masses of host haloes inferred
from clustering measurements may be overestimated by a factor of $\sim 2$
at redshifts $z \simlt 2$ \citep{DeGraf2016}. The masses of the host haloes for the QSOs in our
sample are, therefore, extremely uncertain. The grey band in Fig. \ref{fig:M_Eth} shows 
the wide range of halo masses quoted in the literature.

The results presented in Section~\ref{sec:resqso} suggest thermal energies of 
$\simeq 5 \times 10^{60}~\mathrm{erg}$, 
but the uncertainties are so large that we cannot claim a statistically significant
detection. Comparing our results to the simulations
indicates halo  masses of 
$\sim 4 \times 10^{12}~\hinv\Msol$,
consistent with the clustering or HOD measurements of QSO host masses discussed above.  Our upper limit
on $E_\mathrm{th}$ restricts the allowed mass range for QSO host haloes
to $\lesssim 10^{13}~\hinv\Msol$. Given these large uncertainties
in both the observational constraint on the SZ amplitude and the QSO host halo masses,
it will be difficult to detect the small changes in thermal energy associated with AGN feedback via an SZ signature in the foreseeable future
(unless our simulations have grossly underestimated the effects of AGN feedback, which seems unlikely).

\subsection{Comparison with previous work}
\label{subsec:comparison}

\cite{Ruan2015} found a strong correlation between the \cite{Hill2013}
Compton-$y$ map and the SDSS DR7 QSO sample, inferring a total
thermal energy in the hot halo gas of $E_\mathrm{th} \simeq
10^{62}~\mathrm{erg}$.  
Such a high value is inconsistent with the scaling relations determined from our simulations (see Fig.~\ref{fig:M_Eth}) for any plausible value of the QSO host halo masses.
Our upper limit on the thermal energy associated with the QSO host haloes is approximately an order of magnitude lower than the \cite{Ruan2015} estimate.
Subsequent studies by 
\cite{Verdier2015} and \cite{Crichton2015} showed that
the emission around QSOs at frequencies $\simgt 100 \, {\rm GHz}$ is dominated 
by dust; here we confirm this finding.
The signal detected by \cite{Ruan2015} is therefore most likely related to residual dust contamination in the \cite{Hill2013} Compton-$y$ map.
This dust emission must be subtracted accurately in order to recover a constraint on the
SZ amplitude. 
Our results differ from those of \cite{Verdier2015} and \cite{Crichton2015} in the following important aspects.

For low-redshift QSOs ($z < 1.5$), \cite{Verdier2015} found a $2\sigma$ indication for a
{\it negative} Compton-$y$ parameter,
whereas we obtain a preference for a positive SZ amplitude at all redshifts.
At higher redshifts ($z > 2.5$), they
found a $\simeq 7 \sigma$ detection of an SZ signal, with an
amplitude corresponding to a spherically-integrated Compton-$y$
parameter of $\hat{Y}^s_{500} \simeq 1.1 \times 10^{-5}~\arcm^2$.
This can be converted to the cylindrically-integrated total Compton-$y$ parameter 
we measure by assuming a specific pressure profile for the halo gas. 
For the \cite{Arnaud2010} profile adopted by
\cite{Verdier2015}, the conversion factor is $\simeq 1.8$,
and so their result translates to $E_\mathrm{th} \simeq 3.5 \times
10^{61}~\mathrm{erg}$. This value is almost a factor of two larger than our measurement of $\approx 2 \times 10^{61} ~\mathrm{erg}$ for QSOs at $z > 2.6$.
 
The negative SZ amplitude at $z\sim 1.5$ and the relatively high SZ
at $\sim 2.5$ determined by   \cite{Verdier2015} correlate with large changes in the
parameters of the dust model: low values of $\beta_d \sim 1$ and high dust temperatures $T_d \sim 40 \, {\rm K}$ at $z\sim1.5$ 
and high values of  $\beta_d \sim 2.5$ and low values of $T_d  \sim 20 \, {\rm K}$ at $z\sim 2.5$. 
Although we find qualitatively similar trends for the dust parameters (see Section~\ref{sec:zsplit} and Fig. \ref{fig:zsplit_new}),
the variations of $\beta_d$ and $T_d$ with redshift found in our analysis are less extreme than those found by \cite{Verdier2015}.
The differences between our and their SZ results may therefore arise from instabilities in their dust solutions caused by the lack of far-infrared data in their analysis: only the introduction of the AKARI FIR data allows us to firmly anchor the dust SED.

On the other hand,
\cite{Crichton2015} reported a $3-4\sigma$ evidence for an SZ
signal (averaged over their entire sample spanning the redshift
range $0.5-3.5$), corresponding to a measured thermal energy of
$E_\mathrm{th} = (6.2 \pm 1.7) \times 10^{60}~\mathrm{erg}$.  This is
marginally higher than the value we find, but consistent within the measurement uncertainties.
However, their estimate of the SZ amplitude (as
quantified by the feedback efficiency), is again strongly degenerate with
the dust parameters, $T_d \approx (40 \pm 3) \, {\rm K}$, $\beta_d=1.12 \pm
0.13$, which differ from those found here: $T_d = (26.4 \pm 1.6)~ \,
{\rm K}$, $\beta_d = 2.25 \pm 0.21$.  A shift in the
\cite{Crichton2015} dust parameters towards our values would reduce
the amplitude of their recovered SZ signal (see their Fig. 3). The
interpretation of their results as a detection of SZ signal therefore
depends on the fidelity of their dust solution.
\cite{Crichton2015} then used the $\hat{Y}_{500}$-$M_{500}$ scaling relation
of \cite{Planckint_hotgas} to estimate the expected thermal energy
from purely gravitational heating. Assuming QSO halo masses in the
range of $(1-5)\times 10^{12}~\hinv M_\odot$, they concluded that
gravitational heating accounts for a small fraction ($\simlt 30\%$) of
their measured signal. However, this conclusion is incompatible with
the results from our simulations shown in Fig. \ref{fig:M_Eth}, where
we find that the AGN contribution to $E_\mathrm{th}$ is always a small
fraction of that from gravitational heating. 

Finally, \cite{Spacek2016a,Spacek2016b} investigated constraints
on AGN feedback energetics using samples of massive quiescent
elliptical galaxies with redshifts in the range $0.5 < z < 1.5$, with
the aim of finding signatures of past AGN activity.  By stacking CMB
maps from SPT and ACT, respectively, they
found low-significance indications of an SZ signal associated with
these galaxies, with corresponding thermal energies of $\simeq
6 \times 10^{60}~\mathrm{erg}$, comparable to the values found for
QSOs by \cite{Crichton2015} and in this paper. However, the SZ
signal is consistent (with large uncertainties) with their estimates
for gravitational heating and therefore does not require additional energy input from
AGN feedback.

\section{Conclusions}
\label{sec:concl}
The analysis discussed in this paper was motivated by the results of \cite{Ruan2015} who presented 
evidence for a high-amplitude SZ signal in the vicinity of QSOs, suggestive of strong AGN feedback. 
Subsequent work by \cite{Verdier2015} and \cite{Crichton2015} showed that dust emission dominates at
microwave and sub-mm frequencies and must be subtracted to high accuracy to recover an SZ signal. Both of
these groups found evidence for an SZ signal, though at a much lower amplitude than found by \cite{Ruan2015}.
However, the corresponding thermal energies and dust emission parameters that they recovered differ substantially.

Our main contribution has been to analyse the cross-correlation of QSO catalogues from SDSS/BOSS with maps from
\Planck and AKARI. 
In particular, the inclusion of the AKARI far-infrared data at $90 \, {\mathrm{\upmu m}}$ extends beyond the peak of the dust emission and
helps to break the degeneracy between dust parameters and the amplitude of any SZ signal. We have paid careful attention to
the selection of a  clean QSO sample, removing  objects associated with radio sources, extragalactic point sources, and in areas
of high contamination by Galactic dust. We then applied a filter to the CMB and  sub-mm maps that optimally recovers the flux density of unresolved objects such as the QSO hosts. The emission around the QSO was modelled using two components,
consisting of thermal SZ and dust emission (CIB) approximated by a single-temperature modified blackbody spectrum. We also experimented with adding a synchrotron component at low frequencies, which has little effect on our solutions, consistent with
our expectations based on source counts.

We find indications for an SZ signal at low significance ($\sim~1.6
\sigma$). In particular, we do not reproduce the strong $\sim 7\sigma$
detection of an SZ signal for QSOs at $z>2.5$ found by
\cite{Verdier2015}. We also do not find the strong trends in the dust
parameters with redshift reported by \cite{Verdier2015}. The redshift
dependence of the dust parameters in our analysis is more gentle,
though we do find evidence for a rise in the dust temperature to $T_d
\sim 35 \, {\rm K}$ and a lowering of the spectral index to $\beta_d \sim
1.3$ at $z\sim 1.5$. Averaging over the entire redshift range, the
best-fitting modified blackbody parameters are $T_d = (26.4 \pm
1.6) \, {\rm K}$ and $\beta_d=2.25 \pm 0.21$, similar to the dust
parameters of normal nearby galaxies. Our dust parameters disagree
strongly with those determined by \cite{Crichton2015}. Since the dust
parameters are highly correlated with the SZ amplitude, our results
suggest that the detections of an SZ signal reported by
\cite{Verdier2015} and \cite{Crichton2015} should be treated with
caution.

We have compared our results with hydrodynamic simulations of haloes
run with and without AGN feedback. In these simulations, the effects
of AGN feedback lead to small ($\sim 10-20 \%$) enhancements to the
SZ signal both at the low redshifts of redMaPPer clusters ($z=0.36$)
and at the median redshift of the QSO sample ($z=2.07$). For QSO
hosts, our upper limits to the SZ signal are consistent with the
simulations, provided the typical QSO host halo masses are
$\lesssim 10^{13}~h^{-1} M_\odot$, in agreement with the halo masses
inferred by other techniques.

As an aside, we also analysed the SZ signal of redMaPPer clusters of
galaxies using the \Planck and AKARI data. The results show a clear
($>$$20 \sigma$) detection of a SZ signal, with an amplitude
consistent with theoretical expectations from our
numerical simulations. Interestingly, our analysis also shows a
strongly correlated dust signal in both the high-frequency \Planck and AKARI maps,
with dust parameters of
$T_d = (21 \pm 1) \, {\rm K}$, $\beta_d = 2.04^{+0.14}_{-0.13}$.

In our analysis we have assumed that dust emission is described by a single-temperature
modified blackbody. This assumption provides good fits to the data and we find no evidence
to support using a more complex model. With only 4 frequency bands above $350\,{\rm GHz}$, it is
simply not possible to constrain the parameters of a more complicated model for dust emission
reliably.  

According to our simulations it will be difficult to measure the small ($10-20\%$) enhancement of the SZ signal associated
with AGN feedback. With larger spectroscopic samples of QSOs (or
galaxies that have hosted AGN in the past; see \citealt{Spacek2016a})
and CMB data with higher resolution and sensitivity
\citep[e.g.][]{SPT3G,deBernardis2016}, the detection of a
statistically significant SZ signal (disentangled from dust emission)
is a more realistic goal. Together with numerical simulations, it may
then be possible to accurately determine QSO host halo masses, which
are poorly constrained at present.

\section*{Acknowledgements}
The authors thank Nick Battaglia, Lindsey Bleem, Anthony Challinor, Suet-Ying Mak, and Glenn White for helpful discussions during this work, and further Tom Crawford and Kyle Story for helpful conversations about matched filtering during a previous study.
BS acknowledges support from an Isaac Newton Studentship at the University of Cambridge and from the Science and Technology Facilities Council (STFC).
TG acknowledges support from the Kavli Foundation and STFC grant ST/L000636/1. DS acknowledges support by the STFC and the ERC Starting Grant 638707 ``Black holes and their host galaxies: coevolution across cosmic time''.
This research made use of \textit{Astropy}, a community-developed core Python package for Astronomy \citep{astropy}.
Some of the results in this paper have been derived using the HEALPix \citep{Gorski2005} package. 
Furthermore, the use of the following tools is acknowledged: \textit{healpy} (\url{https://github.com/healpy/healpy}), \textit{CosmoloPy} (\url{http://roban.github.io/CosmoloPy}), and \textit{corner.py} (\url{http://corner.readthedocs.io/en/latest}).

This research is based on observations obtained with Planck (\url{http://www.esa.int/Planck}), an ESA science mission with instruments and contributions directly funded by ESA Member States, NASA, and Canada.
Furthermore, it is in parts based on observations with AKARI, a JAXA project with the participation of ESA.

Funding for SDSS-III has been provided by the Alfred P. Sloan Foundation, the Participating Institutions, the National Science Foundation, and the U.S. Department of Energy Office of Science. The SDSS-III web site is http://www.sdss3.org/.
SDSS-III is managed by the Astrophysical Research Consortium for the Participating Institutions of the SDSS-III Collaboration including the University of Arizona, the Brazilian Participation Group, Brookhaven National Laboratory, Carnegie Mellon University, University of Florida, the French Participation Group, the German Participation Group, Harvard University, the Instituto de Astrofisica de Canarias, the Michigan State/Notre Dame/JINA Participation Group, Johns Hopkins University, Lawrence Berkeley National Laboratory, Max Planck Institute for Astrophysics, Max Planck Institute for Extraterrestrial Physics, New Mexico State University, New York University, Ohio State University, Pennsylvania State University, University of Portsmouth, Princeton University, the Spanish Participation Group, University of Tokyo, University of Utah, Vanderbilt University, University of Virginia, University of Washington, and Yale University.




\bibliographystyle{mnras}
\bibliography{qsotsz} 

\begin{thebibliography}{}
\makeatletter
\relax
\def\mn@urlcharsother{\let\do\@makeother \do\$\do\&\do\#\do\^\do\_\do\%\do\~}
\def\mn@doi{\begingroup\mn@urlcharsother \@ifnextchar [ {\mn@doi@}
  {\mn@doi@[]}}
\def\mn@doi@[#1]#2{\def\@tempa{#1}\ifx\@tempa\@empty \href
  {http://dx.doi.org/#2} {doi:#2}\else \href {http://dx.doi.org/#2} {#1}\fi
  \endgroup}
\def\mn@eprint#1#2{\mn@eprint@#1:#2::\@nil}
\def\mn@eprint@arXiv#1{\href {http://arxiv.org/abs/#1} {{\tt arXiv:#1}}}
\def\mn@eprint@dblp#1{\href {http://dblp.uni-trier.de/rec/bibtex/#1.xml}
  {dblp:#1}}
\def\mn@eprint@#1:#2:#3:#4\@nil{\def\@tempa {#1}\def\@tempb {#2}\def\@tempc
  {#3}\ifx \@tempc \@empty \let \@tempc \@tempb \let \@tempb \@tempa \fi \ifx
  \@tempb \@empty \def\@tempb {arXiv}\fi \@ifundefined
  {mn@eprint@\@tempb}{\@tempb:\@tempc}{\expandafter \expandafter \csname
  mn@eprint@\@tempb\endcsname \expandafter{\@tempc}}}

\bibitem[\protect\citeauthoryear{{Afshordi}, {Lin}, {Nagai}  \&
  {Sanderson}}{{Afshordi} et~al.}{2007}]{Afshordi07}
{Afshordi} N.,  {Lin} Y.-T.,  {Nagai} D.,   {Sanderson} A.~J.~R.,  2007,
  \mn@doi [\mnras] {10.1111/j.1365-2966.2007.11776.x}, \href
  {http://ukads.nottingham.ac.uk/abs/2007MNRAS.378..293A} {378, 293}

\bibitem[\protect\citeauthoryear{{Arnaud}, {Pratt}, {Piffaretti},
  {B{\"o}hringer}, {Croston}  \& {Pointecouteau}}{{Arnaud}
  et~al.}{2010}]{Arnaud2010}
{Arnaud} M.,  {Pratt} G.~W.,  {Piffaretti} R.,  {B{\"o}hringer} H.,  {Croston}
  J.~H.,   {Pointecouteau} E.,  2010, \mn@doi [\aap]
  {10.1051/0004-6361/200913416}, \href
  {http://adsabs.harvard.edu/abs/2010A%26A...517A..92A} {517, A92}

\bibitem[\protect\citeauthoryear{{Astropy Collaboration}}{{Astropy
  Collaboration}}{2013}]{astropy}
{Astropy Collaboration} 2013, \mn@doi [\aap] {10.1051/0004-6361/201322068},
  \href {http://adsabs.harvard.edu/abs/2013A%26A...558A..33A} {558, A33}

\bibitem[\protect\citeauthoryear{{Battaglia}, {Bond}, {Pfrommer}, {Sievers}  \&
  {Sijacki}}{{Battaglia} et~al.}{2010}]{Battaglia2010}
{Battaglia} N.,  {Bond} J.~R.,  {Pfrommer} C.,  {Sievers} J.~L.,   {Sijacki}
  D.,  2010, \mn@doi [\apj] {10.1088/0004-637X/725/1/91}, \href
  {http://adsabs.harvard.edu/abs/2010ApJ...725...91B} {725, 91}

\bibitem[\protect\citeauthoryear{{Benson} et~al.,}{{Benson}
  et~al.}{2014}]{SPT3G}
{Benson} B.~A.,  et~al., 2014, in Millimeter, Submillimeter, and Far-Infrared
  Detectors and Instrumentation for Astronomy VII. p. 91531P (\mn@eprint
  {arXiv} {1407.2973}), \mn@doi{10.1117/12.2057305}

\bibitem[\protect\citeauthoryear{Bhattacharya, Di~Matteo  \&
  Kosowsky}{Bhattacharya et~al.}{2008}]{Bhattacharya2007}
Bhattacharya S.,  Di~Matteo T.,   Kosowsky A.,  2008, \mn@doi [Mon. Not. Roy.
  Astron. Soc.] {10.1111/j.1365-2966.2008.13555.x}, 389, 34

\bibitem[\protect\citeauthoryear{{Binney} \& {Tabor}}{{Binney} \&
  {Tabor}}{1995}]{Binney1995}
{Binney} J.,  {Tabor} G.,  1995, \mn@doi [\mnras] {10.1093/mnras/276.2.663},
  \href {http://adsabs.harvard.edu/abs/1995MNRAS.276..663B} {276, 663}

\bibitem[\protect\citeauthoryear{Birkinshaw}{Birkinshaw}{1999}]{Birkinshaw1998}
Birkinshaw M.,  1999, \mn@doi [Phys.Rept.] {10.1016/S0370-1573(98)00080-5},
  310, 97

\bibitem[\protect\citeauthoryear{{Blain}, {Smail}, {Ivison}, {Kneib}  \&
  {Frayer}}{{Blain} et~al.}{2002}]{Blain2002}
{Blain} A.~W.,  {Smail} I.,  {Ivison} R.~J.,  {Kneib} J.-P.,   {Frayer} D.~T.,
  2002, \mn@doi [\physrep] {10.1016/S0370-1573(02)00134-5}, \href
  {http://adsabs.harvard.edu/abs/2002PhR...369..111B} {369, 111}

\bibitem[\protect\citeauthoryear{{Blandford}}{{Blandford}}{1990}]{Blandford1990}
{Blandford} R.,  1990, in {Courvoisier} T.-L.,  {Mayor} M.,  eds, Active
  Galactic Nuclei. Saas-Fee Advanced Course 20.
pp 161--269

\bibitem[\protect\citeauthoryear{{Bleem}}{{Bleem}}{2013}]{Bleem_thesis}
{Bleem} L.~E.,  2013, PhD thesis, The University of Chicago

\bibitem[\protect\citeauthoryear{{Bleem}, {Stalder}, {Brodwin}, {Busha},
  {Gladders}, {High}, {Rest}  \& {Wechsler}}{{Bleem}
  et~al.}{2015a}]{Bleem2015Blanco}
{Bleem} L.~E.,  {Stalder} B.,  {Brodwin} M.,  {Busha} M.~T.,  {Gladders} M.~D.,
   {High} F.~W.,  {Rest} A.,   {Wechsler} R.~H.,  2015a, \mn@doi [\apjs]
  {10.1088/0067-0049/216/1/20}, \href
  {http://adsabs.harvard.edu/abs/2015ApJS..216...20B} {216, 20}

\bibitem[\protect\citeauthoryear{{Bleem} et~al.,}{{Bleem}
  et~al.}{2015b}]{Bleem2015}
{Bleem} L.~E.,  et~al., 2015b, \mn@doi [\apjs] {10.1088/0067-0049/216/2/27},
  \href {http://adsabs.harvard.edu/abs/2015ApJS..216...27B} {216, 27}

\bibitem[\protect\citeauthoryear{{Boehringer}, {Voges}, {Fabian}, {Edge}  \&
  {Neumann}}{{Boehringer} et~al.}{1993}]{Boehringer1993}
{Boehringer} H.,  {Voges} W.,  {Fabian} A.~C.,  {Edge} A.~C.,   {Neumann}
  D.~M.,  1993, \mn@doi [\mnras] {10.1093/mnras/264.1.L25}, \href
  {http://ukads.nottingham.ac.uk/abs/1993MNRAS.264L..25B} {264, L25}

\bibitem[\protect\citeauthoryear{{Bower}, {Benson}, {Malbon}, {Helly}, {Frenk},
  {Baugh}, {Cole}  \& {Lacey}}{{Bower} et~al.}{2006}]{Bower2006}
{Bower} R.~G.,  {Benson} A.~J.,  {Malbon} R.,  {Helly} J.~C.,  {Frenk} C.~S.,
  {Baugh} C.~M.,  {Cole} S.,   {Lacey} C.~G.,  2006, \mn@doi [\mnras]
  {10.1111/j.1365-2966.2006.10519.x}, \href
  {http://ukads.nottingham.ac.uk/abs/2006MNRAS.370..645B} {370, 645}

\bibitem[\protect\citeauthoryear{Brooks \& Gelman}{Brooks \&
  Gelman}{1998}]{Brooks1998}
Brooks S.~P.,  Gelman A.,  1998, Journal of computational and graphical
  statistics, 7, 434

\bibitem[\protect\citeauthoryear{{Burns}}{{Burns}}{1990}]{Burns1990}
{Burns} J.~O.,  1990, \mn@doi [\aj] {10.1086/115307}, \href
  {http://ukads.nottingham.ac.uk/abs/1990AJ.....99...14B} {99, 14}

\bibitem[\protect\citeauthoryear{{Calzetti}, {Armus}, {Bohlin}, {Kinney},
  {Koornneef}  \& {Storchi-Bergmann}}{{Calzetti} et~al.}{2000}]{Calzetti2000}
{Calzetti} D.,  {Armus} L.,  {Bohlin} R.~C.,  {Kinney} A.~L.,  {Koornneef} J.,
   {Storchi-Bergmann} T.,  2000, \mn@doi [\apj] {10.1086/308692}, \href
  {http://adsabs.harvard.edu/abs/2000ApJ...533..682C} {533, 682}

\bibitem[\protect\citeauthoryear{{Carilli}, {Perley}  \& {Harris}}{{Carilli}
  et~al.}{1994}]{Carilli1994}
{Carilli} C.~L.,  {Perley} R.~A.,   {Harris} D.~E.,  1994, \mn@doi [\mnras]
  {10.1093/mnras/270.1.173}, \href
  {http://ukads.nottingham.ac.uk/abs/1994MNRAS.270..173C} {270, 173}

\bibitem[\protect\citeauthoryear{Carlstrom, Holder  \& Reese}{Carlstrom
  et~al.}{2002}]{Carlstrom2002}
Carlstrom J.~E.,  Holder G.~P.,   Reese E.~D.,  2002, \mn@doi [Ann.Rev.\aap]
  {10.1146/annurev.astro.40.060401.093803}, 40, 643

\bibitem[\protect\citeauthoryear{{Cavaliere} \& {Fusco-Femiano}}{{Cavaliere} \&
  {Fusco-Femiano}}{1976}]{cavaliere76}
{Cavaliere} A.,  {Fusco-Femiano} R.,  1976, \aap, \href
  {http://adsabs.harvard.edu/abs/1976A%26A....49..137C} {49, 137}

\bibitem[\protect\citeauthoryear{Cen \& Safarzadeh}{Cen \&
  Safarzadeh}{2015}]{Cen2015}
Cen R.,  Safarzadeh M.,  2015, \mn@doi [Astrophys. J.]
  {10.1088/2041-8205/809/2/L32}, 809, L32

\bibitem[\protect\citeauthoryear{Chatterjee \& Kosowsky}{Chatterjee \&
  Kosowsky}{2007}]{Chatterjee2007}
Chatterjee S.,  Kosowsky A.,  2007, \mn@doi [Astrophys. J.] {10.1086/518860},
  661, L113

\bibitem[\protect\citeauthoryear{Chatterjee, Di~Matteo, Kosowsky  \&
  Pelupessy}{Chatterjee et~al.}{2008}]{Chatterjee2008}
Chatterjee S.,  Di~Matteo T.,  Kosowsky A.,   Pelupessy I.,  2008, \mn@doi
  [Mon. Not. Roy. Astron. Soc.] {10.1111/j.1365-2966.2008.13784.x}, 390, 535

\bibitem[\protect\citeauthoryear{Chatterjee, Ho, Newman  \&
  Kosowsky}{Chatterjee et~al.}{2010}]{Chatterjee2009}
Chatterjee S.,  Ho S.,  Newman J.~A.,   Kosowsky A.,  2010, \mn@doi [Astrophys.
  J.] {10.1088/0004-637X/720/1/299}, 720, 299

\bibitem[\protect\citeauthoryear{{Chelouche}, {Koester}  \&
  {Bowen}}{{Chelouche} et~al.}{2007}]{Chelouche2007}
{Chelouche} D.,  {Koester} B.~P.,   {Bowen} D.~V.,  2007, \mn@doi [\apjl]
  {10.1086/525251}, \href {http://adsabs.harvard.edu/abs/2007ApJ...671L..97C}
  {671, L97}

\bibitem[\protect\citeauthoryear{{Churazov}, {Br{\"u}ggen}, {Kaiser},
  {B{\"o}hringer}  \& {Forman}}{{Churazov} et~al.}{2001}]{Churazov2001}
{Churazov} E.,  {Br{\"u}ggen} M.,  {Kaiser} C.~R.,  {B{\"o}hringer} H.,
  {Forman} W.,  2001, \mn@doi [\apj] {10.1086/321357}, \href
  {http://ukads.nottingham.ac.uk/abs/2001ApJ...554..261C} {554, 261}

\bibitem[\protect\citeauthoryear{{Churazov}, {Sunyaev}, {Forman}  \&
  {B{\"o}hringer}}{{Churazov} et~al.}{2002}]{Churazov2002}
{Churazov} E.,  {Sunyaev} R.,  {Forman} W.,   {B{\"o}hringer} H.,  2002,
  \mn@doi [\mnras] {10.1046/j.1365-8711.2002.05332.x}, \href
  {http://ukads.nottingham.ac.uk/abs/2002MNRAS.332..729C} {332, 729}

\bibitem[\protect\citeauthoryear{{Clemens} et~al.,}{{Clemens}
  et~al.}{2013}]{Clemens2013}
{Clemens} M.~S.,  et~al., 2013, \mn@doi [\mnras] {10.1093/mnras/stt760}, \href
  {http://adsabs.harvard.edu/abs/2013MNRAS.433..695C} {433, 695}

\bibitem[\protect\citeauthoryear{{Crenshaw}, {Kraemer}  \& {George}}{{Crenshaw}
  et~al.}{2003}]{Crenshaw2003}
{Crenshaw} D.~M.,  {Kraemer} S.~B.,   {George} I.~M.,  2003, \mn@doi [\araa]
  {10.1146/annurev.astro.41.082801.100328}, \href
  {http://ukads.nottingham.ac.uk/abs/2003ARA%26A..41..117C} {41, 117}

\bibitem[\protect\citeauthoryear{{Crichton} et~al.,}{{Crichton}
  et~al.}{2016}]{Crichton2015}
{Crichton} D.,  et~al., 2016, \mn@doi [\mnras] {10.1093/mnras/stw344}, \href
  {http://adsabs.harvard.edu/abs/2016MNRAS.458.1478C} {458, 1478}

\bibitem[\protect\citeauthoryear{{Croom} et~al.,}{{Croom}
  et~al.}{2005}]{Croom2005}
{Croom} S.~M.,  et~al., 2005, \mn@doi [\mnras]
  {10.1111/j.1365-2966.2004.08379.x}, \href
  {http://adsabs.harvard.edu/abs/2005MNRAS.356..415C} {356, 415}

\bibitem[\protect\citeauthoryear{{Croton} et~al.,}{{Croton}
  et~al.}{2006}]{Croton2006}
{Croton} D.~J.,  et~al., 2006, \mn@doi [\mnras]
  {10.1111/j.1365-2966.2005.09675.x}, \href
  {http://ukads.nottingham.ac.uk/abs/2006MNRAS.365...11C} {365, 11}

\bibitem[\protect\citeauthoryear{{Dawson} et~al.,}{{Dawson}
  et~al.}{2013}]{BOSS}
{Dawson} K.~S.,  et~al., 2013, \mn@doi [\aj] {10.1088/0004-6256/145/1/10},
  \href {http://adsabs.harvard.edu/abs/2013AJ....145...10D} {145, 10}

\bibitem[\protect\citeauthoryear{{De Bernardis} et~al.,}{{De Bernardis}
  et~al.}{2016}]{deBernardis2016}
{De Bernardis} F.,  et~al., 2016, preprint, \href
  {http://adsabs.harvard.edu/abs/2016arXiv160702120D} {} (\mn@eprint {arXiv}
  {1607.02120})

\bibitem[\protect\citeauthoryear{{DeGraf} \& {Sijacki}}{{DeGraf} \&
  {Sijacki}}{2016}]{DeGraf2016}
{DeGraf} C.,  {Sijacki} D.,  2016, preprint, \href
  {http://adsabs.harvard.edu/abs/2016arXiv160906727D} {} (\mn@eprint {arXiv}
  {1609.06727})

\bibitem[\protect\citeauthoryear{{Di Matteo}, {Springel}  \& {Hernquist}}{{Di
  Matteo} et~al.}{2005}]{DiMatteo2005}
{Di Matteo} T.,  {Springel} V.,   {Hernquist} L.,  2005, \mn@doi [\nat]
  {10.1038/nature03335}, \href
  {http://ukads.nottingham.ac.uk/abs/2005Natur.433..604D} {433, 604}

\bibitem[\protect\citeauthoryear{{Diego} \& {Partridge}}{{Diego} \&
  {Partridge}}{2010}]{Diego10}
{Diego} J.~M.,  {Partridge} B.,  2010, \mn@doi [\mnras]
  {10.1111/j.1365-2966.2009.15949.x}, \href
  {http://ukads.nottingham.ac.uk/abs/2010MNRAS.402.1179D} {402, 1179}

\bibitem[\protect\citeauthoryear{{Doi} et~al.,}{{Doi} et~al.}{2015}]{akari}
{Doi} Y.,  et~al., 2015, \mn@doi [\pasj] {10.1093/pasj/psv022}, \href
  {http://adsabs.harvard.edu/abs/2015PASJ...67...50D} {67, 50}

\bibitem[\protect\citeauthoryear{{Dubois}, {Devriendt}, {Slyz}  \&
  {Teyssier}}{{Dubois} et~al.}{2010}]{Dubois2010}
{Dubois} Y.,  {Devriendt} J.,  {Slyz} A.,   {Teyssier} R.,  2010, \mn@doi
  [\mnras] {10.1111/j.1365-2966.2010.17338.x}, \href
  {http://ukads.nottingham.ac.uk/abs/2010MNRAS.409..985D} {409, 985}

\bibitem[\protect\citeauthoryear{{Dubois}, {Peirani}, {Pichon}, {Devriendt},
  {Gavazzi}, {Welker}  \& {Volonteri}}{{Dubois} et~al.}{2016}]{Dubois2016}
{Dubois} Y.,  {Peirani} S.,  {Pichon} C.,  {Devriendt} J.,  {Gavazzi} R.,
  {Welker} C.,   {Volonteri} M.,  2016, \mn@doi [\mnras]
  {10.1093/mnras/stw2265}, \href
  {http://ukads.nottingham.ac.uk/abs/2016MNRAS.463.3948D} {463, 3948}

\bibitem[\protect\citeauthoryear{{Eifler}, {Krause}, {Dodelson}, {Zentner},
  {Hearin}  \& {Gnedin}}{{Eifler} et~al.}{2015}]{Eifler2015}
{Eifler} T.,  {Krause} E.,  {Dodelson} S.,  {Zentner} A.~R.,  {Hearin} A.~P.,
  {Gnedin} N.~Y.,  2015, \mn@doi [\mnras] {10.1093/mnras/stv2000}, \href
  {http://adsabs.harvard.edu/abs/2015MNRAS.454.2451E} {454, 2451}

\bibitem[\protect\citeauthoryear{{Eisenstein} et~al.,}{{Eisenstein}
  et~al.}{2011}]{Eisenstein2011}
{Eisenstein} D.~J.,  et~al., 2011, \mn@doi [\aj] {10.1088/0004-6256/142/3/72},
  \href {http://adsabs.harvard.edu/abs/2011AJ....142...72E} {142, 72}

\bibitem[\protect\citeauthoryear{{Fabian}}{{Fabian}}{1999}]{Fabian1999}
{Fabian} A.~C.,  1999, \mn@doi [\mnras] {10.1046/j.1365-8711.1999.03017.x},
  \href {http://ukads.nottingham.ac.uk/abs/1999MNRAS.308L..39F} {308, L39}

\bibitem[\protect\citeauthoryear{{Fabian}}{{Fabian}}{2012}]{Fabian2012}
{Fabian} A.~C.,  2012, \mn@doi [\araa] {10.1146/annurev-astro-081811-125521},
  \href {http://adsabs.harvard.edu/abs/2012ARA%26A..50..455F} {50, 455}

\bibitem[\protect\citeauthoryear{{Fabian}, {Sanders}, {Taylor}, {Allen},
  {Crawford}, {Johnstone}  \& {Iwasawa}}{{Fabian} et~al.}{2006}]{Fabian2006}
{Fabian} A.~C.,  {Sanders} J.~S.,  {Taylor} G.~B.,  {Allen} S.~W.,  {Crawford}
  C.~S.,  {Johnstone} R.~M.,   {Iwasawa} K.,  2006, \mn@doi [\mnras]
  {10.1111/j.1365-2966.2005.09896.x}, \href
  {http://ukads.nottingham.ac.uk/abs/2006MNRAS.366..417F} {366, 417}

\bibitem[\protect\citeauthoryear{{Foreman-Mackey}, {Hogg}, {Lang}  \&
  {Goodman}}{{Foreman-Mackey} et~al.}{2013}]{emcee}
{Foreman-Mackey} D.,  {Hogg} D.~W.,  {Lang} D.,   {Goodman} J.,  2013, \mn@doi
  [\pasp] {10.1086/670067}, \href
  {http://adsabs.harvard.edu/abs/2013PASP..125..306F} {125, 306}

\bibitem[\protect\citeauthoryear{{Forman} et~al.,}{{Forman}
  et~al.}{2007}]{Forman2007}
{Forman} W.,  et~al., 2007, \mn@doi [\apj] {10.1086/519480}, \href
  {http://ukads.nottingham.ac.uk/abs/2007ApJ...665.1057F} {665, 1057}

\bibitem[\protect\citeauthoryear{{Franceschini}}{{Franceschini}}{2000}]{Franceschini2000}
{Franceschini} A.,  2000, ArXiv Astrophysics e-prints, \href
  {http://adsabs.harvard.edu/abs/2000astro.ph..9121F} {}

\bibitem[\protect\citeauthoryear{Gelman \& Rubin}{Gelman \&
  Rubin}{1992}]{Gelman1992}
Gelman A.,  Rubin D.~B.,  1992, Statistical science, pp 457--472

\bibitem[\protect\citeauthoryear{{Genel} et~al.,}{{Genel}
  et~al.}{2014}]{Genel2014}
{Genel} S.,  et~al., 2014, \mn@doi [\mnras] {10.1093/mnras/stu1654}, \href
  {http://adsabs.harvard.edu/abs/2014MNRAS.445..175G} {445, 175}

\bibitem[\protect\citeauthoryear{Goodman \& Weare}{Goodman \&
  Weare}{2010}]{Goodman2010}
Goodman J.,  Weare J.,  2010, \mn@doi [Comm.\ App.\ Math.\ Comp.\ Sci.]
  {10.2140/camcos.2010.5.65}, 5, 65

\bibitem[\protect\citeauthoryear{{G{\'o}rski}, {Hivon}, {Banday}, {Wandelt},
  {Hansen}, {Reinecke}  \& {Bartelmann}}{{G{\'o}rski}
  et~al.}{2005}]{Gorski2005}
{G{\'o}rski} K.~M.,  {Hivon} E.,  {Banday} A.~J.,  {Wandelt} B.~D.,  {Hansen}
  F.~K.,  {Reinecke} M.,   {Bartelmann} M.,  2005, \mn@doi [\apj]
  {10.1086/427976}, \href {http://adsabs.harvard.edu/abs/2005ApJ...622..759G}
  {622, 759}

\bibitem[\protect\citeauthoryear{{Gupta} et~al.,}{{Gupta}
  et~al.}{2016}]{Gupta2016}
{Gupta} N.,  et~al., 2016, preprint, \href
  {http://adsabs.harvard.edu/abs/2016arXiv160505329G} {} (\mn@eprint {arXiv}
  {1605.05329})

\bibitem[\protect\citeauthoryear{{Guti{\'e}rrez} \&
  {L{\'o}pez-Corredoira}}{{Guti{\'e}rrez} \&
  {L{\'o}pez-Corredoira}}{2014}]{Gutierrez2014}
{Guti{\'e}rrez} C.~M.,  {L{\'o}pez-Corredoira} M.,  2014, \mn@doi [\aap]
  {10.1051/0004-6361/201424598}, \href
  {http://adsabs.harvard.edu/abs/2014A%26A...571A..66G} {571, A66}

\bibitem[\protect\citeauthoryear{Haehnelt \& Tegmark}{Haehnelt \&
  Tegmark}{1996}]{Haehnelt1996}
Haehnelt M.~G.,  Tegmark M.,  1996, \mn@doi [\mnras] {10.1093/mnras/279.2.545},
  279, 545

\bibitem[\protect\citeauthoryear{{Hajian}, {Battaglia}, {Spergel}, {Bond},
  {Pfrommer}  \& {Sievers}}{{Hajian} et~al.}{2013}]{Hajian13}
{Hajian} A.,  {Battaglia} N.,  {Spergel} D.~N.,  {Bond} J.~R.,  {Pfrommer} C.,
   {Sievers} J.~L.,  2013, \mn@doi [\jcap] {10.1088/1475-7516/2013/11/064},
  \href {http://ukads.nottingham.ac.uk/abs/2013JCAP...11..064H} {11, 064}

\bibitem[\protect\citeauthoryear{{Hasselfield} et~al.,}{{Hasselfield}
  et~al.}{2013}]{Hasselfield2013}
{Hasselfield} M.,  et~al., 2013, \mn@doi [\jcap]
  {10.1088/1475-7516/2013/07/008}, \href
  {http://adsabs.harvard.edu/abs/2013JCAP...07..008H} {7, 008}

\bibitem[\protect\citeauthoryear{{Helou} \& {Walker}}{{Helou} \&
  {Walker}}{1988}]{IRAS_pointsources}
{Helou} G.,  {Walker} D.~W.,  eds, 1988, {Infrared astronomical satellite
  (IRAS) catalogs and atlases. Volume 7: The small scale structure catalog} ~
  Vol. 7

\bibitem[\protect\citeauthoryear{Hill \& Spergel}{Hill \&
  Spergel}{2014}]{Hill2013}
Hill J.~C.,  Spergel D.~N.,  2014, \mn@doi [JCAP]
  {10.1088/1475-7516/2014/02/030}, 1402, 030

\bibitem[\protect\citeauthoryear{{Ishihara} et~al.,}{{Ishihara}
  et~al.}{2010}]{AKARI_pointsources}
{Ishihara} D.,  et~al., 2010, \mn@doi [\aap] {10.1051/0004-6361/200913811},
  \href {http://adsabs.harvard.edu/abs/2010A%26A...514A...1I} {514, A1}

\bibitem[\protect\citeauthoryear{{Kaiser}}{{Kaiser}}{1986}]{Kaiser1986}
{Kaiser} N.,  1986, \mn@doi [\mnras] {10.1093/mnras/222.2.323}, \href
  {http://adsabs.harvard.edu/abs/1986MNRAS.222..323K} {222, 323}

\bibitem[\protect\citeauthoryear{{Kauffmann} \& {Haehnelt}}{{Kauffmann} \&
  {Haehnelt}}{2000}]{Kauffmann2000}
{Kauffmann} G.,  {Haehnelt} M.,  2000, \mn@doi [\mnras]
  {10.1046/j.1365-8711.2000.03077.x}, \href
  {http://ukads.nottingham.ac.uk/abs/2000MNRAS.311..576K} {311, 576}

\bibitem[\protect\citeauthoryear{{Kellermann}, {Sramek}, {Schmidt}, {Shaffer}
  \& {Green}}{{Kellermann} et~al.}{1989}]{Kellermann1989}
{Kellermann} K.~I.,  {Sramek} R.,  {Schmidt} M.,  {Shaffer} D.~B.,   {Green}
  R.,  1989, \mn@doi [\aj] {10.1086/115207}, \href
  {http://adsabs.harvard.edu/abs/1989AJ.....98.1195K} {98, 1195}

\bibitem[\protect\citeauthoryear{{King}}{{King}}{2003}]{King2003}
{King} A.,  2003, \mn@doi [\apjl] {10.1086/379143}, \href
  {http://ukads.nottingham.ac.uk/abs/2003ApJ...596L..27K} {596, L27}

\bibitem[\protect\citeauthoryear{{Lynden-Bell}}{{Lynden-Bell}}{1969}]{LyndenBell1969}
{Lynden-Bell} D.,  1969, \mn@doi [\nat] {10.1038/223690a0}, \href
  {http://adsabs.harvard.edu/abs/1969Natur.223..690L} {223, 690}

\bibitem[\protect\citeauthoryear{{Madau} \& {Dickinson}}{{Madau} \&
  {Dickinson}}{2014}]{Madau2014}
{Madau} P.,  {Dickinson} M.,  2014, \mn@doi [\araa]
  {10.1146/annurev-astro-081811-125615}, \href
  {http://adsabs.harvard.edu/abs/2014ARA%26A..52..415M} {52, 415}

\bibitem[\protect\citeauthoryear{{Maiolino} et~al.,}{{Maiolino}
  et~al.}{2012}]{Maiolino2012}
{Maiolino} R.,  et~al., 2012, \mn@doi [\mnras]
  {10.1111/j.1745-3933.2012.01303.x}, \href
  {http://ukads.nottingham.ac.uk/abs/2012MNRAS.425L..66M} {425, L66}

\bibitem[\protect\citeauthoryear{{Mak}, {Challinor}, {Efstathiou}  \&
  {Lagache}}{{Mak} et~al.}{2016}]{Mak2016}
{Mak} D.~S.~Y.,  {Challinor} A.,  {Efstathiou} G.,   {Lagache} G.,  2016,
  preprint, \href {http://adsabs.harvard.edu/abs/2016arXiv160908942M} {}
  (\mn@eprint {arXiv} {1609.08942})

\bibitem[\protect\citeauthoryear{{McCarthy}, {Schaye}, {Bower}, {Ponman},
  {Booth}, {Dalla Vecchia}  \& {Springel}}{{McCarthy}
  et~al.}{2011}]{McCarthy2011}
{McCarthy} I.~G.,  {Schaye} J.,  {Bower} R.~G.,  {Ponman} T.~J.,  {Booth}
  C.~M.,  {Dalla Vecchia} C.,   {Springel} V.,  2011, \mn@doi [\mnras]
  {10.1111/j.1365-2966.2010.18033.x}, \href
  {http://adsabs.harvard.edu/abs/2011MNRAS.412.1965M} {412, 1965}

\bibitem[\protect\citeauthoryear{{Melchior} et~al.,}{{Melchior}
  et~al.}{2016}]{Melchior2016}
{Melchior} P.,  et~al., 2016, preprint, \href
  {http://adsabs.harvard.edu/abs/2016arXiv161006890M} {} (\mn@eprint {arXiv}
  {1610.06890})

\bibitem[\protect\citeauthoryear{{Miley} \& {De Breuck}}{{Miley} \& {De
  Breuck}}{2008}]{Miley2008}
{Miley} G.,  {De Breuck} C.,  2008, \mn@doi [\aapr]
  {10.1007/s00159-007-0008-z}, \href
  {http://ukads.nottingham.ac.uk/abs/2008A%26ARv..15...67M} {15, 67}

\bibitem[\protect\citeauthoryear{{Miville-Desch{\^e}nes} \&
  {Lagache}}{{Miville-Desch{\^e}nes} \& {Lagache}}{2005}]{IRIS}
{Miville-Desch{\^e}nes} M.-A.,  {Lagache} G.,  2005, \mn@doi [\apjs]
  {10.1086/427938}, \href {http://adsabs.harvard.edu/abs/2005ApJS..157..302M}
  {157, 302}

\bibitem[\protect\citeauthoryear{{Navarro}, {Frenk}  \& {White}}{{Navarro}
  et~al.}{1996}]{navarro96}
{Navarro} J.~F.,  {Frenk} C.~S.,   {White} S.~D.~M.,  1996, \mn@doi [\apj]
  {10.1086/177173}, \href {http://adsabs.harvard.edu/abs/1996ApJ...462..563N}
  {462, 563}

\bibitem[\protect\citeauthoryear{{Neugebauer} et~al.,}{{Neugebauer}
  et~al.}{1984}]{IRAS}
{Neugebauer} G.,  et~al., 1984, \mn@doi [\apjl] {10.1086/184209}, \href
  {http://adsabs.harvard.edu/abs/1984ApJ...278L...1N} {278, L1}

\bibitem[\protect\citeauthoryear{{Omont}, {Beelen}, {Bertoldi}, {Cox},
  {Carilli}, {Priddey}, {McMahon}  \& {Isaak}}{{Omont}
  et~al.}{2003}]{Omont2003}
{Omont} A.,  {Beelen} A.,  {Bertoldi} F.,  {Cox} P.,  {Carilli} C.~L.,
  {Priddey} R.~S.,  {McMahon} R.~G.,   {Isaak} K.~G.,  2003, \mn@doi [\aap]
  {10.1051/0004-6361:20021652}, \href
  {http://adsabs.harvard.edu/abs/2003A%26A...398..857O} {398, 857}

\bibitem[\protect\citeauthoryear{{P{\^a}ris} et~al.,}{{P{\^a}ris}
  et~al.}{2016}]{Paris2016}
{P{\^a}ris} I.,  et~al., 2016, preprint, \href
  {http://adsabs.harvard.edu/abs/2016arXiv160806483P} {} (\mn@eprint {arXiv}
  {1608.06483})

\bibitem[\protect\citeauthoryear{{Planck Collaboration}}{{Planck
  Collaboration}}{2011}]{Bartlett11}
{Planck Collaboration} 2011, \mn@doi [\aap] {10.1051/0004-6361/201116489},
  \href {http://ukads.nottingham.ac.uk/abs/2011A%26A...536A..12P} {536, A12}

\bibitem[\protect\citeauthoryear{{Planck Collaboration}}{{Planck
  Collaboration}}{2013}]{Planckint_hotgas}
{Planck Collaboration} 2013, \mn@doi [\aap] {10.1051/0004-6361/201220941},
  \href {http://adsabs.harvard.edu/abs/2013A%26A...557A..52P} {557, A52}

\bibitem[\protect\citeauthoryear{{Planck Collaboration}}{{Planck
  Collaboration}}{2014a}]{Planck2013_dust}
{Planck Collaboration} 2014a, \mn@doi [\aap] {10.1051/0004-6361/201323195},
  \href {http://adsabs.harvard.edu/abs/2014A%26A...571A..11P} {571, A11}

\bibitem[\protect\citeauthoryear{{Planck Collaboration}}{{Planck
  Collaboration}}{2014b}]{Planck2013_compactsources}
{Planck Collaboration} 2014b, \mn@doi [\aap] {10.1051/0004-6361/201321524},
  \href {http://adsabs.harvard.edu/abs/2014A%26A...571A..28P} {571, A28}

\bibitem[\protect\citeauthoryear{{Planck Collaboration}}{{Planck
  Collaboration}}{2015a}]{Planck2015_HFI1}
{Planck Collaboration} 2015a, preprint, \href
  {http://adsabs.harvard.edu/abs/2015arXiv150201586P} {} (\mn@eprint {arXiv}
  {1502.01586})

\bibitem[\protect\citeauthoryear{{Planck Collaboration}}{{Planck
  Collaboration}}{2015b}]{Planck2015_HFI2}
{Planck Collaboration} 2015b, preprint, \href
  {http://adsabs.harvard.edu/abs/2015arXiv150201587P} {} (\mn@eprint {arXiv}
  {1502.01587})

\bibitem[\protect\citeauthoryear{{Planck Collaboration}}{{Planck
  Collaboration}}{2015d}]{Planck2015_dust}
{Planck Collaboration} 2015d, preprint, \href
  {http://adsabs.harvard.edu/abs/2015arXiv150201588P} {} (\mn@eprint {arXiv}
  {1502.01588})

\bibitem[\protect\citeauthoryear{{Planck Collaboration}}{{Planck
  Collaboration}}{2015c}]{Planck2015_compactsources}
{Planck Collaboration} 2015c, preprint, \href
  {http://adsabs.harvard.edu/abs/2015arXiv150702058P} {} (\mn@eprint {arXiv}
  {1507.02058})

\bibitem[\protect\citeauthoryear{{Planck Collaboration}}{{Planck
  Collaboration}}{2016b}]{Planck2016_radio}
{Planck Collaboration} 2016b, preprint, \href
  {http://adsabs.harvard.edu/abs/2016arXiv160605120P} {} (\mn@eprint {arXiv}
  {1606.05120})

\bibitem[\protect\citeauthoryear{{Planck Collaboration}}{{Planck
  Collaboration}}{2016a}]{Planckint_dust}
{Planck Collaboration} 2016a, preprint, \href
  {http://adsabs.harvard.edu/abs/2016arXiv160509387P} {} (\mn@eprint {arXiv}
  {1605.09387})

\bibitem[\protect\citeauthoryear{{Planck Collaboration}}{{Planck
  Collaboration}}{2016c}]{Planck2015_overview}
{Planck Collaboration} 2016c, \mn@doi [\aap] {10.1051/0004-6361/201527101},
  \href {http://adsabs.harvard.edu/abs/2016A%26A...594A...1P} {594, A1}

\bibitem[\protect\citeauthoryear{{Planck Collaboration}}{{Planck
  Collaboration}}{2016d}]{Planck2015_LFI}
{Planck Collaboration} 2016d, \mn@doi [\aap] {10.1051/0004-6361/201525818},
  \href {http://adsabs.harvard.edu/abs/2016A%26A...594A...2P} {594, A2}

\bibitem[\protect\citeauthoryear{{Planck Collaboration}}{{Planck
  Collaboration}}{2016e}]{Planckclusters2016}
{Planck Collaboration} 2016e, \mn@doi [\aap] {10.1051/0004-6361/201525823},
  \href {http://adsabs.harvard.edu/abs/2016A%26A...594A..27P} {594, A27}

\bibitem[\protect\citeauthoryear{{Porciani}, {Magliocchetti}  \&
  {Norberg}}{{Porciani} et~al.}{2004}]{Porciani2004}
{Porciani} C.,  {Magliocchetti} M.,   {Norberg} P.,  2004, \mn@doi [\mnras]
  {10.1111/j.1365-2966.2004.08408.x}, \href
  {http://adsabs.harvard.edu/abs/2004MNRAS.355.1010P} {355, 1010}

\bibitem[\protect\citeauthoryear{{Priddey} \& {McMahon}}{{Priddey} \&
  {McMahon}}{2001}]{Priddey2001}
{Priddey} R.~S.,  {McMahon} R.~G.,  2001, \mn@doi [\mnras]
  {10.1046/j.1365-8711.2001.04548.x}, \href
  {http://adsabs.harvard.edu/abs/2001MNRAS.324L..17P} {324, L17}

\bibitem[\protect\citeauthoryear{{Puchwein}, {Sijacki}  \&
  {Springel}}{{Puchwein} et~al.}{2008}]{Puchwein2008}
{Puchwein} E.,  {Sijacki} D.,   {Springel} V.,  2008, \mn@doi [\apjl]
  {10.1086/593352}, \href {http://adsabs.harvard.edu/abs/2008ApJ...687L..53P}
  {687, L53}

\bibitem[\protect\citeauthoryear{{Puchwein}, {Springel}, {Sijacki}  \&
  {Dolag}}{{Puchwein} et~al.}{2010}]{Puchwein2010}
{Puchwein} E.,  {Springel} V.,  {Sijacki} D.,   {Dolag} K.,  2010, \mn@doi
  [\mnras] {10.1111/j.1365-2966.2010.16786.x}, \href
  {http://adsabs.harvard.edu/abs/2010MNRAS.406..936P} {406, 936}

\bibitem[\protect\citeauthoryear{{Quilis}, {Bower}  \& {Balogh}}{{Quilis}
  et~al.}{2001}]{Quilis2001}
{Quilis} V.,  {Bower} R.~G.,   {Balogh} M.~L.,  2001, \mn@doi [\mnras]
  {10.1046/j.1365-8711.2001.04927.x}, \href
  {http://ukads.nottingham.ac.uk/abs/2001MNRAS.328.1091Q} {328, 1091}

\bibitem[\protect\citeauthoryear{{Rees}}{{Rees}}{1984}]{Rees1984}
{Rees} M.~J.,  1984, \mn@doi [\araa] {10.1146/annurev.aa.22.090184.002351},
  \href {http://adsabs.harvard.edu/abs/1984ARA%26A..22..471R} {22, 471}

\bibitem[\protect\citeauthoryear{{Richardson}, {Zheng}, {Chatterjee}, {Nagai}
  \& {Shen}}{{Richardson} et~al.}{2012}]{Richardson2012}
{Richardson} J.,  {Zheng} Z.,  {Chatterjee} S.,  {Nagai} D.,   {Shen} Y.,
  2012, \mn@doi [\apj] {10.1088/0004-637X/755/1/30}, \href
  {http://adsabs.harvard.edu/abs/2012ApJ...755...30R} {755, 30}

\bibitem[\protect\citeauthoryear{{Ross} et~al.,}{{Ross}
  et~al.}{2012}]{Ross2012}
{Ross} N.~P.,  et~al., 2012, \mn@doi [\apjs] {10.1088/0067-0049/199/1/3}, \href
  {http://adsabs.harvard.edu/abs/2012ApJS..199....3R} {199, 3}

\bibitem[\protect\citeauthoryear{Rozo, Rykoff, Bartlett  \& Melin}{Rozo
  et~al.}{2015}]{Rozo2014a}
Rozo E.,  Rykoff E.~S.,  Bartlett J.~G.,   Melin J.~B.,  2015, \mn@doi [\mnras]
  {10.1093/mnras/stv605}, 450, 592

\bibitem[\protect\citeauthoryear{Ruan, McQuinn  \& Anderson}{Ruan
  et~al.}{2015}]{Ruan2015}
Ruan J.~J.,  McQuinn M.,   Anderson S.~F.,  2015, \mn@doi [Astrophys. J.]
  {10.1088/0004-637X/802/2/135}, 802, 135

\bibitem[\protect\citeauthoryear{{Rupke} \& {Veilleux}}{{Rupke} \&
  {Veilleux}}{2011}]{Rupke2011}
{Rupke} D.~S.~N.,  {Veilleux} S.,  2011, \mn@doi [\apjl]
  {10.1088/2041-8205/729/2/L27}, \href
  {http://ukads.nottingham.ac.uk/abs/2011ApJ...729L..27R} {729, L27}

\bibitem[\protect\citeauthoryear{{Rykoff} et~al.,}{{Rykoff}
  et~al.}{2012}]{Rykoff2012}
{Rykoff} E.~S.,  et~al., 2012, \mn@doi [\apj] {10.1088/0004-637X/746/2/178},
  \href {http://adsabs.harvard.edu/abs/2012ApJ...746..178R} {746, 178}

\bibitem[\protect\citeauthoryear{Rykoff et~al.}{Rykoff
  et~al.}{2014}]{Rykoff2013}
Rykoff E.~S.,  et~al., 2014, \mn@doi [Astrophys. J.]
  {10.1088/0004-637X/785/2/104}, 785, 104

\bibitem[\protect\citeauthoryear{{Rykoff} et~al.,}{{Rykoff}
  et~al.}{2016}]{Rykoff2016}
{Rykoff} E.~S.,  et~al., 2016, \mn@doi [\apjs] {10.3847/0067-0049/224/1/1},
  \href {http://adsabs.harvard.edu/abs/2016ApJS..224....1R} {224, 1}

\bibitem[\protect\citeauthoryear{{Saro} et~al.,}{{Saro}
  et~al.}{2015}]{Saro2015}
{Saro} A.,  et~al., 2015, \mn@doi [\mnras] {10.1093/mnras/stv2141}, \href
  {http://adsabs.harvard.edu/abs/2015MNRAS.454.2305S} {454, 2305}

\bibitem[\protect\citeauthoryear{{Saro} et~al.,}{{Saro}
  et~al.}{2016}]{Saro2016}
{Saro} A.,  et~al., 2016, preprint, \href
  {http://adsabs.harvard.edu/abs/2016arXiv160508770S} {} (\mn@eprint {arXiv}
  {1605.08770})

\bibitem[\protect\citeauthoryear{{Scannapieco}, {Thacker}  \&
  {Couchman}}{{Scannapieco} et~al.}{2008}]{Scannapieco2008}
{Scannapieco} E.,  {Thacker} R.~J.,   {Couchman} H.~M.~P.,  2008, \mn@doi
  [\apj] {10.1086/528948}, \href
  {http://adsabs.harvard.edu/abs/2008ApJ...678..674S} {678, 674}

\bibitem[\protect\citeauthoryear{{Sch{\"a}fer}, {Pfrommer}, {Hell}  \&
  {Bartelmann}}{{Sch{\"a}fer} et~al.}{2006}]{Schaefer2006}
{Sch{\"a}fer} B.~M.,  {Pfrommer} C.,  {Hell} R.~M.,   {Bartelmann} M.,  2006,
  \mn@doi [\mnras] {10.1111/j.1365-2966.2006.10622.x}, \href
  {http://adsabs.harvard.edu/abs/2006MNRAS.370.1713S} {370, 1713}

\bibitem[\protect\citeauthoryear{{Schaye} et~al.,}{{Schaye}
  et~al.}{2010}]{Schaye2010}
{Schaye} J.,  et~al., 2010, \mn@doi [\mnras]
  {10.1111/j.1365-2966.2009.16029.x}, \href
  {http://ukads.nottingham.ac.uk/abs/2010MNRAS.402.1536S} {402, 1536}

\bibitem[\protect\citeauthoryear{{Schaye} et~al.,}{{Schaye}
  et~al.}{2015}]{Schaye2015}
{Schaye} J.,  et~al., 2015, \mn@doi [\mnras] {10.1093/mnras/stu2058}, \href
  {http://adsabs.harvard.edu/abs/2015MNRAS.446..521S} {446, 521}

\bibitem[\protect\citeauthoryear{{Schneider} et~al.,}{{Schneider}
  et~al.}{2010}]{Schneider2010}
{Schneider} D.~P.,  et~al., 2010, \mn@doi [\aj] {10.1088/0004-6256/139/6/2360},
  \href {http://adsabs.harvard.edu/abs/2010AJ....139.2360S} {139, 2360}

\bibitem[\protect\citeauthoryear{{Sembolini} et~al.,}{{Sembolini}
  et~al.}{2016}]{Sembolini2016}
{Sembolini} F.,  et~al., 2016, \mn@doi [\mnras] {10.1093/mnras/stw800}, \href
  {http://adsabs.harvard.edu/abs/2016MNRAS.459.2973S} {459, 2973}

\bibitem[\protect\citeauthoryear{{Semboloni}, {Hoekstra}, {Schaye}, {van
  Daalen}  \& {McCarthy}}{{Semboloni} et~al.}{2011}]{Semboloni2011}
{Semboloni} E.,  {Hoekstra} H.,  {Schaye} J.,  {van Daalen} M.~P.,   {McCarthy}
  I.~G.,  2011, \mn@doi [\mnras] {10.1111/j.1365-2966.2011.19385.x}, \href
  {http://adsabs.harvard.edu/abs/2011MNRAS.417.2020S} {417, 2020}

\bibitem[\protect\citeauthoryear{{Sijacki}, {Springel}, {Di Matteo}  \&
  {Hernquist}}{{Sijacki} et~al.}{2007}]{Sijacki2007}
{Sijacki} D.,  {Springel} V.,  {Di Matteo} T.,   {Hernquist} L.,  2007, \mn@doi
  [\mnras] {10.1111/j.1365-2966.2007.12153.x}, \href
  {http://adsabs.harvard.edu/abs/2007MNRAS.380..877S} {380, 877}

\bibitem[\protect\citeauthoryear{{Sijacki}, {Vogelsberger}, {Genel},
  {Springel}, {Torrey}, {Snyder}, {Nelson}  \& {Hernquist}}{{Sijacki}
  et~al.}{2015}]{Sijacki2015}
{Sijacki} D.,  {Vogelsberger} M.,  {Genel} S.,  {Springel} V.,  {Torrey} P.,
  {Snyder} G.~F.,  {Nelson} D.,   {Hernquist} L.,  2015, \mn@doi [\mnras]
  {10.1093/mnras/stv1340}, \href
  {http://adsabs.harvard.edu/abs/2015MNRAS.452..575S} {452, 575}

\bibitem[\protect\citeauthoryear{{Silk} \& {Rees}}{{Silk} \&
  {Rees}}{1998}]{Silk1998}
{Silk} J.,  {Rees} M.~J.,  1998, \aap, \href
  {http://adsabs.harvard.edu/abs/1998A%26A...331L...1S} {331, L1}

\bibitem[\protect\citeauthoryear{{Simet}, {McClintock}, {Mandelbaum}, {Rozo},
  {Rykoff}, {Sheldon}  \& {Wechsler}}{{Simet} et~al.}{2016}]{Simet2016}
{Simet} M.,  {McClintock} T.,  {Mandelbaum} R.,  {Rozo} E.,  {Rykoff} E.,
  {Sheldon} E.,   {Wechsler} R.~H.,  2016, preprint, \href
  {http://adsabs.harvard.edu/abs/2016arXiv160306953S} {} (\mn@eprint {arXiv}
  {1603.06953})

\bibitem[\protect\citeauthoryear{{Smith} et~al.,}{{Smith}
  et~al.}{2012}]{Smith2012}
{Smith} M.~W.~L.,  et~al., 2012, \mn@doi [\apj] {10.1088/0004-637X/748/2/123},
  \href {http://adsabs.harvard.edu/abs/2012ApJ...748..123S} {748, 123}

\bibitem[\protect\citeauthoryear{{Soergel} et~al.,}{{Soergel}
  et~al.}{2016}]{Soergel2016}
{Soergel} B.,  et~al., 2016, \mn@doi [\mnras] {10.1093/mnras/stw1455}, \href
  {http://adsabs.harvard.edu/abs/2016MNRAS.461.3172S} {461, 3172}

\bibitem[\protect\citeauthoryear{{Spacek}, {Scannapieco}, {Cohen}, {Joshi}  \&
  {Mauskopf}}{{Spacek} et~al.}{2016a}]{Spacek2016b}
{Spacek} A.,  {Scannapieco} E.,  {Cohen} S.,  {Joshi} B.,   {Mauskopf} P.,
  2016a, preprint, \href {http://adsabs.harvard.edu/abs/2016arXiv161002068S} {}
  (\mn@eprint {arXiv} {1610.02068})

\bibitem[\protect\citeauthoryear{{Spacek}, {Scannapieco}, {Cohen}, {Joshi}  \&
  {Mauskopf}}{{Spacek} et~al.}{2016b}]{Spacek2016a}
{Spacek} A.,  {Scannapieco} E.,  {Cohen} S.,  {Joshi} B.,   {Mauskopf} P.,
  2016b, \mn@doi [\apj] {10.3847/0004-637X/819/2/128}, \href
  {http://adsabs.harvard.edu/abs/2016ApJ...819..128S} {819, 128}

\bibitem[\protect\citeauthoryear{{Springel}}{{Springel}}{2005}]{Springel2005}
{Springel} V.,  2005, \mn@doi [\mnras] {10.1111/j.1365-2966.2005.09655.x},
  \href {http://adsabs.harvard.edu/abs/2005MNRAS.364.1105S} {364, 1105}

\bibitem[\protect\citeauthoryear{{Springel}}{{Springel}}{2010}]{Springel2010}
{Springel} V.,  2010, \mn@doi [\mnras] {10.1111/j.1365-2966.2009.15715.x},
  \href {http://adsabs.harvard.edu/abs/2010MNRAS.401..791S} {401, 791}

\bibitem[\protect\citeauthoryear{{Springel}, {White}  \&
  {Hernquist}}{{Springel} et~al.}{2001}]{Springel2001}
{Springel} V.,  {White} M.,   {Hernquist} L.,  2001, \mn@doi [\apj]
  {10.1086/319473}, \href {http://adsabs.harvard.edu/abs/2001ApJ...549..681S}
  {549, 681}

\bibitem[\protect\citeauthoryear{{Sturm} et~al.,}{{Sturm}
  et~al.}{2011}]{Sturm2011}
{Sturm} E.,  et~al., 2011, \mn@doi [\apjl] {10.1088/2041-8205/733/1/L16}, \href
  {http://ukads.nottingham.ac.uk/abs/2011ApJ...733L..16S} {733, L16}

\bibitem[\protect\citeauthoryear{{Sunyaev} \& {Zeldovich}}{{Sunyaev} \&
  {Zeldovich}}{1970}]{SZ1970}
{Sunyaev} R.~A.,  {Zeldovich} Y.~B.,  1970, \mn@doi [\apss]
  {10.1007/BF00653471}, \href
  {http://adsabs.harvard.edu/abs/1970Ap%26SS...7....3S} {7, 3}

\bibitem[\protect\citeauthoryear{{Sunyaev} \& {Zeldovich}}{{Sunyaev} \&
  {Zeldovich}}{1972}]{SZ1972}
{Sunyaev} R.~A.,  {Zeldovich} Y.~B.,  1972, Comments on Astrophysics and Space
  Physics, \href {http://adsabs.harvard.edu/abs/1972CoASP...4..173S} {4, 173}

\bibitem[\protect\citeauthoryear{{Takita} et~al.,}{{Takita}
  et~al.}{2015}]{Akari_calibration}
{Takita} S.,  et~al., 2015, \mn@doi [\pasj] {10.1093/pasj/psv033}, \href
  {http://adsabs.harvard.edu/abs/2015PASJ...67...51T} {67, 51}

\bibitem[\protect\citeauthoryear{{Tegmark} \& {de Oliveira-Costa}}{{Tegmark} \&
  {de Oliveira-Costa}}{1998}]{Tegmark1998}
{Tegmark} M.,  {de Oliveira-Costa} A.,  1998, \mn@doi [\apjl] {10.1086/311410},
  \href {http://adsabs.harvard.edu/abs/1998ApJ...500L..83T} {500, L83}

\bibitem[\protect\citeauthoryear{{Tombesi}, {Mel{\'e}ndez}, {Veilleux},
  {Reeves}, {Gonz{\'a}lez-Alfonso}  \& {Reynolds}}{{Tombesi}
  et~al.}{2015}]{Tombesi2015}
{Tombesi} F.,  {Mel{\'e}ndez} M.,  {Veilleux} S.,  {Reeves} J.~N.,
  {Gonz{\'a}lez-Alfonso} E.,   {Reynolds} C.~S.,  2015, \mn@doi [\nat]
  {10.1038/nature14261}, \href
  {http://ukads.nottingham.ac.uk/abs/2015Natur.519..436T} {519, 436}

\bibitem[\protect\citeauthoryear{{Tormen}, {Bouchet}  \& {White}}{{Tormen}
  et~al.}{1997}]{Tormen1997}
{Tormen} G.,  {Bouchet} F.~R.,   {White} S.~D.~M.,  1997, \mn@doi [\mnras]
  {10.1093/mnras/286.4.865}, \href
  {http://adsabs.harvard.edu/abs/1997MNRAS.286..865T} {286, 865}

\bibitem[\protect\citeauthoryear{{Verdier}, {Melin}, {Bartlett}, {Magneville},
  {Palanque-Delabrouille}  \& {Y{\`e}che}}{{Verdier}
  et~al.}{2016}]{Verdier2015}
{Verdier} L.,  {Melin} J.-B.,  {Bartlett} J.~G.,  {Magneville} C.,
  {Palanque-Delabrouille} N.,   {Y{\`e}che} C.,  2016, \mn@doi [\aap]
  {10.1051/0004-6361/201527431}, \href
  {http://adsabs.harvard.edu/abs/2016A%26A...588A..61V} {588, A61}

\bibitem[\protect\citeauthoryear{{Vogelsberger} et~al.,}{{Vogelsberger}
  et~al.}{2014}]{Vogelsberger2014}
{Vogelsberger} M.,  et~al., 2014, \mn@doi [\nat] {10.1038/nature13316}, \href
  {http://adsabs.harvard.edu/abs/2014Natur.509..177V} {509, 177}

\bibitem[\protect\citeauthoryear{{White}, {Becker}, {Helfand}  \&
  {Gregg}}{{White} et~al.}{1997}]{FIRST}
{White} R.~L.,  {Becker} R.~H.,  {Helfand} D.~J.,   {Gregg} M.~D.,  1997, \apj,
  \href {http://adsabs.harvard.edu/abs/1997ApJ...475..479W} {475, 479}

\bibitem[\protect\citeauthoryear{{White} et~al.,}{{White}
  et~al.}{2012}]{White2012}
{White} M.,  et~al., 2012, \mn@doi [\mnras] {10.1111/j.1365-2966.2012.21251.x},
  \href {http://adsabs.harvard.edu/abs/2012MNRAS.424..933W} {424, 933}

\bibitem[\protect\citeauthoryear{{York} et~al.,}{{York}
  et~al.}{2000}]{York2000}
{York} D.~G.,  et~al., 2000, \mn@doi [\aj] {10.1086/301513}, \href
  {http://adsabs.harvard.edu/abs/2000AJ....120.1579Y} {120, 1579}

\bibitem[\protect\citeauthoryear{{de Haan} et~al.,}{{de Haan}
  et~al.}{2016}]{deHaan2016}
{de Haan} T.,  et~al., 2016, preprint, \href
  {http://adsabs.harvard.edu/abs/2016arXiv160306522D} {} (\mn@eprint {arXiv}
  {1603.06522})

\bibitem[\protect\citeauthoryear{{van Daalen}, {Schaye}, {Booth}  \& {Dalla
  Vecchia}}{{van Daalen} et~al.}{2011}]{vanDaalen2011}
{van Daalen} M.~P.,  {Schaye} J.,  {Booth} C.~M.,   {Dalla Vecchia} C.,  2011,
  \mn@doi [\mnras] {10.1111/j.1365-2966.2011.18981.x}, \href
  {http://adsabs.harvard.edu/abs/2011MNRAS.415.3649V} {415, 3649}

\makeatother
\end{thebibliography}


\appendix

\section{Additional data and data quality tests}
\label{app:dataquality}
In this appendix, we describe the additional FIR data that we have used for testing purposes.  
We further provide details about the tests that we have performed on the microwave and FIR data at the various frequencies before including them into our analysis. 

\subsection{IRAS/IRIS data}
IRAS has surveyed the infrared sky in four bands, with band centres between 12 and 100 ${\mathrm{\upmu m}}$, and a resolution of approximately $4\arcmin$. 
Here we make use of the Improved Reprocessing of the IRAS Survey (IRIS) maps by \cite{IRIS}, which --- amongst others --- feature improvements in calibration and destriping over the original IRAS maps.\footnote{The IRIS maps are available at \url{https://www.cita.utoronto.ca/~mamd/IRIS/IrisOverview.html} both as $12.5~\deg \times 12.5~\deg$ cutouts and in the \HEALPIX~format with $N_\mathrm{side}=2048$.}
The IRAS 100~${\mathrm{\upmu m}}$ band is already measuring the falling part of the dust spectrum. 
The higher-frequency bands of IRIS are more sensitive to dust at higher temperatures and therefore do not add additional information to our measurement.
As additionally the 60, 25 and 12~${\mathrm{\upmu m}}$ bands show a stronger contamination by residual zodiacal light, we only consider the IRIS 100~${\mathrm{\upmu m}}$ map.

\subsection{Random points null test}
\label{app:randompoints}
We perform a null test by stacking on random positions in the same footprint.
The latter are generated as follows:
(1) We generate a low-resolution \HEALPIX~map of the QSO density, similar to the one in Fig.~\ref{fig:qsodensity}. (2) Next, we create a large number of `candidate points' which are randomly distributed on the sphere.
(3) We finally select $N_\mathrm{QSO}$ final points randomly from the candidates. The probability of any given candidate point to be selected is given by the appropriately normalized QSO surface density at its position in the original catalogue.

\begin{figure}
	\centering
	\includegraphics[width=\columnwidth]{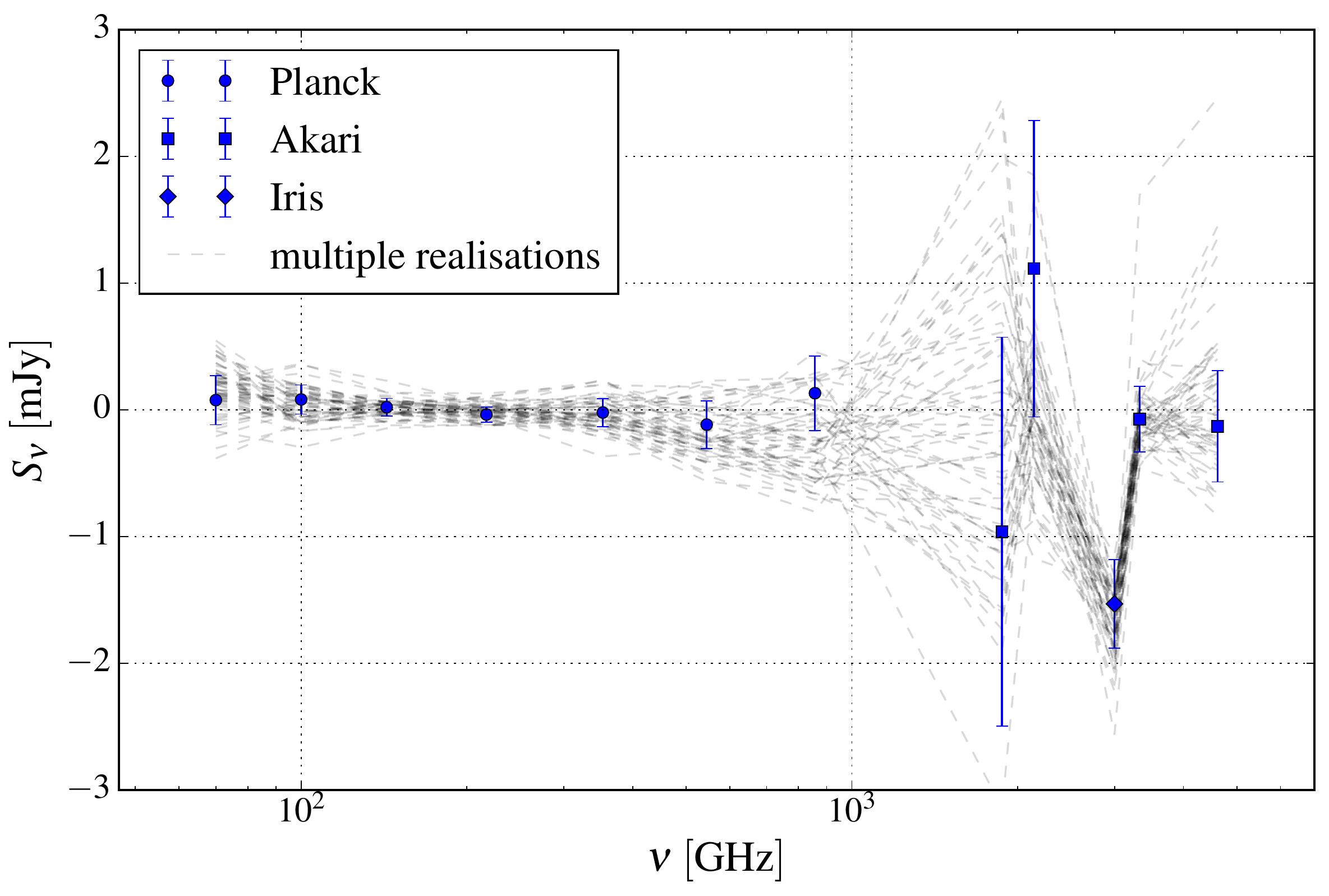}
	\caption{Random points null test: we show here the results of the stacking on random positions in the map. The blue points show one representative realization, with uncertainties estimated as in the original measurement. The shaded dashed lines indicate the results of 60 further random realizations.}
	\label{fig:nulltest}
\end{figure}

We then process the random points through the same measurement pipeline as the original catalogue and show the resulting flux density measurement in Fig.~\ref{fig:nulltest}.
From the results of this tests, we draw the following conclusions: 
all \Planck frequency points pass the null test; the same holds true for the AKARI 90~${\mathrm{\upmu m}}$ band. 
The results from the AKARI 160, 140, and 65~${\mathrm{\upmu m}}$ bands are also consistent with no bias, but these bands show large scatter between individual realizations, reflecting their high noise levels and significant calibration uncertainties.
This provides additional justification to our decision not to include them in the main analysis.

Most notably, however, the IRIS 100~${\mathrm{\upmu m}}$ band fails the null test, revealing a statistically significant offset of around $-1.5~\mathrm{mJy}$.
Amongst the possible reasons for this result are residual striping errors or calibration uncertainties.
However, a more in-depth investigation of this, for example by a detailed comparison of AKARI and IRIS at the map level, is beyond the scope of this paper.
We therefore do not include the IRIS data in our main analysis.
This does not have a significant impact on our ability to constrain the dust parameters,
as there is significant overlap in the instrumental bandpasses of the IRIS 100~${\mathrm{\upmu m}}$ and AKARI 90~${\mathrm{\upmu m}}$ bands (see Fig.~\ref{fig:bandpasses}). 

It is worth noting that the successful null test at low frequencies also demonstrates that our result for the Compton-$y$ parameter is not significantly affected by a stacking bias.
The latter can, in principle, arise because the measured $y$ receives contributions from all objects along the same line of sight, not only from the QSO (or cluster) in our catalogue. 
However, this would result in a signal also in sightlines that do not contain such an object.
Our null test demonstrates that this bias is well below the statistical uncertainties in our analysis.

\subsection{Impact of discarded data}
\label{app:impact}
We next demonstrate the impact that the discarded data can have on our measurement. 
As a reminder, we have excluded the AKARI 160, 140 and 90 ${\mathrm{\upmu m}}$ bands because of their uncertain calibration at low intensities (see Table~\ref{tab:maps} and \citealt{Akari_calibration}).
Furthermore, we have not used the IRIS 100 ${\mathrm{\upmu m}}$ band because it does not pass the stacking null test.
Here we now repeat our analysis including these frequencies.

\begin{figure}
	\centering
	\includegraphics[width=\columnwidth]{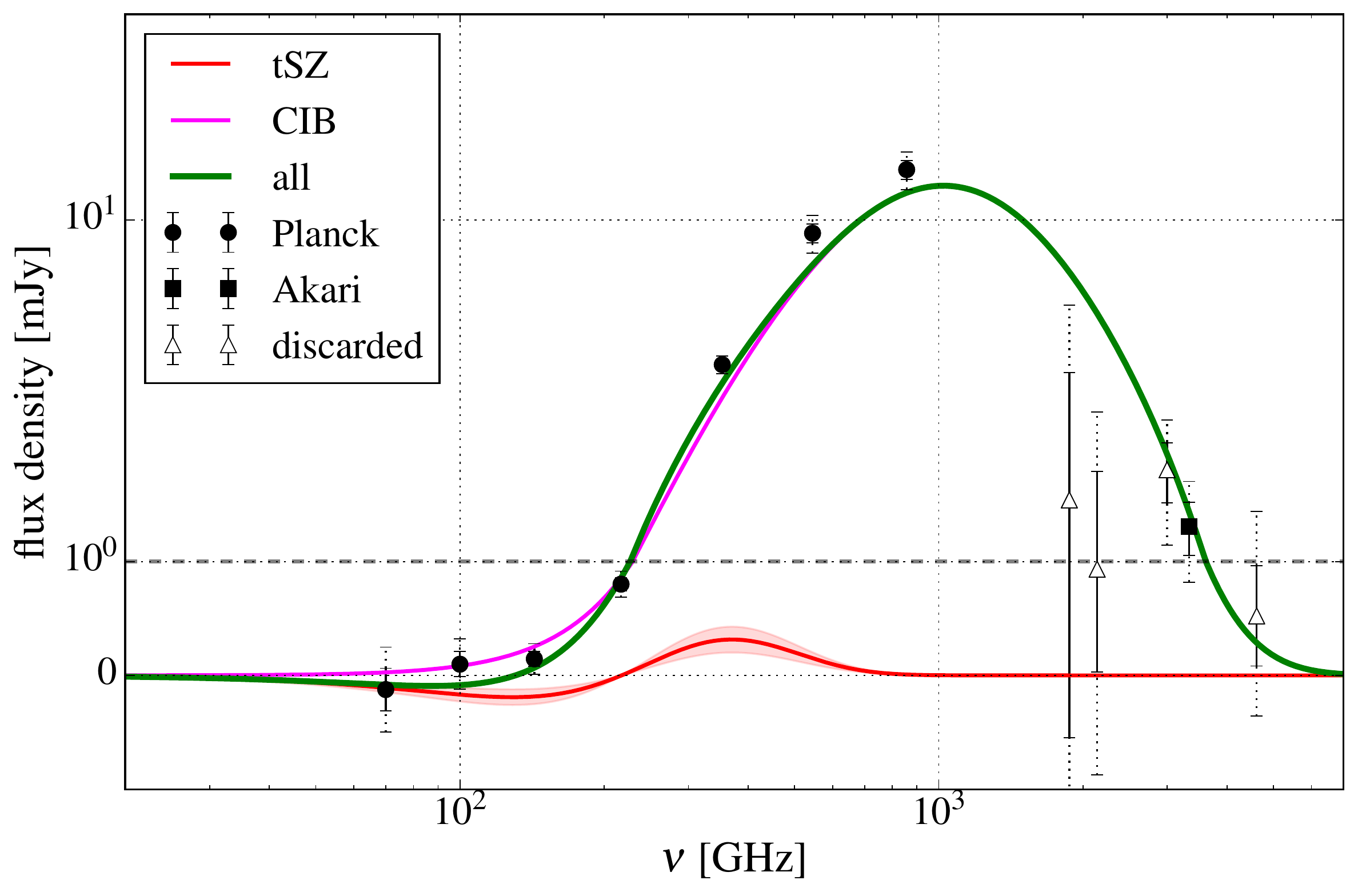}
	\caption{Impact of the discarded data on our analysis: we show here the best-fitting SED if the discarded frequencies are included in the analysis. We denote the latter by the open triangles, from low to high frequencies they are: AKARI 160 and 140 ${\mathrm{\upmu m}}$, IRIS 100 ${\mathrm{\upmu m}}$, and AKARI 65 ${\mathrm{\upmu m}}$. Including these frequencies significantly affects the dust solution, and leaves more room for an SZ component. However, the quality of the fit is poor.}
	\label{fig:discarded_bestfit}
\end{figure}

Adding only the IRIS 100 ${\mathrm{\upmu m}}$ band only marginally affects our result, as it is consistent with the SED we have determined from the seven \Planck bands and AKARI 90 ${\mathrm{\upmu m}}$.
When additionally adding the three discarded AKARI bands, we assume that their flux calibration uncertainty is twice the lower limit we quote in Table~\ref{tab:maps}.
Despite their large uncertainties, we find that including these three frequencies pulls the dust solution away from the best fit in the main analysis:
we now obtain $\beta_d = 1.44^{+0.27}_{-0.24}$ and \mbox{$T_d = 33.9^{+3.5}_{-3.3}~\mathrm{K}$}.
This changes the shape of the dust spectrum at low frequencies, so that now there is room for an SZ component with \mbox{$\hat{Y} = (10.4 \pm 3.8) \times 10^{-6}~\mathrm{arcmin}^2$}, or \mbox{$E_\mathrm{th} = (14.6 \pm 5.3) \times 10^{60}~\mathrm{erg}$}.
At the same time, however, the fit both to the higher-frequency \Planck points and the lowest-frequency AKARI points is poor, which would already cause us to reject this solution on goodness-of-fit considerations only. We show this in Fig.~\ref{fig:discarded_bestfit}.
Therefore, we stress that this should not be interpreted as a significant detection of an SZ signal when adding additional data. 
Rather, we consider it a demonstration how sensitive the SZ result is on the dust solution, and how easily the latter is affected by the quality of the used data.

\bsp	
\label{lastpage}
\end{document}